\documentclass{aa}

\usepackage{graphicx}
\usepackage{txfonts}
\usepackage{aalongtable}

\begin{document}

	\title{A detailed study of the L1641N star formation region
		\thanks{Based on observations made with the Nordic Optical Telescope, operated on the island of
			La Palma jointly by Denmark, Finland, Iceland, Norway, and Sweden, in the Spanish Observatorio
			del Roque de los Muchachos of the Instituto de Astrofisica de Canarias.}\fnmsep
		\thanks{This work is based in part on observations made with the Spitzer Space Telescope, which is operated
			by the Jet Propulsion Laboratory, California Institute of Technology under a contract with NASA.}\fnmsep
		\thanks{Based on observations with ISO, an ESA project with instruments funded by ESA Member
	        States (especially the PI countries: France, Germany, the Netherlands and the United Kingdom)
		and with the participation of ISAS and NASA.}
	}

	\author{M.~G\aa lfalk \inst{1} \and
		G.~Olofsson \inst{1}
		}
    
	\offprints{\\M. G\aa lfalk, \email{magnusg@astro.su.se}}

	\institute{Stockholm Observatory, Sweden				
	      }
	\date{}

\abstract
{}
{We search for young stellar objects (YSOs) in the L1641N cluster and characterize the star formation activity through determination of
the age distribution, mass function, spatial distribution, and the star formation history.
\vspace{2mm}
}
{Multi-wavelength broad band photometry both from space and the ground are used to look for IR excess in order to separate field stars from YSOs
and to sample the spectral energy distributions. Space-based observations were obtained using the ISO satellite (ISOCAM)
in two filters, centred at 6.7 and 14.3\,$\mu$m, and Spitzer (IRAC) at 3.6, 4.5, 5.8, and 8.0\,$\mu$m.
Our ground-based observations were made with the Nordic Optical Telescope (NOT) using ALFOSC ($I$ band), NOTCam ($J$, $K_S$ and 2.12\,$\mu$m H$_2$), and
SIRCA (L'). More than 50 of the brightest $I$-band sources were then studied with follow-up optical spectroscopy (5780--8340 \AA) to check
for signs of accretion (H$\alpha$ in emission) and youth (Li\,I\,$\lambda6707$ in absorption) and to determine their effective temperatures.
By comparing theoretical evolution tracks with our YSO sample in the H-R diagram, we calculated an age, luminosity, and mass distribution.
\vspace{2mm}
}
{We detect a total of 216 (Spitzer or $I$ band) sources in L1641N, 89 of which are YSO candidates.
Most of the spectra are of M-type with H$\alpha$ strongly in emission, and many have Li\,6707 in absorption. The four brightest $I$ band sources
(F and G stars) are suggested as foreground stars, and the L1641N IRAS source is shown to be the combined flux of at least four sources.
We find that the interstellar extinction is well-fit in the optical and near-IR by a power law
with an exponent of 1.58, although in the mid-IR the Spitzer observations show a higher extinction than expected from theory. 
The median age of the YSO sample is $\sim$1\,Myr and the resulting MF has a flat distribution for low masses down to the completeness limit.
There is evidence of a constant star formation rate of one star in 3.7\,$\times$\,10$^4$\,yr during the past few Myr.
We find 11 sources older than 10\,Myr and a spatial separation between younger and older YSOs, suggesting that many of the older stars formed in
L1641N could have left the cluster, giving the appearance of an increased star formation rate with time.
}
{}


\maketitle

\section{Introduction}

Lynds 1641 (L1641) is a dark molecular cloud, located in the southern part of our nearest giant molecular cloud Orion A (d$\sim$450\,pc).
It is a site of active star formation with a lot of very young stars. L1641 is therefore ideal for studying both the stars themselves and
the related Herbig-Haro (HH) outflows.
 
Attention was drawn to the northern part of the cloud, simply called L1641N, in the mid 80s with the discovery of
a molecular outflow (Fukui et al.~\cite{fukui86}) at the position of the mid-IR bright source IRAS 05338-0624. Many studies have
investigated the nature of this source, Fukui et al. (\cite{fukui88}) found well-separated lobes to the north (blue-shifted)
and south (red-shifted) of the IRAS source using CO observations of higher resolution. Further observations have
been made through molecular line studies (e.g. Chen et al.~\cite{chen96}, Sakamoto et al.~\cite{sakamoto}, Stanke \& Williams~\cite{stanke07}), Optical
and near-IR surveys (e.g. Strom et al.~\cite{strom89}, Hodapp \& Deane~\cite{hodapp93}, Chen et al.~\cite{chen93}) and in the mid-IR (Ali et al.~\cite{ali}).

L1641N is presently the most active site of low-mass star formation in the L1641 molecular cloud and has one of
the very highest concentrations of HH objects known anywhere in the sky (Reipurth el al.~\cite{reipurth}).
HH shocks in these flows have been observed in both the near-IR 2.12 $\mu$m line of H$_2$
(Stanke et al.~\cite{stanke}) and in the usual optical shock lines H$\alpha$ and the [SII] doublet (Reipurth et al.~\cite{reipurth} and
Mader et al.~\cite{mader}).

In this article we make the most detailed mid-IR study yet of L1641N using both our ISO observations and the much more sensitive and higher
resolution Spitzer data to survey the region for YSO candidates and determine the mass function in L1641N. Even though the Spitzer
observations cover a much larger region, we have restricted ourselves to the region we selected for our ISOCAM observations in order to get
a more complete wavelength coverage and because L1641N is the most dense, active part of L1641. We have also made additional near-IR and optical
ground-based imaging and taken optical spectra of many of the objects, to be able to confirm the YSO status and to estimate their
effective temperatures. Finally, as part of this study, we have also made deep narrow band observations using a 2.12\,$\mu$m H$_2$S(1) filter
to study the numerous H$_2$ sources in the region (G\aa lfalk \& Olofsson \cite{galfalk07}).

\section{Observations and reductions}

\subsection{Space based}

Mid-IR observations were carried out using six filters centred in the wavelength range 3.6 - 14.3 $\mu$m using ISOCAM onboard
the ISO satellite and IRAC onboard the Spitzer Space Telescope.

\subsubsection{ISO satellite}

We have used the ISOCAM instrument onboard the ISO satellite (60\,cm primary mirror) to observe L1641N at mid-IR wavelengths as part of
a mid-IR star formation survey (principal investigator L.~Nordh). Our observations were made using two broad-band filters at 6.7 and 14.3\,$\mu$m.
Centred on the IRAS source 05338-0624 they cover an area of 8\farcm35 $\times$ 7\farcm60 $\sim$\,0.0176\,sq.deg.~(63.5\,sq.arcmin).

All imaging was done at a PFOV of 3$\arcsec$ and the integration time of each transmitted $32\times32$ pixel frame was 2.1\,s.
For both filters the full 10$\times$9 mosaic of 90 images uses 1550 frames, thus at least 17 frames are combined at each position, forming images
with a temporal history which is used to e.g. remove cosmic rays and source transients (ghost sources) from previous mosaic positions. There
is also a spatial redundancy in both equatorial directions due to image overlaps, which is necessary to avoid having a source imaged on the same array
pixels, especially on the unfortunate dead column (No\,24). This of course also means that we have an increased total exposure time and temporal
history in overlapping regions.

For the data reduction we used the CIA V4.0 package (Ott et al. \cite{ott}; Delaney et al. \cite{delaney}) and
the SLICE package (Simple \& Light ISOCAM Calibration Environment) accessed from inside CIA.
The CIA reduction steps consisted of (in order): Extracting useful observations, dark correction (Vilspa dark 
model, Biviano et al. \cite{biviano}), glitch removal (multiresolution median transform, Starck et al. 
\cite{starck}), short transient correction (Fouks-Schubert model, Coulais \& Abergel \cite{coulais}), 
flatfielding (constant median flatfield from observations). We also used a set of our own programs for source
detection and photometry.

As a last step, all frames were projected into mosaics in both filters, followed by point source detection and aperture photometry.
All point sources were traced back to their corresponding original frames for temporal and spatial 
verification (in order to exclude remaining artefacts). For the photometry, the point spread function was used
to correct for the flux outside of each aperture. In total 41 sources were detected, with photometry possible for all sources
at 6.7\,$\mu$m and for 31 of these at 14.3\,$\mu$m.

Even though the integration time was 2.1\,s for each frame and there are about 17 frames for each image, the 
total exposure time for a given pixel in the mosaic varies between 36\,s and 143\,s due to the varying number of overlaps.
Thus, it is expected that the faintest sources will generally be detected away from the mosaic edges. 
From mean temporal and sky noise calculations of all detected ISOCAM sources in L1641N, the photometric uncertainties
are estimated to be ($1\sigma$) 2.9\,mJy at 6.7\,$\mu$m and 5.4\,mJy at 14.3\,$\mu$m which also approximately 
represent the detection limits. However, such faint sources are only detected in low nebulosity 
regions that are free of artefacts. On the other hand the surveyed region is not 
completely mapped down to the $1\sigma$ level, since it is possible that even sources brighter than 
$3\sigma$ (8.7\,mJy and 16.2\,mJy respectively) may have been unnoticed due to varying nebulosity, 
glitches, memory effects, uncovered dead columns and source confusion close to very bright sources.

In order to convert mJy fluxes into 6.7\,$\mu$m and 14.3\,$\mu$m magnitudes, we have used the zero-magnitude flux densities
given in Blommaert et al.~(\cite{blommaert}) which gives the following relations:

\vspace{2mm}

\( \begin{array}{lclcl}
	m_{6.7}  &  =  &  12.39  &  -  &  2.5\log_{10} F_{6.7}  \\
	m_{14.3} &  =  &  10.74  &  -  &  2.5\log_{10} F_{14.3}
   \end{array} \)

\vspace{2mm}

\subsubsection{Spitzer Space Telescope}


\begin{figure}
	\centering
	\includegraphics[width=8.8cm]{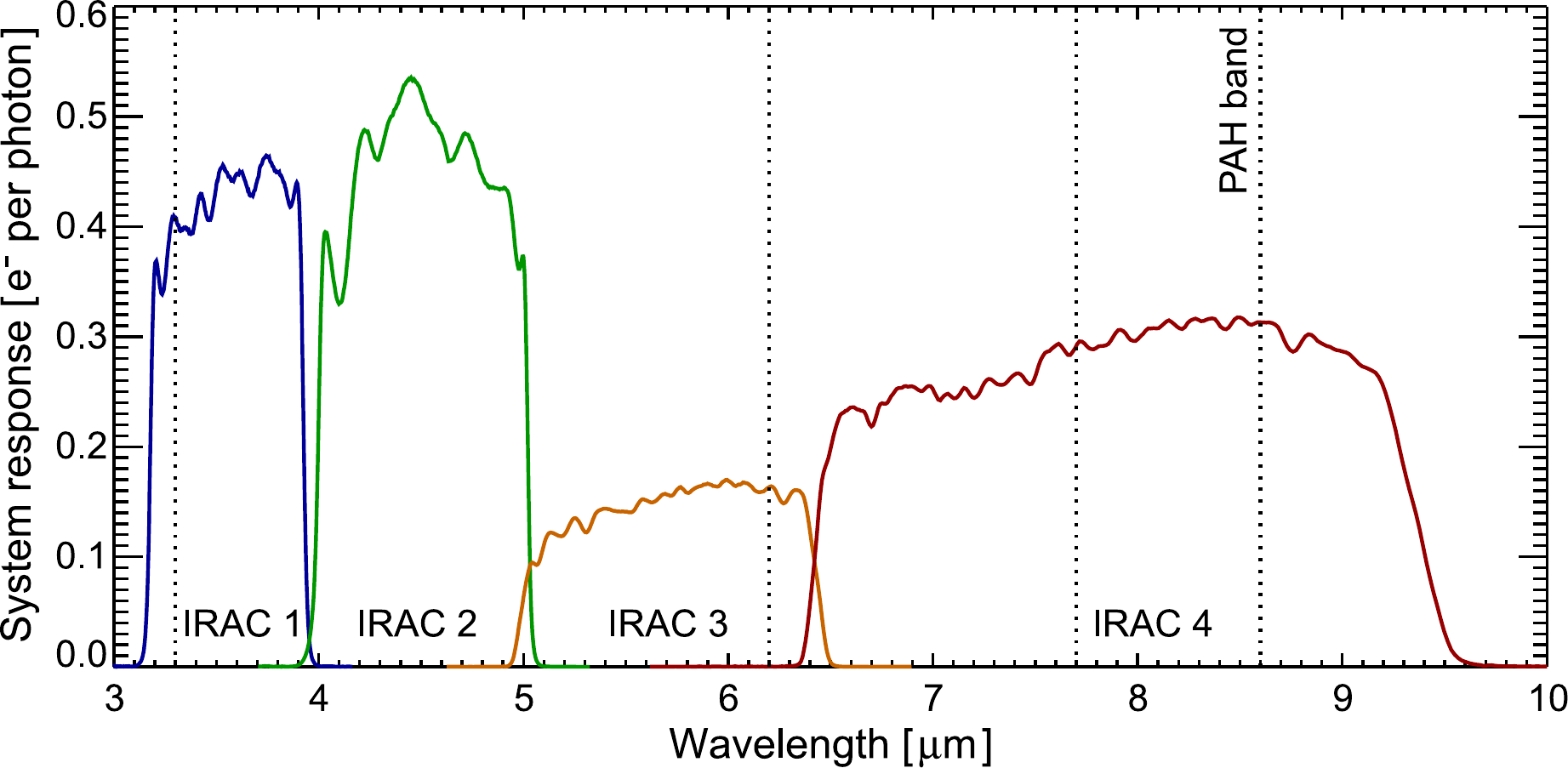}
	\caption{System transmission curves of IRAC channels 1-4 (Spitzer). The four major bands of PAH emission are marked by dotted
		 lines. IRAC channel 4 (centred at 8.0\,$\mu$m) contains two strong bands while images in channel 2 have very smooth
		 backgrounds as they are essentially free from PAH emission.
	}
	\label{IRACPAHs}
\end{figure}


Spitzer is an IR space telescope that carries a 85\,centimeter cryogenic telescope and three cryogenically cooled science instruments, one
of these is the Infrared Array Camera (IRAC) that provides simultaneous 5\farcm2\,$\times$\,5\farcm2 images in four
channels, centred at 3.6, 4.5, 5.8 and 8.0 $\mu$m (Fig.~\ref{IRACPAHs}).
Each channel is equipped with a 256 $\times$ 256 pixel detector array with a pixel size of about 1\farcs2\,$\times$\,1\farcs2.

The Spitzer data used in this paper (see Table~\ref{Spitzer}) was obtained from the Spitzer Science Archive
using the Leopard software. All the data is part of the program ``An IRAC Survey of the L1630 and L1641 (Orion) Molecular
Clouds'' (Prog.ID 43) with G.\,Fazio being the P.I. All data had been reduced to the Post-Basic Calibrated Data (pbcd) level,
however since there was still a lot of artefacts, background variations and varying orientation in the mosaics we used
our own in-house routines to reduce the data further. This was possible since we had eight overlapping mosaics to merge in each
channel, including cosmic ray removal, de-striping (most likely pick-up noise), removing the large background variations between regions
in the same mosaic and also between all mosaics,
marking bad pixels not to be used further, removing ghost-effects from bright sources, de-rotation, shifting, adding and making a
composite colour image. This required some work, but since we only needed to reduce the L1641N region (and these mosaics are huge) we
could concentrate on just that part - which because of all mosaics having different rotations meant first finding L1641N in each one and
then cropping them to manageable sizes.

Table~\ref{Spitzerfm} gives the zero-magnitude flux densities, $F(0)$, that have been used for the flux-magnitude conversion in the four channels
(also see e.g. Reach et al. \cite{reach}). The conversion for a measured flux $F_{i}$ in channel i is then $m_{i}~=~2.5~log_{10}(F(0)/F_{i})$

\begin{table}
	\caption{Spitzer archive data used.}
	\label{Spitzer}
	\begin{tabular}{cccc}
	\hline
	\noalign{\vspace{0.5mm}}
	Key & Type & Released & Scheduled \\
	\noalign{\vspace{0.5mm}}
	\hline
	4101888 & IRAC Mapping & 2005-07-27 & 2004-02-16 \\
	4102912	& IRAC Mapping & 2005-07-27 & 2004-02-17 \\
	4101376 & IRAC Mapping & 2005-07-27 & 2004-02-18 \\
	4102400 & IRAC Mapping & 2005-07-27 & 2004-02-18 \\
	4101632 & IRAC Mapping & 2005-10-27 & 2004-10-08 \\
	4102656	& IRAC Mapping & 2005-10-27 & 2004-10-08 \\
	4102144 & IRAC Mapping & 2005-11-17 & 2004-10-27 \\
	4103168 & IRAC Mapping & 2005-11-17 & 2004-10-27 \\
	\hline
	\end{tabular}
\end{table}

\begin{table}
	\caption{Spitzer zero-magnitude flux densities.}
	\label{Spitzerfm}
	\begin{tabular}{cc}
	\hline
	\noalign{\vspace{0.5mm}}
	Channel & F(0) [Jy] \\
	\noalign{\vspace{0.5mm}}
	\hline
	Channel 1 (3.6 $\mu$m) &  280.9 $\pm$ 4.10 \\
	Channel 2 (4.5 $\mu$m) &  179.7 $\pm$ 2.60 \\
	Channel 3 (5.8 $\mu$m) &  115.0 $\pm$ 1.70 \\
	Channel 4 (8.0 $\mu$m) &  64.13 $\pm$ 0.94 \\
	\hline
	\end{tabular}
\end{table}

\subsection{Ground based}

All photometry and spectroscopy were carried out at the 2.56\,m Nordic Optical Telescope (NOT) located at 2382\,m (7815 feet) above sea level on
the island of La Palma, Canary Islands. We used three different cameras: ALFOSC (Optical), NOTCam (Near-IR), SIRCA (Near and mid-IR). For
details, see the subsections below.

\subsubsection{Optical photometry ($I$ band)}

The $I$ band observations were obtained on Dec.~04 2003 using the ALFOSC (Andalucia Faint  Object Spectrograph and Camera) with an
average seeing of about 1\farcs1 (in the $I$ band).
This instrument has a $2048 \times 2048$ CCD and at a PFOV of 0\farcs188/pixel it has a FOV of
about 6\farcm4\,$\times$\,6\farcm4. A $5\times5$ mosaic was made with individual exposure times of 60\,s using step sizes
of $23\arcsec$ and $30\arcsec$ in RA and Dec respectively, meaning a total exposure time of 25 minutes throughout most of the mosaic.
Aperture photometry was then carried out on all visible sources using our own ALFOSC IDL-scripts.

\subsubsection{Optical spectroscopy (5780--8340 \AA)}

ALFOSC long-slit spectra were obtained during six nights at the NOT, Dec.~02-04 2003 and Jan.~08-10 2005, using grism \#8 which yields spectra
in the wavelength range 5780--8340 \AA.  We used a fairly wide slit (1\farcs2) in order to minimize slit
losses (the seeing was on average $\sim$1\farcs0).

The selected wavelength range covers some very interesting lines for low-mass star formation, as they are signatures
of youth (Li\,6707 in absorption), accretion (H$\alpha$ in emission) and a number of TiO and VO bands that can be effectively used in
the spectral classification of late-type stars.
Since these sources are relatively faint in the optical we had to use total
exposure times ranging from 40--120\,min for many sources, but whenever possible we tried to put 2--3 stars on the same slit. The total exposure time
used for the spectroscopy is about 35\,hours.

Standard subtraction of bias and dark current was carried out, however, the spectroscopic flatfielding was complicated by strong fringing. Great care had
to be taken in order to correctly remove this fringing, which otherwise becomes a problem in the red part of the spectrum. Halogen flats were taken
at each source position directly after each observation since the pattern varies with position and time. Also, a constant was fit and added prior
to flatfielding in order to fine-tune the pattern removal because of background variations. For full-size frames this constant was found from
the overscan region, and iterated in sub-frame exposures.

For wavelength calibration we used 29 Ne lines in long and short exposures of
a Ne calibration lamp to be able to measure both bright and faint lines. A polynomial fit with all these Ne lines included was made at each spatial
position along the slit (meaning 2048 solutions). These fits were also used to unbend the spectra in the wavelength direction. Distortion
correction was made in the spatial direction using calibration stars at different positions on the slit. For each spectrum a constant shift was also
applied to correct for small shifts of the whole wavelength scale with different alt-az positions of the telescope (for this we used several skylines).

\subsubsection{Near-IR photometry ($J$, $K_S$, $L$' and 2.12\,$\mu$m H$_2$)}

The $J$ band images were observed with NOTCam (near-IR camera) on the night of Dec.~10 2003 using 70 exposures of 48\,s with random
small-step dithering in between. These observations cover the central part of L1641N as illustrated by the green square
in Fig.~\ref{Spitzer124} (NOTCam has a square FOV of 4$\arcmin$ and a pixel size of 0\farcs235) and was made as a
complement to the 2MASS $J$ photometry to measure fainter sources. There were thin cirrus covering the sky that night so we have used
relative photometry with 2MASS sources for calibration.

We have selected two regions for $K_S$ band imaging (48 and 24 minutes total exposure times, respectively). 
These observations were made together with deep 2.12\,$\mu$m H$_2$ S(1) observations (175 and 100 minutes) as part of a related project to
measure the proper motions of H$_2$ objects in L1641N (G\aa lfalk \& Olofsson \cite{galfalk07}).
In the present paper we focus on the $K_S$ photometry, for which we detect point-sources down to $K_S$\,$\sim$\,19.5\,mag. The observations are also used to detect wide double sources and for aperture photometry of these.
The observations were made on two photometric nights, Dec.~13-15 2005, with an average seeing of 0\farcs75 (0\farcs60--0\farcs85) using
NOTCam with the newly installed science-grade array. NOTCam is an HAWAII 1024\,$\times$\,1024\,$\times$\,18.5\,$\mu$m pixels HgCdTe array with a
field-of-view of 4\farcm0\,$\times$\,4\farcm0.
Both regions include the centre of L1641N, which thus has an increased total exposure time. The first field is centred on a
position ($05^{h}36^{m}24.07^{s}$, $-06^{\circ}23\arcmin01\farcs9$, Epoch 2000) close to the brightest mid-IR source in L1641N (No.\,172 in
this paper).
The second field, centred on ($05^{h}36^{m}13.94^{s}$, $-06^{\circ}20\arcmin52\farcs5$, Epoch 2000) images the NW and central part of L1641N.

Besides the usual reduction steps of near-IR imaging, we have used our NOTCam model (G\aa lfalk \cite{notcamdist}) to correct for image distortion
and some other in-house routines (written in IDL) to find and remove bad pixels, shift-add images and to remove all the dark stripes that results from
lowered sensitivity after a bright source has been read out of the detector.


\begin{figure}
	\centering
	\includegraphics[width=8.8cm]{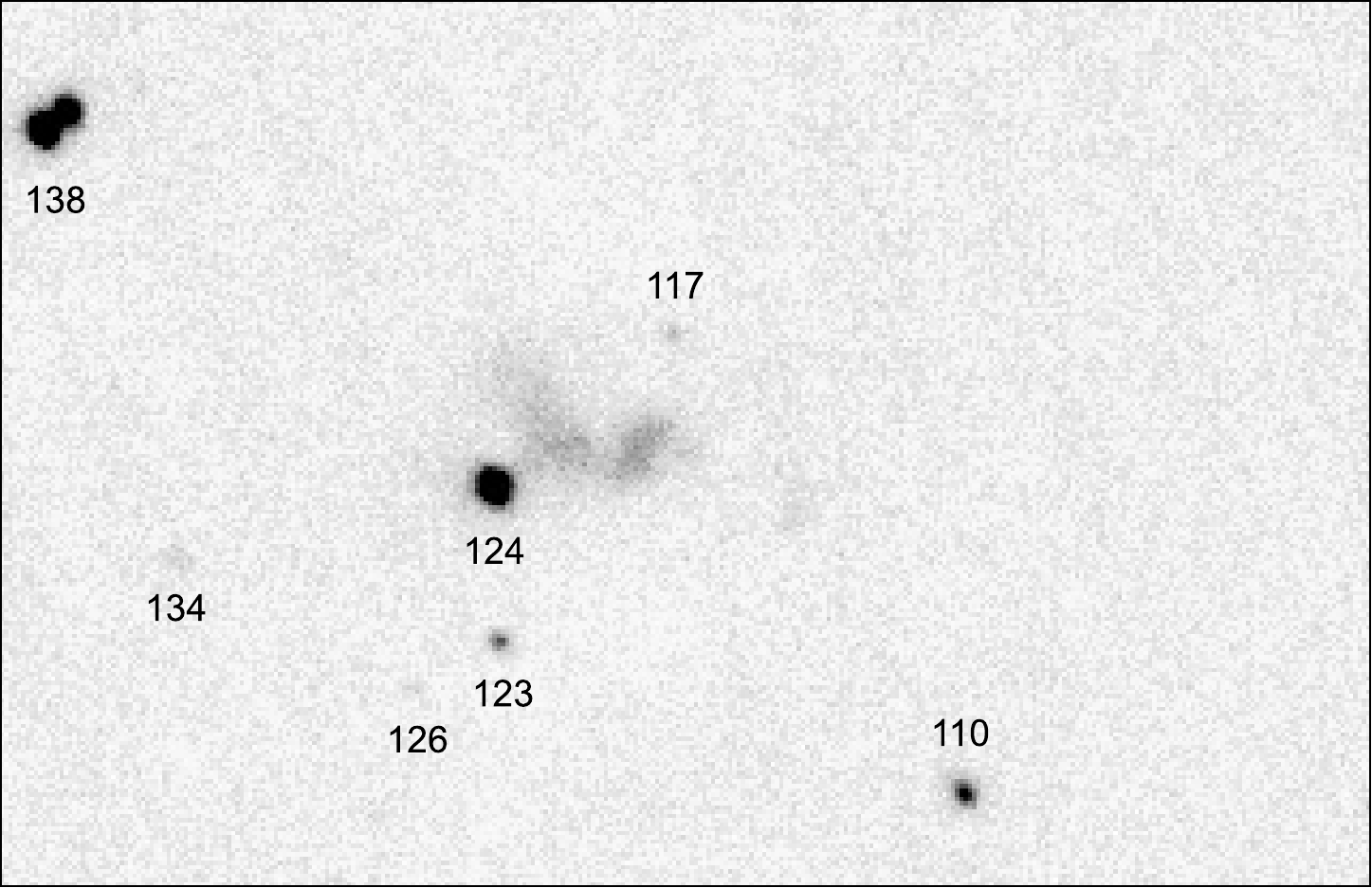}
	\caption{SIRCA L' band image of central L1641N. The field shown has a size of 73$\arcsec$ $\times$ 47$\arcsec$.
	}
	\label{Limage}
\end{figure}



\begin{table}
	\caption{L' band photometry of the central region.}
	\label{Ltable}
	\begin{tabular}{lcl}
	  \hline
        \noalign{\vspace{0.5mm}}
        No. & L' (mag) & Comment \\
 	\noalign{\vspace{0.5mm}}
        \hline

	\noalign{\vspace{1.0mm}}
	{\bf 110} & 11.27 $\pm$ 00.05 & Double source \\
	\noalign{\vspace{1.0mm}}
	{\bf 117} & 13.50 $\pm$ 00.43 & \\
	\noalign{\vspace{1.0mm}}
	{\bf 123} & 12.34 $\pm$ 00.13 & Deeply embedded ($K_S - L' = 5.05$) \\
	\noalign{\vspace{1.0mm}}
	{\bf 124} & 09.56 $\pm$ 00.02 & Deeply embedded ($K_S - L' = 6.84$) \\
	\noalign{\vspace{1.0mm}}
	{\bf 126} & 14.8: & \\
	\noalign{\vspace{1.0mm}}
	{\bf 134} & 13.45 $\pm$ 00.40 & \\
	\noalign{\vspace{1.0mm}}
	{\bf 138 a} & 09.66 $\pm$ 00.02 & \\
	{\bf 138 b} & 10.44 $\pm$ 00.03 & \\

	\noalign{\vspace{1.0mm}}
	\hline

	\end{tabular}
\end{table}


To probe the central $\sim$\,1$\arcmin$ of L1641N (Fig.~\ref{Limage}) we have used our own instrument SIRCA (Stockholm IR Camera) for observations
in the $L'$ band (3.8\,$\mu$m). Observations were carried out 20-21 December 2002. SIRCA uses a 320$\times$256 pixel InSb array, 
yielding a FOV of $70\arcsec \times 56\arcsec$ when mounted on the NOT.
In these observations we used an individual exposure time of 200 ms, 5 readouts per chopper
position, 5 chopper cycles and 54 ABBA nodding cycles of the telescope. This gives a total exposure time of
0.2$\times$5$\times$2$\times$5$\times$4$\times$54 = 2160 seconds (36 minutes). The weather was not fully photometric that night so we had to use
relative photometry from a previous test night (which of course had a much shorter total exposure time) to be on the safe side.
Therefore the uncertainties may seem larger than the exposure time would suggest. Flatfielding was done differentially by chopping between the cold
interior of the camera and the much hotter mirror cover of the telescope.

\subsubsection{Additional observations}

We have used the 2MASS archive (Cutri et al. \cite{cutri}) for regions of L1641N where we lacked $J$ or $K_S$ photometry.
Given that this is an all-sky survey, it cannot have the high sensitivity and resolution obtainable from target specific observations. It
still does a fairly good job filling in photometry of sources outside of our selected $J$ and $K_S$ fields.


\section{Results and discussion}


\subsection{General impression and mosaics}


\begin{figure*}
	\centering
	\includegraphics[width=18cm]{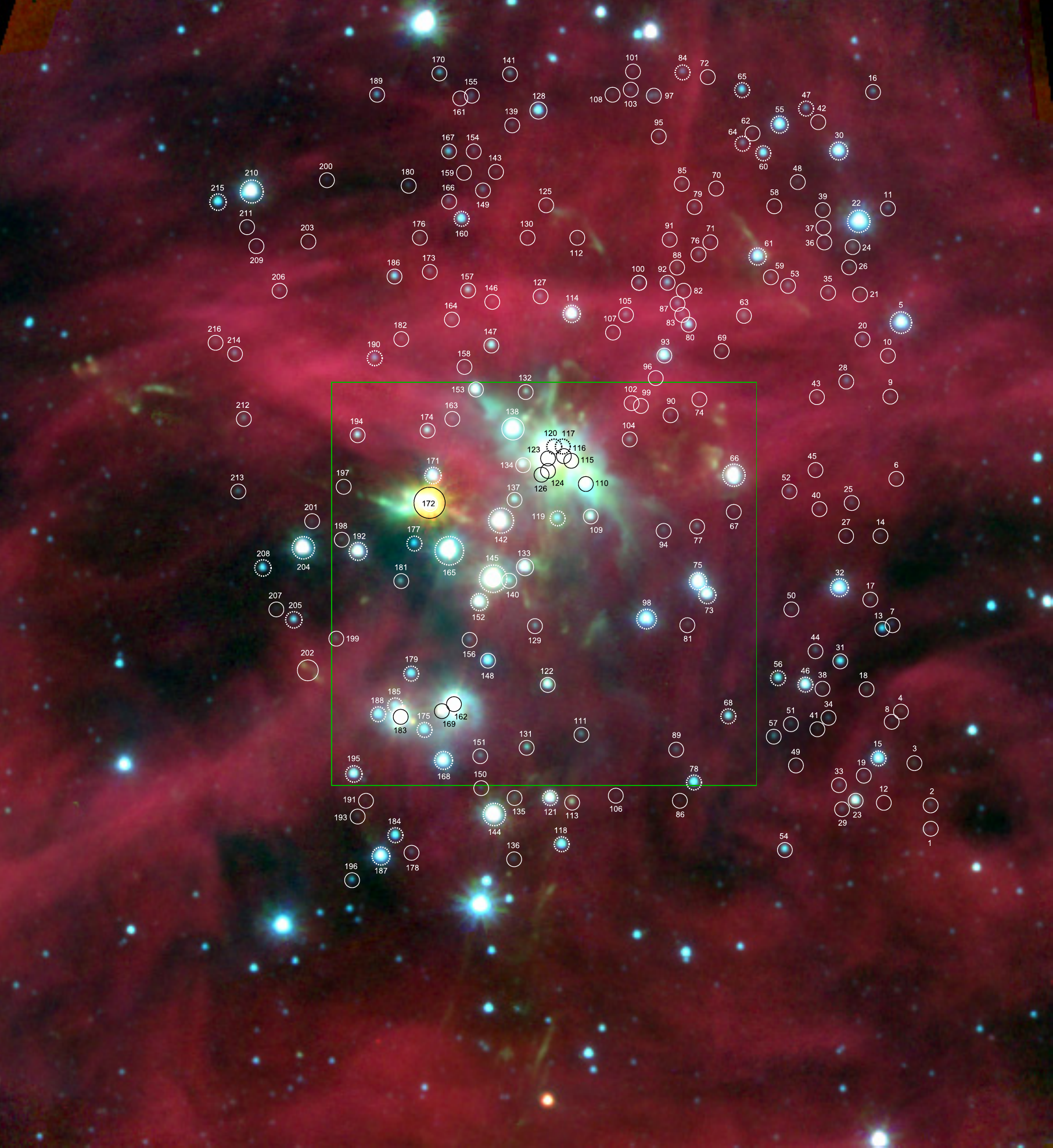}
	\caption{Spitzer composite of 3.6\,$\mu$m (blue), 4.5\,$\mu$m (green) and 8.0\,$\mu$m (red). All our
	sources are marked by circles. Sources for which we have spectroscopy have thick, dashed circles.
	The surveyed region has a size of
	8\farcm45\,$\times$\,7\farcm38 (1.11\,$\times$\,0.97\,pc at a distance of 450\,pc), although the field
	size in this Figure has been made larger to give a better overview of the region. The green square outlines our $J$ band observations.
		}
	\label{Spitzer124}
\end{figure*}


The final reduced Spitzer mosaic (using our own routines) is presented in Fig.\,\ref{Spitzer124} as a colour image using
three of the four IRAC channels (blue = 3.6\,$\mu$m, green = 4.5\,$\mu$m and red = 8.0\,$\mu$m). One outstanding feature in this
image is the very bright extended emission at 8.0\,$\mu$m (also present in the 5.8\,$\mu$m channel but not as bright). 
The reason for this emission can be seen in Figure\,\ref{IRACPAHs}, it is caused by Polycyclic Aromatic Hydrocarbons (PAHs), molecules built
up of benzene rings which emit most strongly in IRAC 4, followed by IRAC 3, IRAC 1  and finally IRAC 2 which is essentially PAH free.

Since the PAH emission in L1641N is very bright and seen across the whole mosaic (see Fig.\,\ref{Spitzer124}) the UV radiation causing
the excitation must be illuminating the cloud uniformly. We suggest that the illuminating source is $\iota$~Orionis (HD\,37043), located
about 30$\arcmin$ to the north. This star has a O9~III spectral type and a parallax of $2.46 \pm 0.77$\,mas according to the Hipparcos
catalogue, corresponding to a distance of about 400\,pc (310-592\,pc). This fits in very well with the assumed distance to L1641N of
about 450\,pc and puts the suggested star at a similar or somewhat closer distance than L1641N. Their separation in the plane of the
sky, if equal distances are assumed, is about 3.5\,pc.

Our two K$_S$ mosaics (and shocked 2.12\,$\mu$m H$_2$) are presented in Figures~\ref{Kmap1} and \ref{Kmap2} with all K$_S$ counterparts to
Spitzer sources marked by circles. There is a large amount of shocked H$_2$ in these mosaics, much of which can also be seen in the IRAC images
(especially in the 4.5\,$\mu$m channel). For a further discussion on the H$_2$ flows see
G\aa lfalk \& Olofsson (\cite{galfalk07}) and Stanke et al. (\cite{stanke}, \cite{stanke2000}).


\subsection{Photometry and source catalogue}


\begin{figure}
	\centering
	\includegraphics[width=8.8cm]{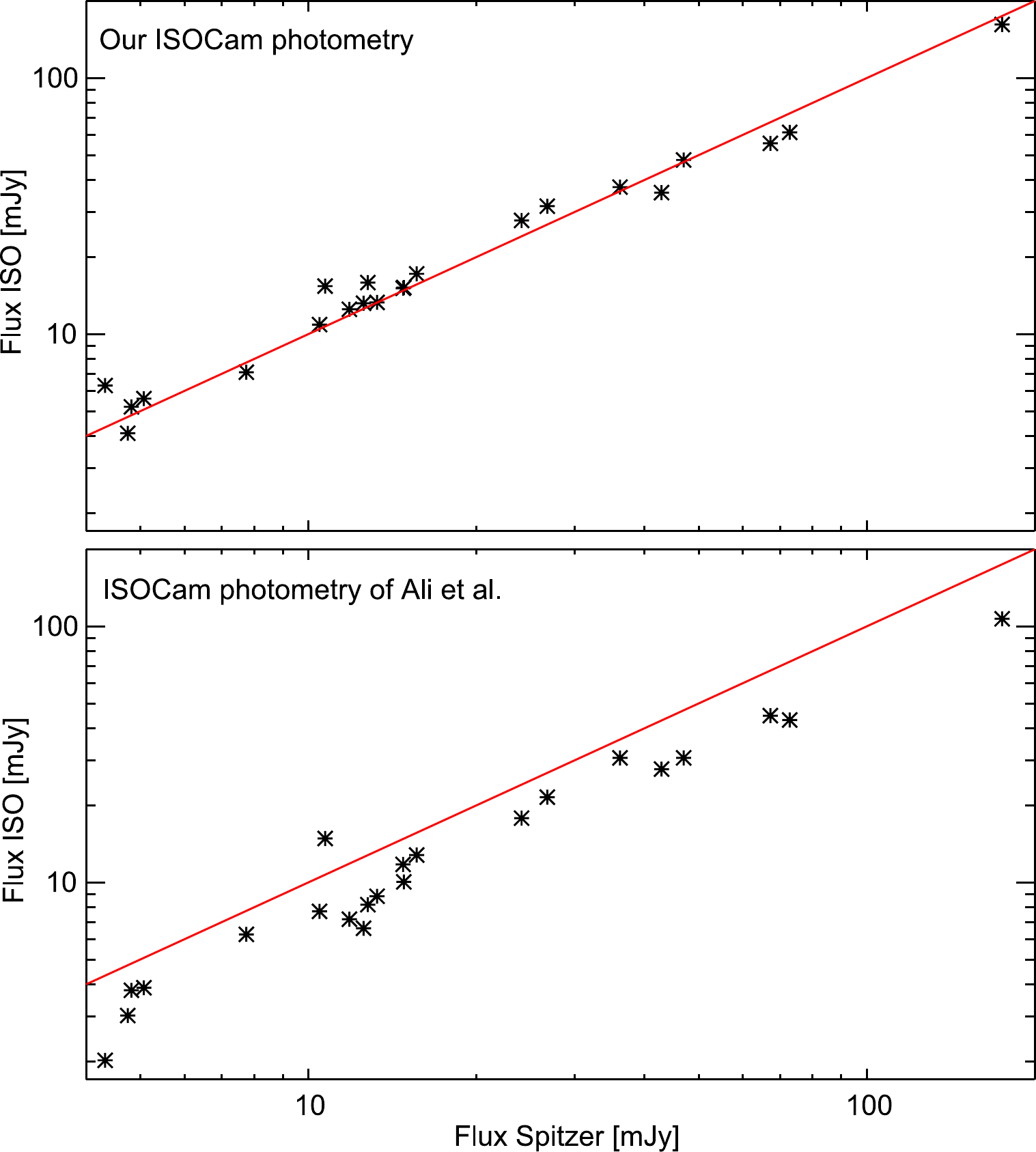}
	\caption{Comparison of ISO and Spitzer photometry. The ISOCAM 6.7\,$\mu$m channel is compared
	to the mean flux of the Spitzer (IRAC) 5.8 and 8.0\,$\mu$m channels. The upper panel shows our ISOCam
	photometry and the lower panel that of Ali et al. (\cite{ali}).
	}
	\label{isospitzer}
\end{figure}


We have carried out photometry in all filters of the ground-based ($I$, $J$, $K_S$, $L$') and
space-based (3.6, 4.5, 5.8, 6.7, 8.0 and 14.3\,$\mu$m) observations within the area covered by the ISO observations.
The full source list is basically the full Spitzer sample in L1641N (any sources seen with IRAC are included in the list), with
(a few) additional sources added from the $I$ band mosaic. 

In total we include 216 sources in our L1641N catalogue. In order to keep the Table width down we have divided our photometry into two Tables.
Table~\ref{L1641N_space} contains space-based photometry from the ISO and Spitzer telescopes and Table~\ref{L1641N_ground} the ground-based
photometry. We have also observed a small number of these sources in the L' band (see Table~\ref{Ltable}).
Source numbering is by a common {\it cloud number}, i.e. the same for all filters and increasing from west to east.

Spitzer sources that appear double in our ground based observations (cf Sect.~\ref{binary_sec}), will in the following
be called double sources, including both gravitationally bound and optical binary sources.
In all Tables we use {\bf bold source numbers} for sources where we have found intrinsic IR excess.

One way of checking the ISO and Spitzer photometry, although observed using different filters, is to plot the ISOCAM LW2 channel (6.7\,$\mu$m)
versus the mean of IRAC channels 3 (5.8\,$\mu$m) and 4 (8.0\,$\mu$m). Points should then roughly follow a line with fluxes being
equal on both axes. This is shown in Fig.~\ref{isospitzer} using both our ISOCAM photometry and that of Ali et al.~(\cite{ali}).
Our ISOCAM photometry agrees very well with the more accurate Spitzer photometry, whereas the ISOCAM photometry of Ali et al. seems to systematically
underestimate the fluxes by a large factor (even more so at 14.3\,$\mu$m) and as an average their flux densities needs to be
multiplied by a factor of 1.41 (6.7\,$\mu$m) and 1.88 (14.3\,$\mu$m) respectively to agree with the Spitzer data.

Our ISOCAM circular aperture photometry uses well tested routines which have been used in several previous papers
(Kaas et al. \cite{kaas99}, Olofsson et al. \cite{olofsson}, Persi et al. \cite{persi}, Bontemps et al. \cite{bontemps}, Kaas et al. \cite{kaas04}
and G\aa lfalk et al. \cite{galfalk}), our Spitzer photometry uses another set of routines that we wrote especially for the Spitzer data. That fact that they agree very well with each
other, although using different routines and calibrations (not done by the same people), indicates that both our ISO and Spitzer photometry
are reliable. In total we detect 211 sources in the Spitzer mosaics and 43 of these in the ISOCAM images (Ali et al. \cite{ali} detected 34 sources).


\subsection{The central region and IRAS 05338-0624}


\begin{figure*}
	\centering
	\includegraphics[width=18cm]{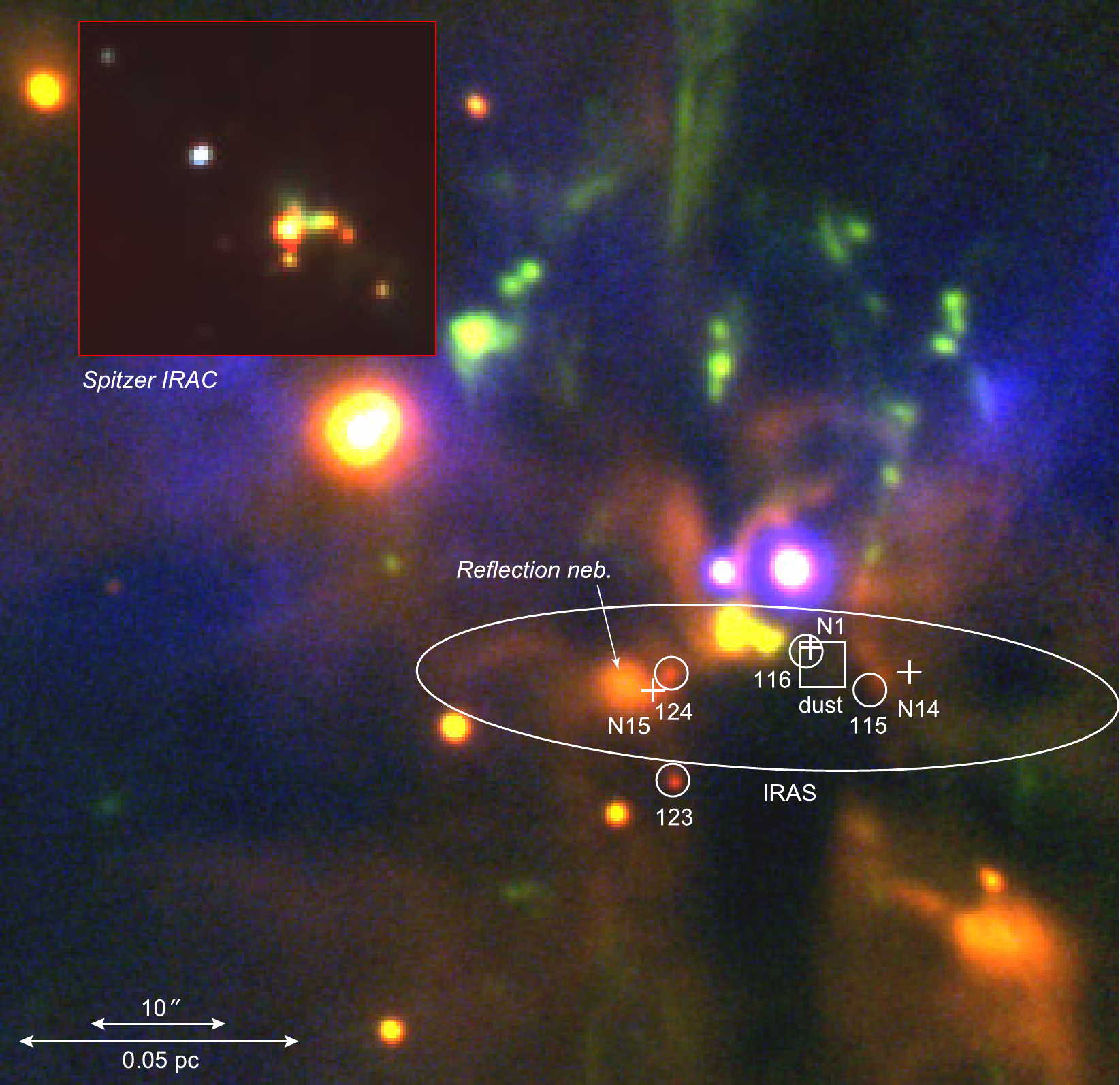}
	\caption{The central region of L1641N in a composite using $I$ (blue colours), 2.12\,$\mu$m\,H$_2$\,S(1) (green) and $K_S$ (red) filters.
		The big ellipse marks IRAS beam centred on IRAS 05338-0624, circles denote our sources and plus signs are corresponding
		sources of Chen et al. (\cite{chen93}). The square is centred on the 2\,mm dust peak (see text). The field shown has a size of
		83\farcs0 $\times$ 80\farcs5 for both images.
		}
	\label{fig_IRAS}
\end{figure*}


The optically invisible source IRAS\,05338-0624 is one of the most luminous IRAS sources detected in the region
and in fact the only IRAS source in L1641N. It has been suggested that its counterpart is N15 or N1 of Chen et al. (\cite{chen93}), corresponding
to L1641N-124 and 116 in our numbering scheme. Given the uncertainty ellipse of the IRAS measurements and the new sources we see in our
$K_S$ and Spitzer observations, it can be seen in Fig.~\ref{fig_IRAS} that the situation is probably much more complicated than just one
source being its counterpart.

Within the IRAS beam we detect four sources: L1641N-115, 116, 123 and 124, all fairly bright in the mid-IR, so we conclude that the IRAS source is in
fact the added flux of a number of sources (with No.\,124 clearly being the brightest in the mid-IR). There is also some mid-IR reflection nebulosity in
the central region that can be part of this flux, as can be seen in our L' image (Fig.\,\ref{Limage}). 

Chen et al. (\cite{chen95}) observed a dust peak in the 2\,mm continuum (3\arcsec beam) close to the IRAS position and
suggested that this could be a circumstellar dust disk. This scenario fits very well with our observations, as a jet
with a chain of H$_2$ knots seems to emanate from this position. Our L' image only shows extended nebulosity at
L1641N-116 (N1), suggesting that only reflection nebulosity is seen in the 3.6\,$\mu$m Spitzer data. This could also be the case for the
4.5\,$\mu$m channel (possibly also some H$_2$ emission) and the $M$ band observations of Chen et al.~(\cite{chen93}) and at even longer wavelengths given the ISO and Spitzer fluxes (ISO
has a larger beam and thus measures more extended flux).

There is one more bright mid-IR source in the vicinity of the central region, and that is L1641N-172. However,
given that it is located more than an arc minute (86\arcsec) from IRAS\,05338-0624, and the fact that it has brightened
considerably the last decade (G\aa lfalk \& Olofsson \cite{galfalk07}) makes this a very unlikely counterpart.
Since only one IRAS source was detected in L1641N, L1641N-172 was probably much fainter at the time of the IRAS observations and thus not detected.

\begin{figure*}
	\centering
	\includegraphics[width=18cm]{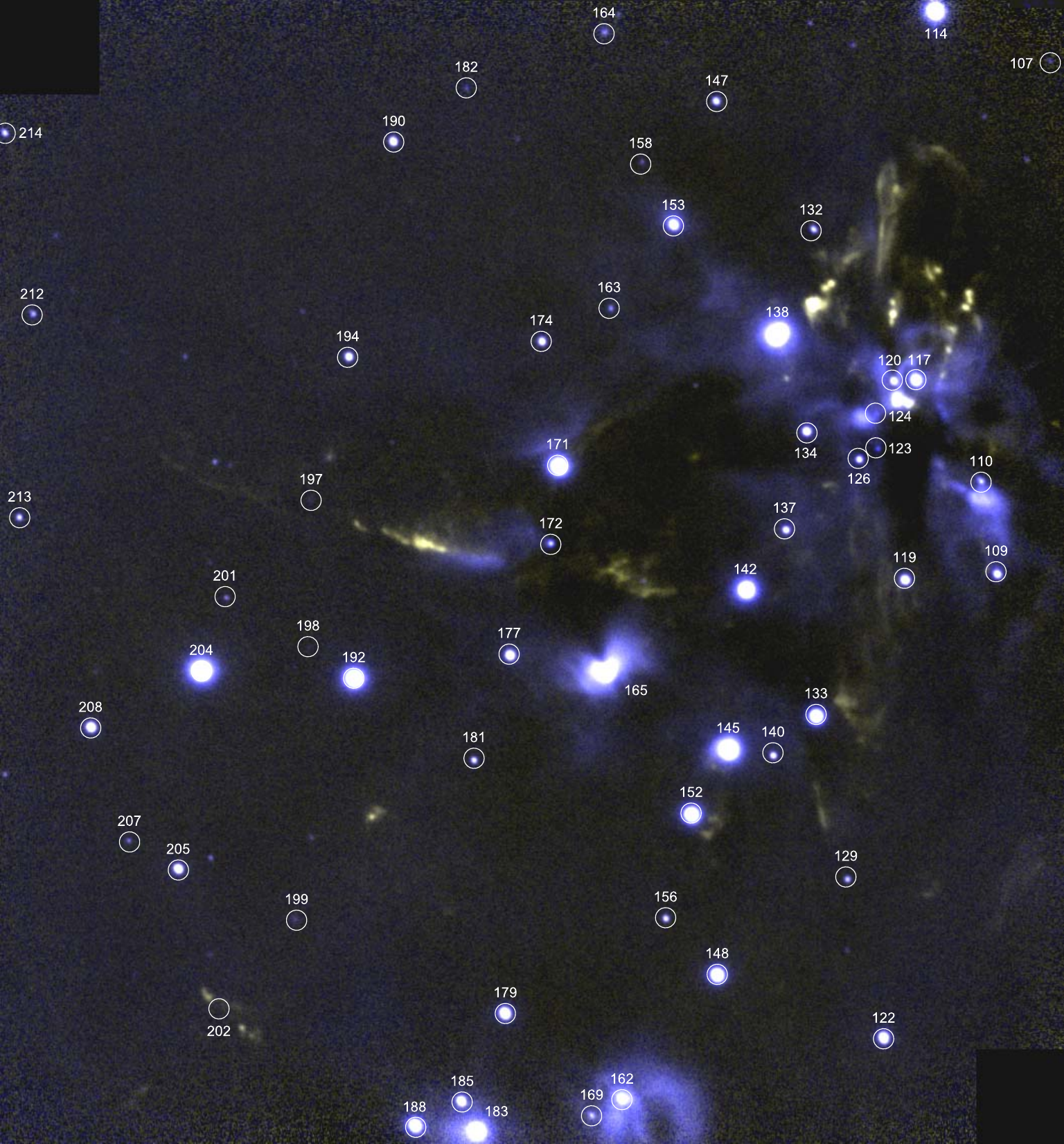}
	\caption{Deep K$_S$ and 2.12\,$\mu$m H$_2$ colour composite of the central and south east region of L1641N. All
	visible sources in our sample are marked by a circle. The field shown has a size of 4\farcm05 $\times$ 4\farcm35.
	}
	\label{Kmap1}
\end{figure*}

\begin{figure*}
	\centering
	\includegraphics[width=18cm]{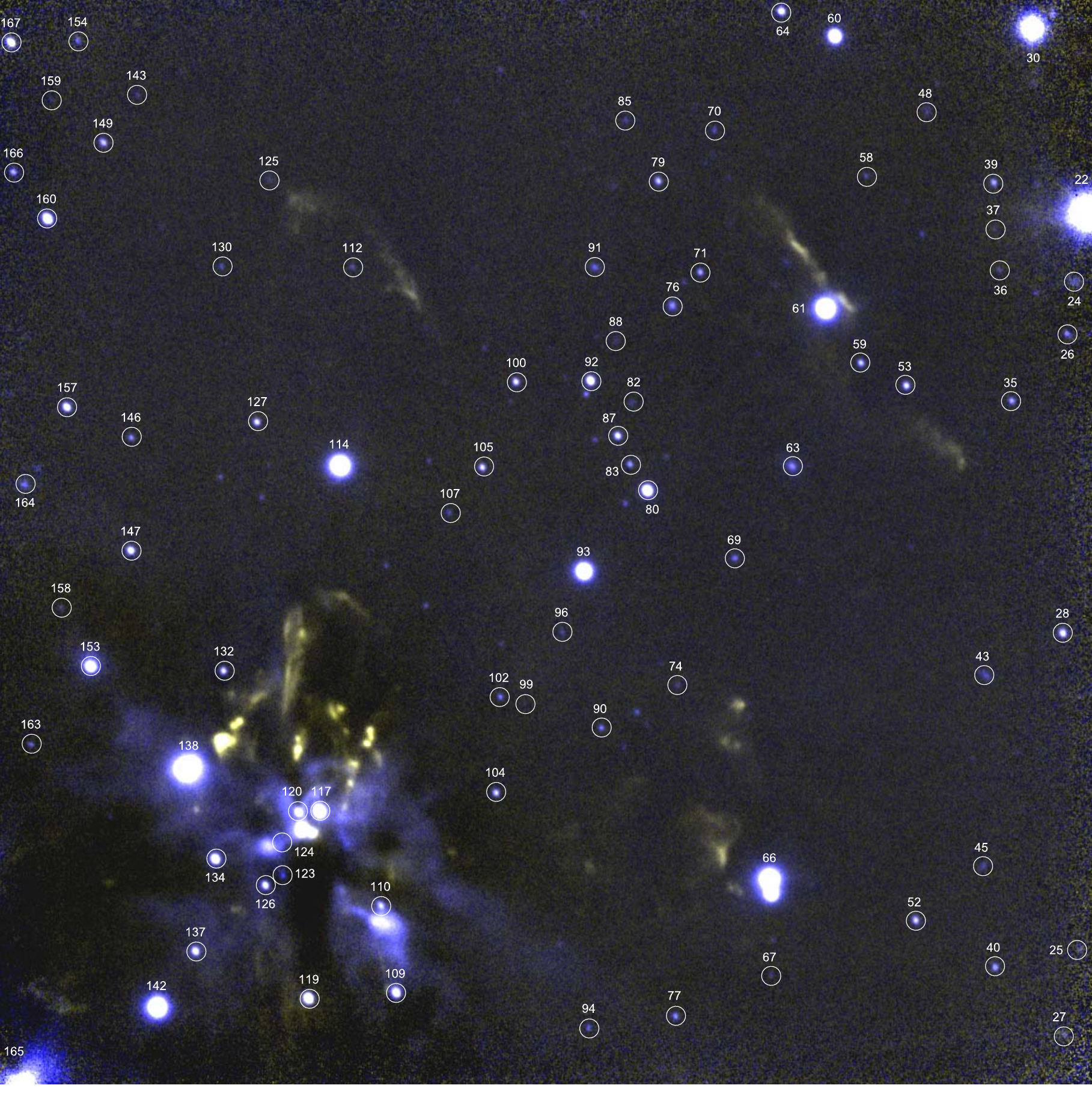}
	\caption{Deep K$_S$ and 2.12\,$\mu$m H$_2$ colour composite of the central and north west region of L1641N. All
	visible sources in our sample are marked by a circle. The field shown has a size of 4\farcm40 $\times$ 4\farcm37.
	}
	\label{Kmap2}
\end{figure*}

The source density is clearly much higher at the centre of L1641N than in the outer parts, and judging from the H$_2$ flows (Figures~\ref{Kmap1} and \ref{Kmap2}) there should be several very young (Class 0) YSOs in that region. These are however much too deeply embedded to be included in our MF calculations (which needs detections in the $I$ and $J$ bands). For further studies of the central region we refer to Chen et al. (\cite{chen96}) and Stanke \& Williams (\cite{stanke07}).


\subsection{Identifying stars with intrinsic IR excesses}

\label{excess_sec}


\begin{figure*}
	\centering
	\includegraphics[width=18cm]{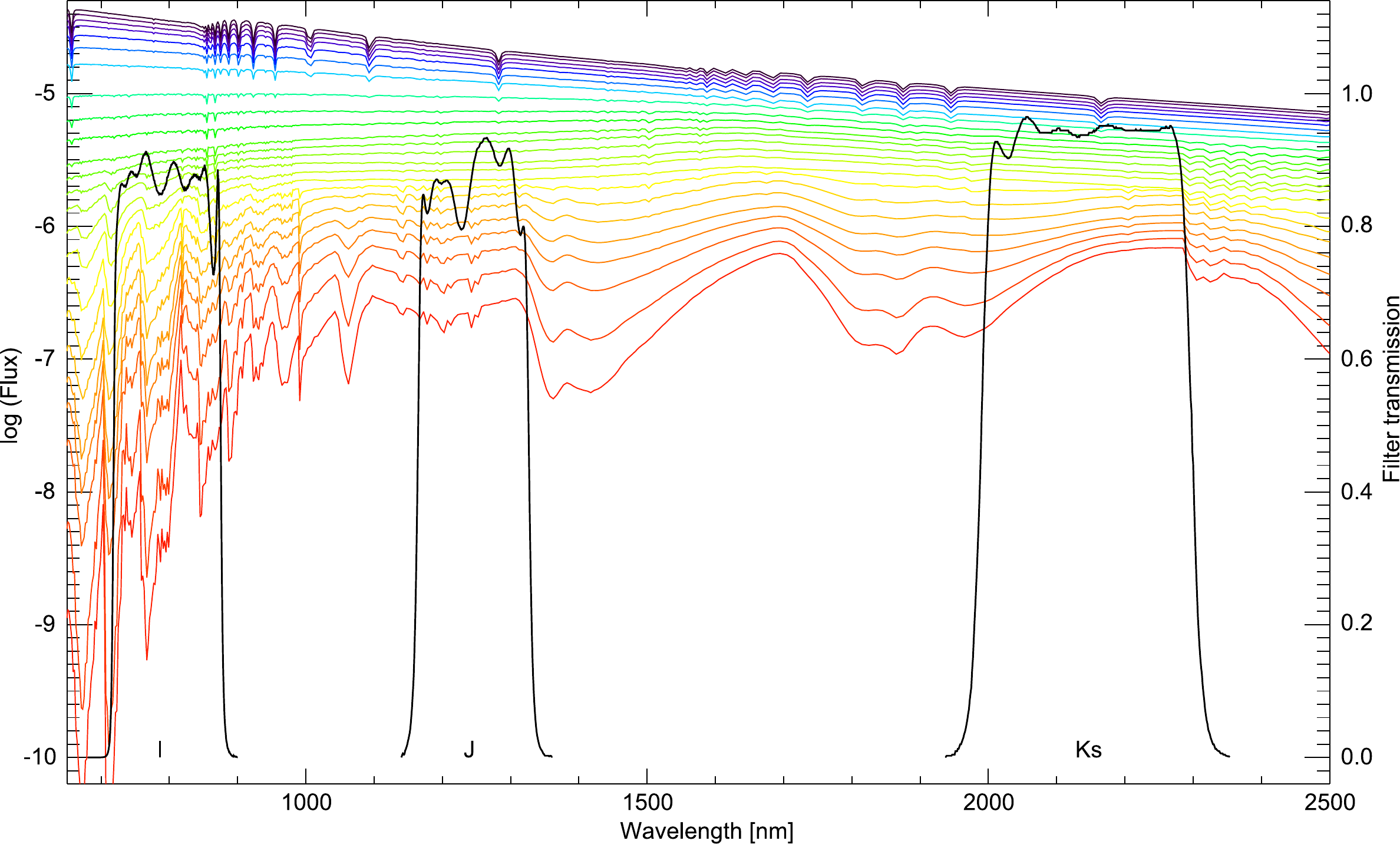}
	\caption{Stellar models (BaSeL v2.2) for log\,g = 4.5 and T$_{\textnormal{eff}}$ = 2\,000 -- 10\,000 K.
	Overplotted are the ground based filter transmission curves ($I$, $J$ and $K_S$) used in our observations.
		}
	\label{Trans_IJKs}
\end{figure*}



\begin{figure*}
	\centering
	\includegraphics[width=18cm]{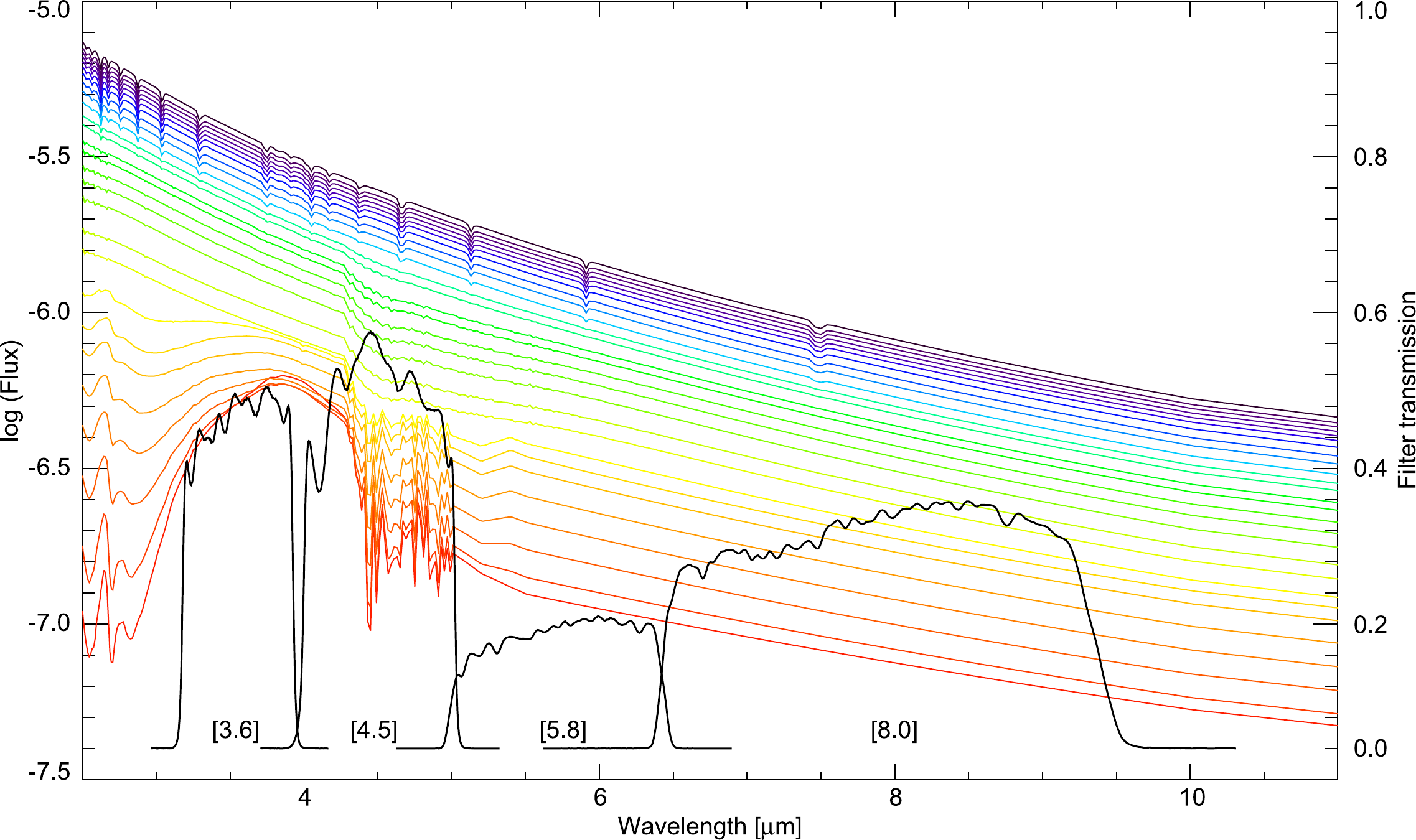}
	\caption{Stellar models (BaSeL v2.2) for log\,g = 4.5 and T$_{\textnormal{eff}}$ = 2\,000 -- 10\,000 K.
	Overplotted are the IRAC filter transmission curves.
		}
	\label{Trans_Spitzer}
\end{figure*}



\begin{table*}
	\caption{Calculated colours of normal stars. The model atmospheres (BaSeL v2.2) and transmission curves used in the calculations
	are shown in Figures \ref{Trans_IJKs} and \ref{Trans_Spitzer}.
	}
	\label{stellar_colours}
	\begin{tabular}{rrrrrrrrrrrrr}
	  \hline
        \noalign{\vspace{0.5mm}}
        T$_{\textnormal{eff}}$ & I-J & J-K$_S$ & K$_S$-3.6 & K$_S$-4.5 & K$_S$-5.8 & K$_S$-8.0 & I-J & J-K$_S$ & K$_S$-3.6 & K$_S$-4.5 & K$_S$-5.8 & K$_S$-8.0 \\
 	(K) & \multicolumn{6}{c}{Main sequence, log\,g = 4.5} & \multicolumn{6}{c}{Supergiants, log\,g = 0.0} \\
 	\noalign{\vspace{0.5mm}}
        \hline

	\noalign{\vspace{1.0mm}}

    	2400 & 2.502 & 1.060 & 0.646 & 0.589 & 0.361 & 0.502 & 3.794 & 0.741 & 0.003 & 0.568 & 0.704 & 0.838 \\
	2600 & 2.244 & 0.927 & 0.531 & 0.497 & 0.363 & 0.501 & 3.136 & 1.194 & 0.397 & 0.775 & 0.905 & 1.035 \\
	2800 & 2.076 & 0.846 & 0.455 & 0.441 & 0.377 & 0.513 & 3.054 & 1.270 & 0.387 & 0.689 & 0.807 & 0.926 \\
	3000 & 1.879 & 0.810 & 0.371 & 0.369 & 0.338 & 0.466 & 2.969 & 1.308 & 0.329 & 0.528 & 0.638 & 0.745 \\
	3200 & 1.647 & 0.804 & 0.290 & 0.263 & 0.236 & 0.358 & 2.524 & 1.282 & 0.253 & 0.320 & 0.421 & 0.519 \\
      	3400 & 1.382 & 0.822 & 0.212 & 0.144 & 0.163 & 0.278 & 1.869 & 1.188 & 0.225 & 0.217 & 0.304 & 0.422 \\
      	3600 & 1.133 & 0.814 & 0.173 & 0.129 & 0.222 & 0.325 & 1.395 & 1.092 & 0.186 & 0.005 & 0.069 & 0.246 \\
      	3800 & 1.031 & 0.800 & 0.156 & 0.103 & 0.183 & 0.276 & 1.146 & 0.971 & 0.158 & -0.022 & 0.032 & 0.201 \\
      	4000 & 0.861 & 0.776 & 0.151 & 0.078 & 0.145 & 0.229 & 1.053 & 0.841 & 0.139 & -0.033 & 0.011 & 0.173 \\
      	4200 & 0.877 & 0.717 & 0.124 & 0.037 & 0.093 & 0.169 & 0.926 & 0.745 & 0.131 & -0.028 & 0.008 & 0.161 \\
      	4400 & 0.871 & 0.670 & 0.122 & 0.032 & 0.079 & 0.148 & 0.858 & 0.658 & 0.124 & -0.022 & 0.009 & 0.149 \\
      	4600 & 0.840 & 0.623 & 0.118 & 0.032 & 0.072 & 0.135 & 0.785 & 0.587 & 0.119 & -0.012 & 0.016 & 0.141 \\
      	4800 & 0.781 & 0.572 & 0.108 & 0.028 & 0.063 & 0.120 & 0.706 & 0.531 & 0.115 & 0.004 & 0.031 & 0.137 \\
        5000 & 0.693 & 0.523 & 0.099 & 0.027 & 0.056 & 0.107 & 0.648 & 0.485 & 0.112 & 0.023 & 0.050 & 0.135 \\
        5200 & 0.627 & 0.490 & 0.092 & 0.029 & 0.053 & 0.097 & 0.601 & 0.427 & 0.130 & 0.068 & 0.094 & 0.155 \\
      	5400 & 0.560 & 0.443 & 0.086 & 0.032 & 0.052 & 0.088 & 0.537 & 0.378 & 0.131 & 0.095 & 0.119 & 0.160 \\
      	5600 & 0.504 & 0.392 & 0.080 & 0.036 & 0.051 & 0.080 & 0.483 & 0.330 & 0.125 & 0.110 & 0.129 & 0.157 \\
      	5800 & 0.466 & 0.346 & 0.074 & 0.040 & 0.050 & 0.071 & 0.449 & 0.286 & 0.117 & 0.116 & 0.131 & 0.150 \\
     	6000 & 0.450 & 0.316 & 0.070 & 0.044 & 0.050 & 0.064 & 0.433 & 0.257 & 0.113 & 0.122 & 0.133 & 0.147 \\
      	6200 & 0.398 & 0.277 & 0.062 & 0.045 & 0.048 & 0.056 & 0.421 & 0.232 & 0.110 & 0.126 & 0.135 & 0.144 \\
      	6400 & 0.347 & 0.236 & 0.055 & 0.044 & 0.044 & 0.049 & 0.351 & 0.187 & 0.102 & 0.122 & 0.132 & 0.143 \\
     	6600 & 0.295 & 0.197 & 0.048 & 0.043 & 0.040 & 0.042 & 0.279 & 0.144 & 0.096 & 0.118 & 0.129 & 0.142 \\
      	6800 & 0.252 & 0.163 & 0.041 & 0.039 & 0.034 & 0.034 & 0.230 & 0.112 & 0.091 & 0.115 & 0.126 & 0.139 \\
      	7000 & 0.230 & 0.137 & 0.034 & 0.032 & 0.027 & 0.026 & 0.203 & 0.090 & 0.084 & 0.108 & 0.119 & 0.132 \\
      	7200 & 0.183 & 0.127 & 0.027 & 0.027 & 0.022 & 0.021 & 0.148 & 0.082 & 0.077 & 0.101 & 0.112 & 0.125 \\
      	7400 & 0.141 & 0.116 & 0.022 & 0.024 & 0.019 & 0.017 & 0.093 & 0.073 & 0.070 & 0.095 & 0.105 & 0.116 \\
      	7600 & 0.105 & 0.106 & 0.016 & 0.019 & 0.014 & 0.012 & 0.058 & 0.068 & 0.062 & 0.086 & 0.095 & 0.106 \\
      	7800 & 0.077 & 0.094 & 0.011 & 0.012 & 0.008 & 0.006 & 0.038 & 0.063 & 0.054 & 0.077 & 0.084 & 0.094 \\
      	8000 & 0.070 & 0.077 & 0.009 & 0.009 & 0.006 & 0.004 & 0.015 & 0.052 & 0.052 & 0.071 & 0.077 & 0.086 \\
      	8200 & 0.060 & 0.058 & 0.011 & 0.011 & 0.009 & 0.008 & -0.006 & 0.043 & 0.052 & 0.066 & 0.071 & 0.080 \\
      	8400 & 0.050 & 0.042 & 0.011 & 0.011 & 0.011 & 0.011 & -0.025 & 0.051 & 0.051 & 0.063 & 0.066 & 0.074 \\
      	8600 & 0.035 & 0.032 & 0.008 & 0.008 & 0.008 & 0.009 & -0.035 & 0.043 & 0.042 & 0.051 & 0.053 & 0.060 \\
      	8800 & 0.017 & 0.025 & 0.002 & 0.002 & 0.003 & 0.003 & -0.042 & 0.024 & 0.027 & 0.035 & 0.035 & 0.040 \\
      	9000 & 0.003 & 0.019 & -0.002 & -0.003 & -0.002 & -0.002 & -0.060 & 0.026 & 0.021 & 0.027 & 0.026 & 0.029 \\
      	9200 & 0.002 & 0.010 & -0.003 & -0.004 & -0.003 & -0.003 & -0.048 & 0.009 & 0.014 & 0.019 & 0.016 & 0.018 \\
      	9400 & -0.001 & 0.004 & -0.002 & -0.003 & -0.003 & -0.003 & -0.051 & 0.003 & 0.012 & 0.016 & 0.012 & 0.013 \\
      	9600 & -0.007 & -0.001 & -0.001 & -0.003 & -0.003 & -0.003 & -0.060 & 0.002 & 0.012 & 0.014 & 0.010 & 0.009 \\
      	9800 & -0.016 & -0.005 & -0.001 & -0.003 & -0.004 & -0.004 & -0.069 & 0.002 & 0.012 & 0.013 & 0.008 & 0.007 \\
      	10000 & -0.026 & -0.010 & -0.001 & -0.003 & -0.004 & -0.005 & -0.080 & 0.000 & 0.013 & 0.014 & 0.008 & 0.006 \\

	\noalign{\vspace{1.0mm}}
	\hline

	\end{tabular}
\end{table*}


\subsubsection{Intrinsic colours for normal stars}

The main problem of using available empirical colour information for normal stars is the differences between the detailed response functions, which
may result in relatively large errors, in particular for late M stars. For this reason we use theoretical colours calculated by convolution of the
filter response functions and synthetic spectra. We have chosen the BaSeL 2.2 spectral library with corrected
SEDs (Lejeune et al.~\cite{lejeune98}, \cite{lejeune02}) and limited the parameter space
to T$_{\textnormal{eff}}$ = 2\,000--10\,000\,K, log\,g = 0, 3 and 4.5 and solar metal abundance.
We use a A0\,V model with T$_{\textnormal{eff}}$ = 9500\,K, log\,g = 4.0 and [Fe/H] = 0.0 to define zero-colours between all filters.

In Figures~\ref{Trans_IJKs} and \ref{Trans_Spitzer} we have plotted all filter curves used for the colour calculations, together with synthetic
stellar atmospheres of different temperatures (log\,g = 4.5). In Table~\ref{stellar_colours} we summarize these calculations.

\subsubsection{Finding excess stars from Spitzer and ISO data}


\begin{figure*}
	\centering
	\includegraphics[width=18cm]{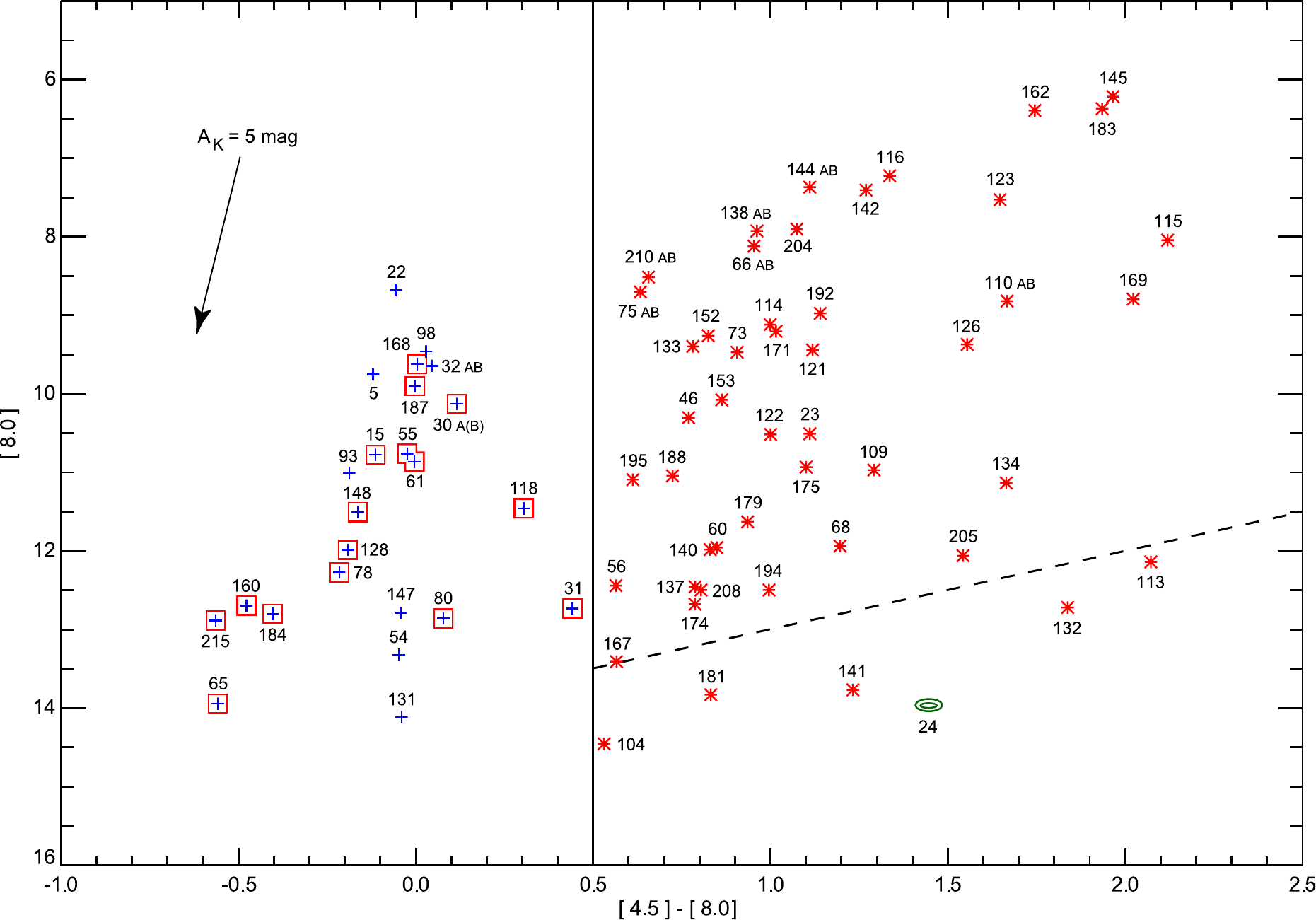}
	\caption{Colour-magnitude diagram using only Spitzer photometry (4.5 and 8.0\,$\mu$m). Sources marked with an asterisk
	show clear IR excess at 8\,$\mu$m compared to photospheres reddened by extinction alone. Sources that show no or very little
	excess are marked with plus signs. Double sources are indicated by an AB suffix. The solid line separates non-excess
	and excess sources while some sources below the dashed line could possibly be an extragalactic contribution.
	Source No.\,24 appears to be an extended background source in our $K_S$ image and is located in the part of the diagram expected for a galaxy.
	The arrow indicates the expected interstellar extinction vector ($A_K = 5$ corresponds to $A_V$ = 47.4).
	YSO candidates without excess are marked with a red square.
	}
	\label{Colcol1}
\end{figure*}


The interstellar extinction is certainly less in the IRAC bands than in the near-IR, making it possible to separate normal and excess stars without
first calculating the extinction to each source.
In Fig.~\ref{Colcol1} we have used a Spitzer-only (4.5 and 8.0\,$\mu$m) colour-magnitude diagram to find YSO candidates, taking into account
the exclusion of both non-excess stellar and extragalactic sources. This method has been shown to be very
efficient (see e.g. J\o rgensen et al.~\cite{jorgensen} and Harvey et al.~\cite{harvey}).

The first criterion assumes that the colour index $[4.5] - [8.0]$ is insensitive to interstellar extinction.
Thus, stars with $[4.5] - [8.0]$ less than the largest colour index for normal stars (we get 0.27 for giants and 0.20 for MS stars) plus some
margin, we chose 0.5, are assumed to have no intrinsic IR excess. This also means that we identify all the stars
with $[4.5] - [8.0] > 0.5$ as stars with intrinsic IR excesses and therefore YSO candidates.

The second criterion, $[8.0] > 14 - ([4.5]-[8.0])$, suggests possible remaining extragalactic sources in the sample (sources that
are obvious galaxies in our $I$ or $K_S$ mosaics have already been excluded). As can be seen in Fig.~\ref{Colcol1}, five of our sources are
located below this (dashed) line. Interstellar extinction, although low at 8.0\,$\mu$m, may have pushed these sources below the
line. Nevertheless, they must be considered less reliable YSO candidates than the others.

Due to the sensitivity limit of the 8.0\,$\mu$m observations (high contrast background from PAH emission bands that effectively
masks faint point sources at these wavelengths), there are sources in our list that lack detection at 8.0\,$\mu$m, but still may exhibit intrinsic
mid-IR excesses. 

We find an excess at 8.0\,$\mu$m for exactly 2/3 of the sources detected at 4.5 and 8.0\,$\mu$m.
It is however also noted that several of the non-excess sources are in fact YSOs, as confirmed by our spectroscopy.

\begin{figure}
	\centering
	\includegraphics[width=8.8cm]{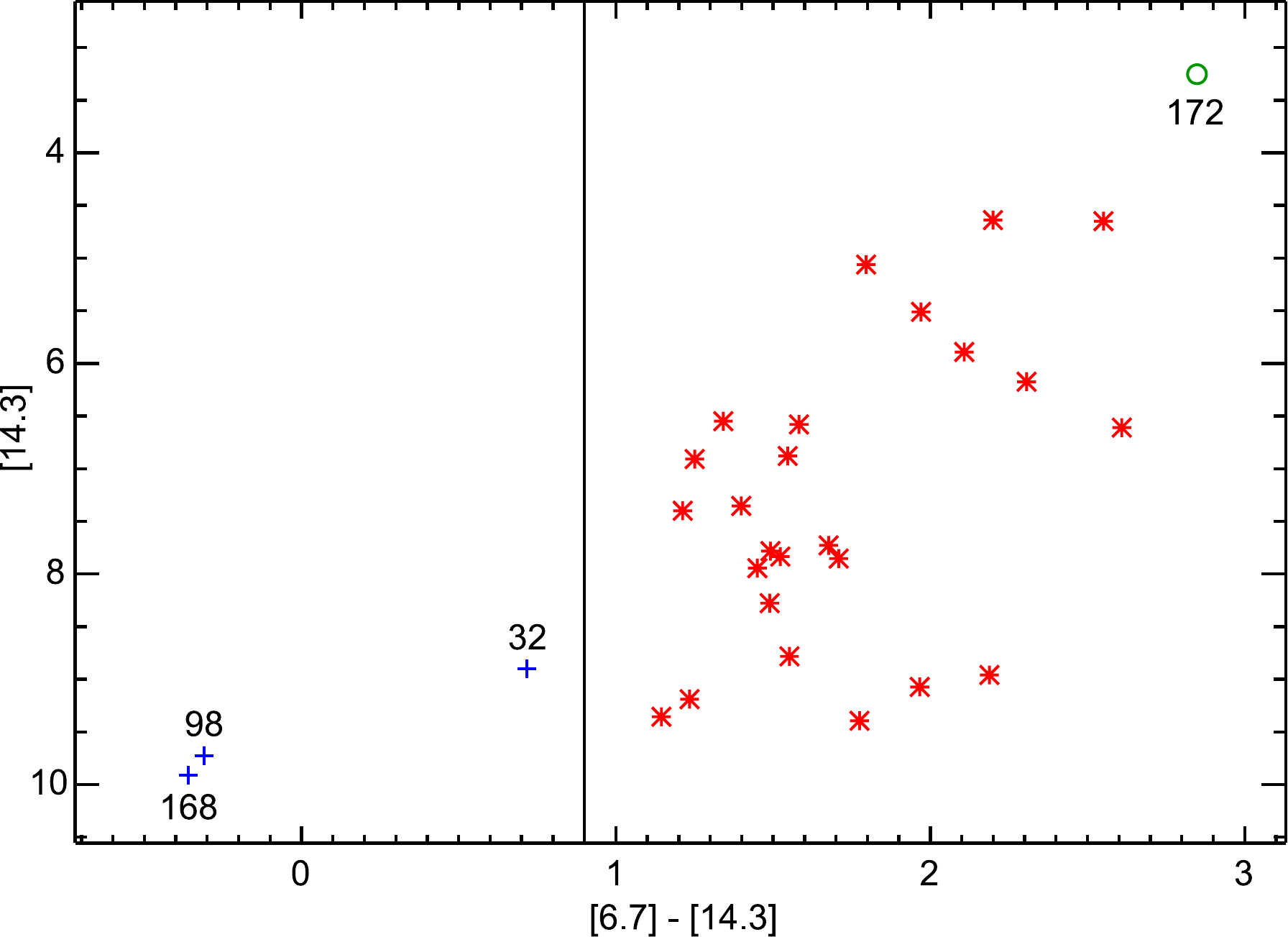}
	\caption{Colour-magnitude diagram showing that most of the sources detected in L1641N by the ISO satellite in fact have
	excess at 14.3\,$\mu$m. Symbols are the same as in Fig.~\ref{Colcol1}. 
	The most luminous and red source in this plot, L1641N-172, is an outflow source and the brightest source at mid-IR wavelengths in the
	entire L1641N. It is marked with a circle since it could not be shown to have excess in the Spitzer data because of saturation.
	}
	\label{ISOcolmag}
\end{figure}

Figure~\ref{ISOcolmag} shows our ISOCAM colour-magnitude diagram for all sources detected at 6.7 and 14.3\,$\mu$m. The symbols are the same as
in the Spitzer plots. There is a very clear separation between non-excess and excess sources at these longer wavelengths and an excellent
agreement between excess sources at 8.0 and 14.3\,$\mu$m. Source L1641N-172, in the upper right corner of Fig.\ref{ISOcolmag}, stands out
as being the brightest and most red source. This is the newly discovered outflow source (G\aa lfalk \& Olofsson \cite{galfalk07}) that
could not be shown to have excess in Fig.\,\ref{Colcol1} because of saturation in the Spitzer data.


\subsubsection{Defining the interstellar reddening and identifying additional YSOs}


\begin{figure*}
	\centering
	\includegraphics[width=17cm]{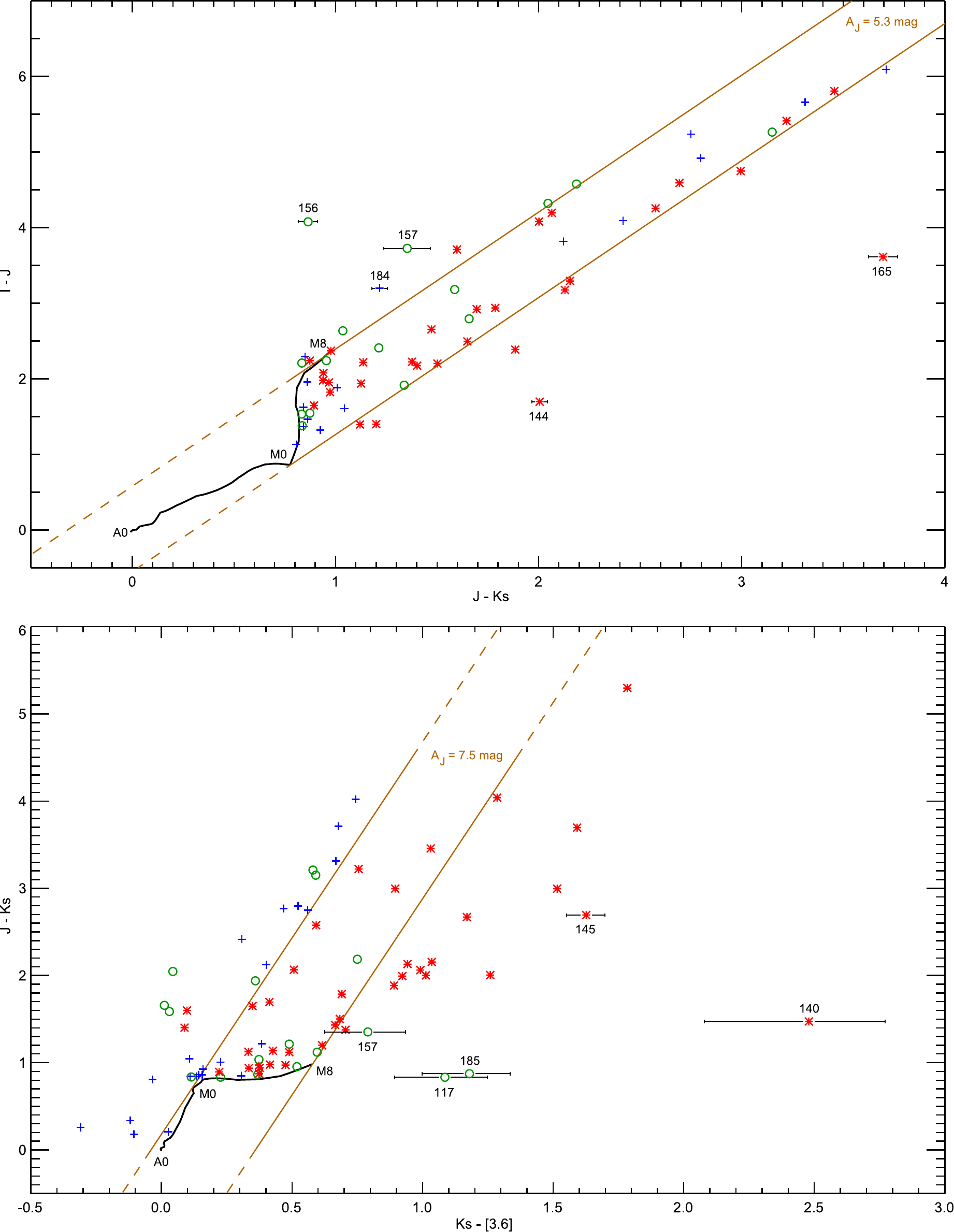}
	\caption{Colour-colour diagrams using ground based and Spitzer photometry. The solid black curve represents model stellar atmospheres
	($log\,g = 4.5$, T$_{\textnormal{eff}} = 2\,500-10\,000\,K$) and the brown lines the reddening band due to interstellar extinction.
	The slopes of the reddening vectors were found by fitting the 'blue' sources.
	Symbols are the same as in Fig.~\ref{Colcol1} with the addition of green circles that denote sources not detected at 8.0\,$\mu$m and
	thus having unknown red/blue status in the [4.5]-[8.0] index. Error bars are shown for selected sources located on the red side of the
	reddening band (all such green circles with error bars are added to our list of YSO candidates). It is noted that very few of the sources
	show clear intrinsic excess at near-IR wavelengths (K$_S$). ($A_J = 5$ corresponds to $A_V$ = 19.4)
	}
	\label{colourdiag_1}
\end{figure*}



\begin{figure*}
	\centering
	\includegraphics[width=16.9cm]{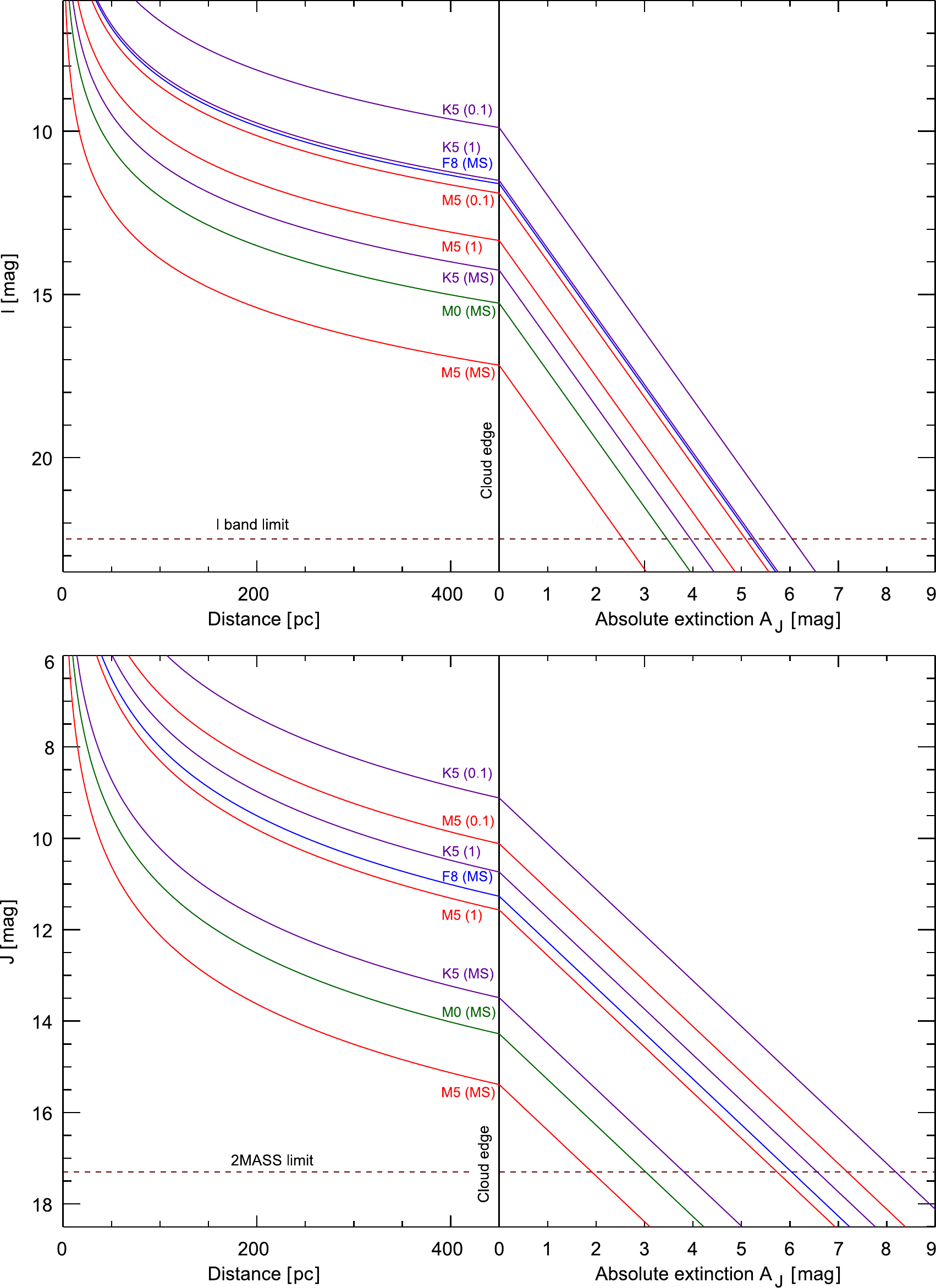}
	\caption{The completeness of the survey as determined by source absolute magnitude, distance and extinction in the $I$ and $J$ band observations.
	The age of the stars are shown in parenthesis following the spectral type. The horizontal axis represents distance and $J$ band extinction.
	Limiting magnitudes are marked by red dashed lines. ($A_J = 5$ corresponds to $A_V$ = 19.4)}
	\label{IJ_completeness}
\end{figure*}



\begin{figure*}
	\centering
	\includegraphics[width=17cm]{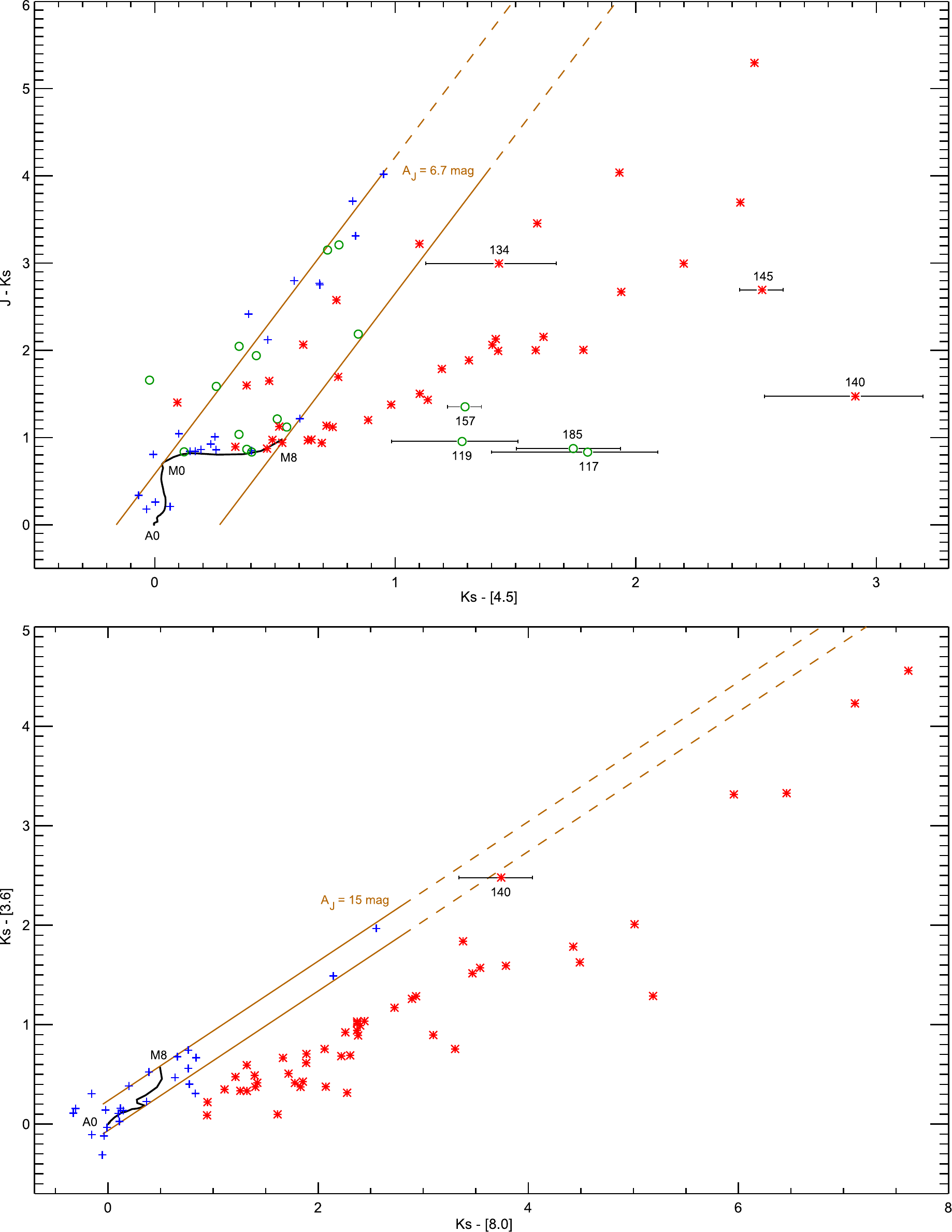}
	\caption{Additional colour-colour diagrams using ground based and Spitzer photometry. The symbols and curve/line styles
	are the same as in Figures~\ref{Colcol1} and \ref{colourdiag_1}. There is a clear separation between intrinsic non-excess and excess
	sources at both 4.5 and 8.0\,$\mu$m, with the non-excess sources agreeing well with the stellar model atmospheres and interstellar
	reddening. The blue sources are somewhat bluer that expected in the K$_S$-[4.5] and K$_S$-[3.6] indices (Fig.~\ref{colourdiag_1}), this is
	probably due to a small systematic shift in the Spitzer 3.6 and 4.5\,$\mu$m magnitude calibrations. ($A_J = 5$ corresponds to $A_V$ = 19.4)
		}
	\label{colourdiag_2}
\end{figure*}



We first consider the $I-J$ vs $J-K_S$ diagram (Fig.~\ref{colourdiag_1}). Red asterisks mark sources with a clear excess at 8\,$\mu$m, and green
circles denote sources not detected in the $[8.0]$ band.
In this Figure, both the $I$ and $J$ band sensitivity limits the sample we can use. This is shown in Figure~\ref{IJ_completeness} where the
observed $I$ and $J$ band fluxes have been calculated for increasing distances towards the cloud edge and increasing extinctions within the
cloud. For late MS M-type stars both $I$ and $J$ limits the sample to a few magnitudes of extinction in the $J$ band. However, for earlier spectral
types and younger PMS stars it is the $I$ band that sets the limit to which stars can be used (because of the higher extinction at 790\,nm).

The 'normal' stars in the $I-J$ vs $J-K_S$ diagram (blue plus signs) occupy a band defined by
the reddening vector and the curve of model stellar atmospheres. The lower border of the reddening band is relatively well defined. 
The slope is 1.81, meaning that $E(I-J) = 1.81 E(J-K_S)$. It is interesting to note that only a few of the stars with excess
at 8\,$\mu$m (red asterisks in the Figure) have significant excess emission at $K_S$.

We next turn to the $J-K_S$ vs $K_S-[3.6]$ diagram (Fig.~\ref{colourdiag_1}, lower panel). Again, we can relatively well define the reddening vector
from the 'normal' stars, and we get $E(J-K_S) = 4.5 E(K_S-[3.6])$. The reason for the 'blue' sources (plus signs) being slightly offset to the left
of the reddening band (by about 0.2--0.3 magnitudes) is probably due to the calibration of the 3.6\,$\mu$m observations.
We note that many of the stars with excess emission at 8.0\,$\mu$m also
show excess emission at 3.6\,$\mu$m. It is also clear that some of the objects not detected at 8\,$\mu$m have significant intrinsic excess
emission at 3.6\,$\mu$m. Taking the photometric accuracy for the individual sources into account, we can add these sources to the list of YSO candidates
having intrinsic IR excess emission.
The $J-K_S$ vs $K_S-[4.5]$ diagram (Fig.~\ref{colourdiag_2}) essentially confirms what is shown in the previous diagram. The slope of the reddening
vector is 3.64 and thus $E(J-K_S) = 3.64 E(K_S-[4.5])$.

Finally, we consider the $K_S-[3.6]$ vs $K_S-[8.0]$ diagram (Fig.~\ref{colourdiag_2}, lower panel). Also in this case there is a well-defined
reddening vector, $E(K_S-[3.6]) = 0.70 E(K_S-[8.0])$.
It is interesting to note that the stars with mid-IR excess emission are spread along a locus roughly parallel to the reddening vector, indicating
that these excess stars could well have intrinsic colours that differ far less than the observed colours.

Sources not detected at 8.0\,$\mu$m that are shown to have intrinsic mid-IR excess using colour-colour diagrams like these have been added to our list of YSO candidates. All sources with intrinsic mid-IR excess are marked by bold style source numbers in the Tables.


\begin{figure*}
	\centering
	\includegraphics[width=16.0cm]{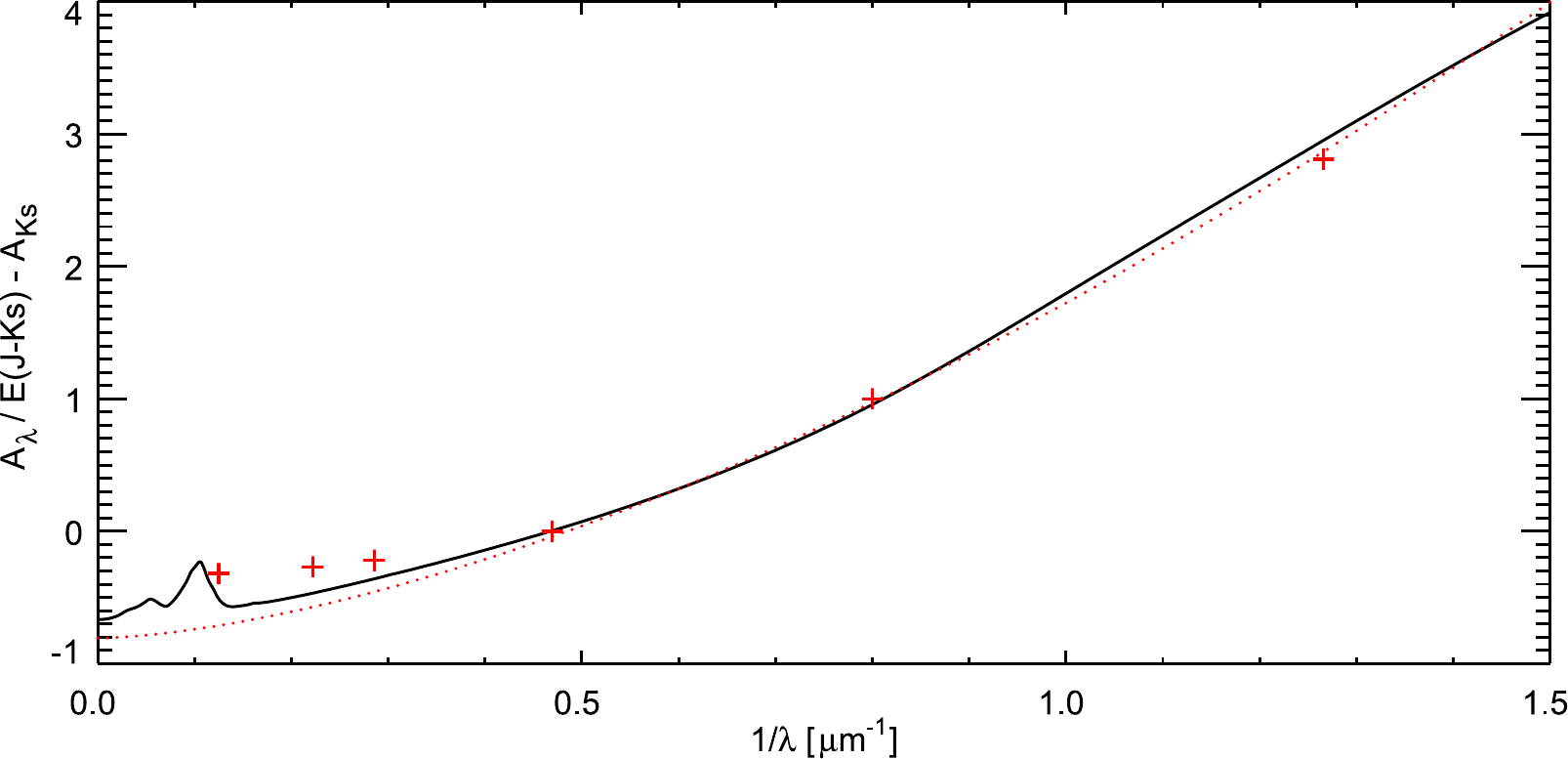}
	\caption{Interstellar extinction curve from the optical (0.67\,$\mu$m) through IR. The red plus signs represents measurements from
	our observations ($I$, $J$, $K_S$, $[3.6]$, $[4.5]$ and $[8.0]$). The dotted line is a power law fit to the observed
	E($I-J$)/E($J-K_S$) and the solid line is a theoretical $R=A_V/E(B-V)=5.5$ extinction curve from Draine
	et al. (\cite{draine}).
	}
	\label{ExtCurve}
\end{figure*}


In Fig.~\ref{ExtCurve} we summarize the reddening information derived above and as a comparison we show the theoretical extinction curve calculated
by Draine et al. (\cite{draine}) for $R = A_V/E(B-V) = 5.5$. The reason why the calculated extinction in the $[3.6]$ and $[4.5]$ bands are
significantly lower than the observed ones is not quite clear, but at least part of the explanation must be absorption bands of ices which are
not included in the model.
There may also be an additional dust component, not included in the model, which adds significantly to the mid-IR extinction, as a similar
lack of a deep extinction minimum around 5-–6 microns was observed by ISO in the diffuse interstellar medium towards the
Galactic Centre (Lutz et al.~\cite{lutz}).

We note that the ratio $E(I-J)/E(J-K_S) = 1.81$ corresponds to an exponential extinction law:

\begin{equation}
	\label{exteq1}
	\frac{A_{\lambda}}{A_J} = \left(\frac{1.25}{\lambda}\right)^{\beta}
\end{equation}

where

\begin{equation}
	\label{exteq2}
	1.81 = \frac{(A_I-A_J)}{(A_J-A_{Ks})} = \frac{A_I/A_J - 1}{1 - A_{Ks}/A_J} = \frac{(1.25/0.80)^{\beta}-1}{1 - (1.25/2.14)^{\beta}}
\end{equation}

solving this equation gives $\beta = 1.58$ and since

\begin{equation}
	\label{exteq3}
	E(I-J) = A_I-A_J = A_J\left(\frac{1.25}{0.80}\right)^{\beta} - A_J = A_J\left(\left(\frac{1.25}{0.80}\right)^{\beta}-1\right)
\end{equation}

we get an empirical absolute $J$-band extinction of

\begin{equation}
	\label{exteq4}
	A_J = 0.965 E(I-J)
\end{equation}

Equation \ref{exteq4} is used to correct our $J$ band photometry for extinction, where E($I-J$)$_0$ has been found from relating observed
and model stellar colours in the $I-J$ vs. $J-K_S$ diagram.

An extrapolation of this power law ($\beta = 1.58$) to zero wave number gives $A_{Ks} = 0.8 E(J-K_S)$ (see Fig.~\ref{ExtCurve}).
The theoretical value is similar, namely $A_{Ks} = 0.67 E(J-K_S)$. Even though there is a
discrepancy between the observed and the theoretical extinction in the $[3.6]$ and $[4.5]$ bands, we adopt this latter value and we get the
following extinction law: 

\vspace{2mm}

\( \begin{array}{lcl}
	A(I)	 & = & 3.48 E(J-K_S)\\
	A(J)	 & = & 1.67 E(J-K_S)\\
	A(Ks)	 & = & 0.67 E(J-K_S)\\
	A([3.6]) & = & 0.45 E(J-K_S)\\
	A([4.5]) & = & 0.40 E(J-K_S)\\
	A([8.0]) & = & 0.35 E(J-K_S)
   \end{array} \)


\subsection{Photometric classification and corrections for extinction}

We return to the $I-J$ vs $J-K_S$ diagram (Fig.~\ref{colourdiag_1}) and note that, assuming no excess emission in these bands, it is possible to
de-redden M-type stars and both determine their temperature and the (absolute) extinction. As the spectral energy distribution
peaks in this wavelength region for these stars we can determine their luminosity with a reasonable accuracy. On the other
hand, for hotter stars we need independent spectroscopy for the extinction correction.


\begin{figure*}
	\centering
	\includegraphics[width=18cm]{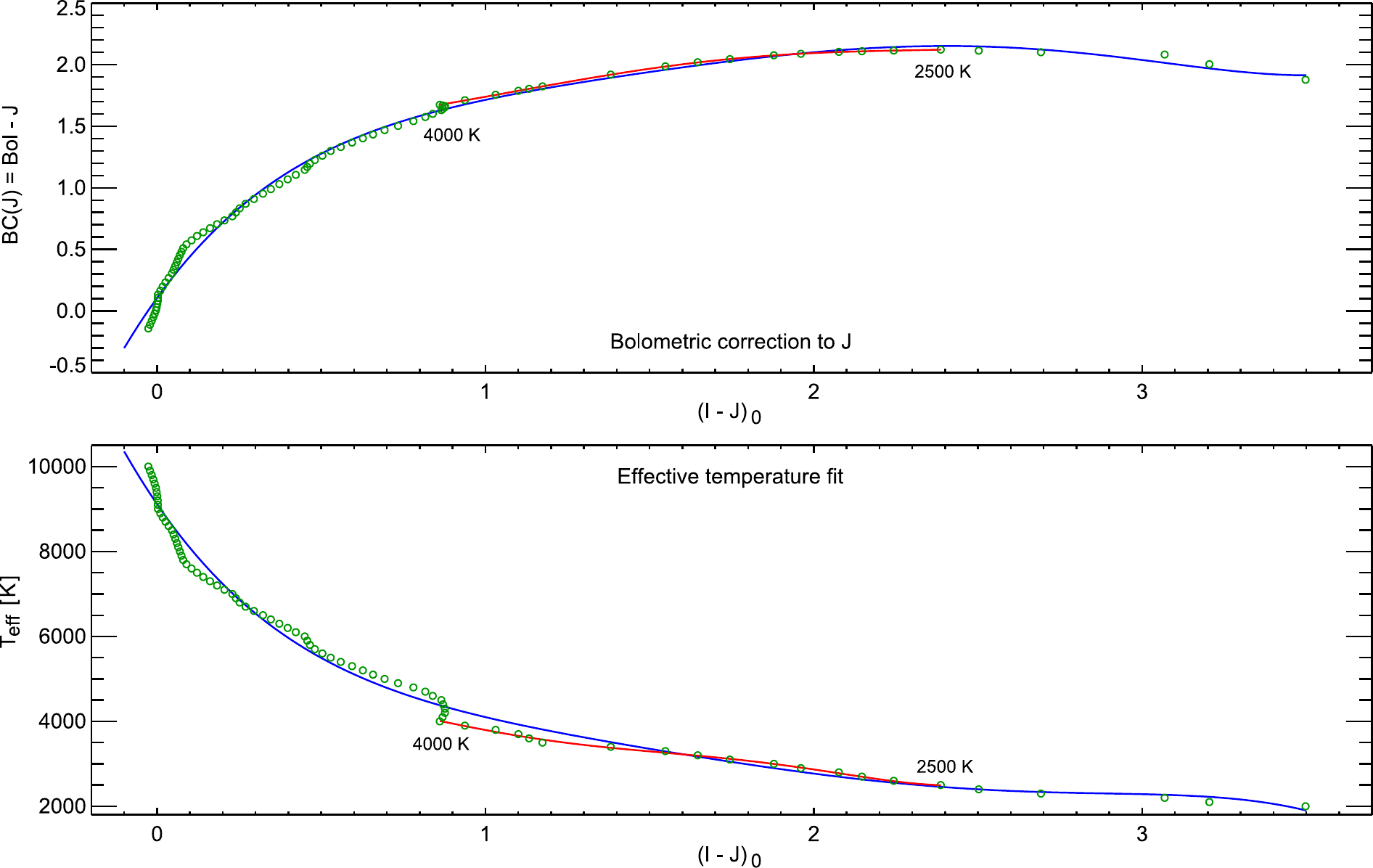}
	\caption{Polynomial fits (solid curves) of bolometric correction BC($J$) and effective temperature T$_{\textnormal{eff}}$ to intrinsic
	colour index $(I-J)_0$.
	Model atmospheres (BaSeL v2.2) with a surface gravity of $log\,g = 4.5$ have been used for calculations in the temperature
	range 2\,000--10\,000\,K with a step size of 100\,K (circles). The blue curves are fits made using all effective temperatures, while the
	red curves are more accurate fits in the 2500--4000\,K region (which we use in our photometric temperature determinations). 
	}
	\label{Bolfit}
\end{figure*}


After de-reddening the M stars in the $I-J$ vs $J-K_S$ diagram we obtain the intrinsic $(I-J)_0$ colours from where the de-reddening lines
crosses the model curve. Using polynomial fits of bolometric correction BC($J$) and T$_{\textnormal{eff}}$ to intrinsic colours
$(I-J)_0$, as shown in Figure~\ref{Bolfit}, we then calculate the luminosities and effective temperatures from photometry.
The results are presented in Table~\ref{tab_phot_mass_age}.


\begin{table}
	\caption{Photometric mass and age estimates.}
	\label{tab_phot_mass_age}
	\begin{tabular}{lcccccl}
	  \hline
        \noalign{\vspace{0.5mm}}
        No. & A$_J$ & T$_{\textnormal{eff}}$ & m$_{bol}$ & L & M & Age \\
            & (mag) & (K)                    & (mag)     & (L$_{\sun}$) & (M$_{\sun}$) & (Myr) \\
 	\noalign{\vspace{0.5mm}}
        \hline

	\noalign{\vspace{1.0mm}}

	{\bf 23}    & 1.69 & 3560 & 6.47 & 0.21 & 0.43 & 3.14 \\
	28    & 1.37 & 3110 & 9.54 & 0.012 & 0.13 & 18.71 \\
	31    & 2.79 & 3550 & 6.76 & 0.16 & 0.43 & 4.98 \\
	47$^{\mathrm b}$    &       1.46 & 3470 & 8.78 & 0.024 & 0.40 & 80.41 \\
	{\bf 64$^{\mathrm b}$}    & 0.71 & 3180 & 9.96 & 0.008 & 0.16 & 52.89 \\
	80    &       3.41 & 3150 & 6.74 & 0.16 & 0.19 & 0.77 \\
	84    & $\sim$0.41 & .... & .... & .... & .... & .... \\
	92    & 2.11 & $\sim$2500 & 8.68 & 0.027 & $<0.1$ & out$^{\mathrm a}$ \\
	118   	    & 2.28 & 3330 & 5.76 & 0.39 & 0.25 & 0.37 \\
	128   	    & 3.45 & 3410 & 5.36 & 0.57 & 0.28 & 0.26 \\
	{\bf 133}   & 4.66 & 3840 & 4.03 & 1.93 & 0.43 & 0.13 \\
	{\bf 138 a} & $\sim$4.61 & .... & .... & .... & .... & .... \\
	{\bf 138 b} & $\sim$3.46 & .... & .... & .... & .... & .... \\
	{\bf 140$^{\mathrm b}$}   & 1.14 & 3320 & 9.74 & 0.010 & 0.26 & 115.45 \\
	{\bf 141}   & $\sim$1.40 & .... & .... & .... & .... & .... \\
	148   	    & 5.05 & 4000 & 4.24 & 1.59 & 0.55 & 0.31 \\
	{\bf 156}   & $\sim$0.11 & .... & .... & .... & .... & .... \\
	{\bf 157}   & $\sim$0.97 & .... & .... & .... & .... & .... \\
	{\bf 167$^{\mathrm b}$}   & 3.11 & 3750 & 7.69 & 0.07 & 0.58 & 52.56 \\
	168   	    & 4.37 & 3630 & 2.94 & 5.25 & 0.31 & 0.01 \\
	170   & 1.87 & $\sim$2500 & 9.00 & 0.020 & $<0.1$ & out$^{\mathrm a}$ \\
	{\bf 174}   & 2.07 & 2940 & 8.80 & 0.024 & $<0.1$ & $\sim$2$^{\mathrm a}$ \\
	186   & 4.12 & 3810 & 6.67 & 0.17 & 0.69 & 12.21 \\

	\noalign{\vspace{1.0mm}}
	\hline

	\end{tabular}

	\begin{list}{}{}
		\item[$^{\mathrm{a}}$] Outside model grid (mass lower than 0.10\,M$_{\sun}$)
		\item[$^{\mathrm{b}}$] Close to MS
	\end{list}

\end{table}



\subsection{Spectral classification}

\label{spec_sec}

\begin{figure*}
	\centering
	\includegraphics[width=17cm]{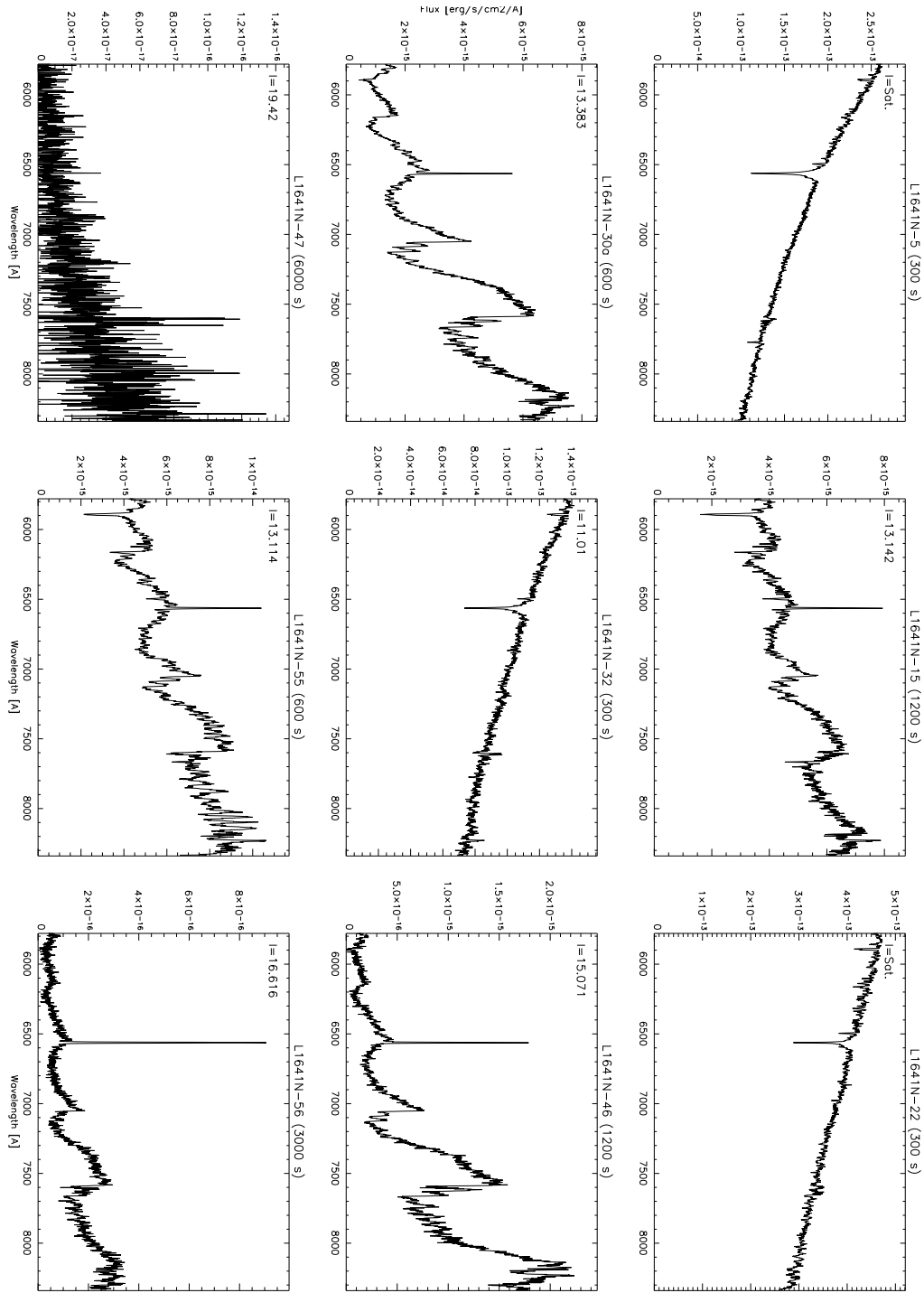}
	\caption{Optical spectra - Exposure times and $I$ band magnitudes are given for each source.
		}
	\label{Spectra_opt}
\end{figure*}

In total we have taken optical spectra (5780--8340 \AA) of 53 sources in L1641N, the observation log and results are presented in
Table~\ref{tab_spectra} and the spectra themselves in Fig.~\ref{Spectra_opt}.
These stars were selected from our $I$ band image and time constraints of our two spectroscopic observing runs, starting with spectra of the brightest
$I$ band sources and observing progressively fainter stars until the total exposure time needed for a single spectrum reached about 2\,hours.

\begin{table*}
	\caption{Optical spectroscopy - observation log and results.}
	\label{tab_spectra}
	\begin{tabular}{lcclccccrrl}
	  \hline
        \noalign{\vspace{0.5mm}}
        No. & t   & A$_J$ & T$_{\textnormal{eff}}$ & m$_{bol}$ & L           & M            & Age   & W$_{H \alpha}$ & W$_{Li}$ & Comment \\
            & (s) & (mag) & (K)                    & (mag)  & (L$_{\sun}$) & (M$_{\sun}$) & (Myr) & (\AA)          & (\AA) & \\

	\noalign{\vspace{0.5mm}}
        \hline

	\noalign{\vspace{1.0mm}}

	5    & 300 & $\sim$0 & 6500 & 2.47 & 8.10 & 1.61 & 9.62 & 4.13 & .... & Early (F5) \\
	15  & 1200 & 0.20 & 3900 (3640) & 5.07 & 0.74 & 0.56 & 0.93 & -4.80 & 0.58 & \\	
	22   & 300 & 0.05 & 5800 & 1.75 & 15.64 & 2.59 & 2.11 & 2.15 & .... & Early (G4) \\
	30 a & 600 & 0.33 & 3200 (3270) & 4.92 & 0.85 & 0.19 & 0.03 & -9.66 & .... & \\
	32   & 300 & 0.00 & 6200 & 2.85 & 5.70 & 1.54 & 10.04 & 3.07 & .... & Early (F8) \\
	{\bf 46}  & 1200 & 0.00 & 2900 (2510) & 6.53 & 0.19 & 0.11 & 0.08 & -27.43 & .... \\
	47$^{\mathrm d}$  & 6000 & .... & .... & ... & .... & .... & .... & .... & .... & Too faint \\
	55   & 600 & 0.03 & 3700 (3420) & 5.23 & 0.64 & 0.42 & 0.61 & -6.21 & 0.40 & \\	
	{\bf 56}  & 3000 & 0.24 & 3200 (3130) & 8.16 & 0.04 & 0.20 & 6.49 & -66.10 & .... & \\
	{\bf 60$^{\mathrm d}$}  & 3600 & 1.23 & 4600 & 6.36 & 0.22 & 0.75 & 51.64 & .... & .... & \\
	61  & 900 & 0.01 & 4100 & 5.10 & 0.72 & 0.75 & 1.93 & -2.29 & 0.32 & \\	
	{\bf 64$^{\mathrm d}$}  & 3600 & .... & .... & .... & .... & .... & .... & .... & .... & Too faint \\
	65  & 6000 & 1.38 & 3000 & 7.07 & 0.12 & 0.13 & 0.47 & -22.42 & .... & \\
	{\bf 66 a}& 3600 & 1.55 & 4050 & 4.85 & 0.91 & 0.66 & 1.03 & -21.56 & .... & \\
	{\bf 66 b}& 3600 & 2.08 & 3400 (3650) & 5.69 & 0.42 & 0.28 & 0.44 & -56.56 & .... & \\	
	{\bf 68}  & 3000 & 0.28 & 3050 (3190) & 8.26 & 0.04 & 0.13 & 2.78 & -11.25 & .... & \\	
	{\bf 73}  & 1200 & 0.97 & 3400 (3530) & 5.41 & 0.54 & 0.27 & 0.27 & -39.32 & .... & \\	
	{\bf 75 a}& 2700 & 1.39 & 3750 (3900) & 3.86 & 2.25 & 0.37 & 0.07 & -5.01 & 0.20 & \\	
	{\bf 75 b}& 1500 & 1.80 & 3750 (3640) & 4.70 & 1.04 & 0.41 & 0.29 & -7.25 & .... & \\	
	78  & 1200 & 0.07 & 3500 (3370) & 6.26 & 0.25 & 0.37 & 1.73 & -3.44 & 0.53 & \\	
	84  & 6000 & .... & .... & .... & .... & .... & .... & .... & .... & Too faint \\	
	98   & 300 & $\sim$0 & 5700 & 2.71 & 6.51 & 2.02 & 3.90 & 2.03 & .... & Early (G6) \\
	{\bf 114} & 1200 & 0.18 & 3800 & 5.50 & 0.50 & 0.53 & 1.41 & -44.80 & 0.49 & \\
	{\bf 117} & 3600 & 0.03 & 3500 (3300) & 6.81 & 0.15 & 0.39 & 4.31 & -5.27 & .... & \\	
	118 & 1200 & .... & .... & .... & .... & .... & .... & .... & .... & Too faint \\
	{\bf 119} & 1200 & 0.20 & 2900 (2830) & 8.53 & 0.03 & $<0.1$ & out$^{\mathrm a}$ & -13.95 & .... & \\
	120 & 3600 & 0.46 & 3300 (3940) & 8.90 & 0.02 & 0.26 & 30.92 & -10.39 & .... & \\
	{\bf 121} & 1200 & 1.26 & 3200 (3960) & 5.65 & 0.43 & 0.20 & 0.18 & -123.0 & .... & \\	
	{\bf 142} & 7200 & 3.84 & 3800 & 3.51 & 3.12 & 0.39 & 0.05 & -70.69 & .... & \\
	{\bf 144$^{\mathrm c}$ a} & 1200 & 0.53 & 4000 & 5.90 & 0.34 & 0.79 & 6.08 & -18.29 & 0.66 & \\	
	{\bf 144$^{\mathrm c}$ b} & 1200 & 0.47 & 4000 & 6.70 & 0.16 & 0.76 & 25.53 & -40.46 & .... & \\
	{\bf 145} & 6000 & 3.28 & 3300 (3550) & 3.83 & 2.32 & $\sim$0.20 & out$^{\mathrm a}$ & -12.67 & .... & \\	
	{\bf 152} & 6000 & 1.85 & 3800 & 5.49 & 0.50 & 0.52 & 1.38 & -243.34 & 1.30 & \\
	160  & 600 & 0.68 & 3600 (3250) & 6.08 & 0.29 & 0.43 & 1.88 & -8.14 & 0.88 & \\	
	{\bf 165} & 3000 & .... & .... & .... & .... & .... & .... & .... & .... & Refl.Neb. \\
	168 & 5400 & .... & .... & .... & .... & .... & .... & .... & .... & Too faint \\
	{\bf 171} & 1200 & 1.52 & 3400 (3410) & 4.80 & 0.94 & 0.25 & 0.09 & -16.46 & .... & \\	
	{\bf 175} & 5400 & 2.18 & 2650 (2940) & 6.39 & 0.22 & $<0.1$ & out$^{\mathrm a}$ & .... & .... & \\
	177 & 1800 & 0.06 & 2900 & 8.44 & 0.03 & $<0.1$ & $\sim$1 & -13.70 & 0.86 & \\
	{\bf 179} & 1800 & 0.13 & 3000 & 7.58 & 0.07 & 0.12 & 0.86 & -17.62 & .... & \\
	184 & 1200 & 0.73 & 2650 & 7.34 & 0.09 & $<0.1$ & out$^{\mathrm a}$ & -74.22 & .... & \\	
	{\bf 185} & 3000 & 0.09 & 3300 (3330) & 8.33 & 0.04 & 0.26 & 14.71 & .... & .... & \\
	187 & 1200 & 0.39 & 4000 (3550) & 4.06 & 1.87 & 0.53 & 0.23 & -3.18 & 0.40 & \\
	{\bf 188} & 3000 & 0.28 & 3100 (3270) & 6.74 & 0.16 & 0.17 & 0.60 & -16.91 & 0.58 & \\
	190$^{\mathrm d}$ & 3000 & 1.16 & 3600 (3400) & 7.94 & 0.05 & 0.80 & 42.52 & .... & .... & \\
	{\bf 192} & 1200 & 0.58 & 3100 (3220) & 5.18 & 0.67 & 0.17 & 0.03 & -36.55 & .... & \\	
	{\bf 195} & 1800 & 0.62 & 3400 (3290) & 5.95 & 0.33 & 0.29 & 0.69 & -21.94 & 0.18 & \\
	{\bf 204} & 1200 & 0.34 & 4200 & 5.25 & 0.63 & 0.90 & 3.54 & -7.31 & 0.50 & \\
	{\bf 205} & 1800 & 0.24 & 3000 (3050) & 8.65 & 0.03 & 0.10 & 2.85 & -36.12 & .... & \\	
	{\bf 208} & 1800 & 0.53 & 3050 (3380) & 8.21 & 0.04 & 0.13 & 2.57 & -36.96 & .... & \\	
	{\bf 210 a}& 1200 & 0.24 & 5000 & 3.99 & 2.00 & 1.55 & 3.85 & -1.7$^{\mathrm b}$ & 0.30 & \\	
	{\bf 210 b}& 2400 & 0.06 & 2800 & 8.75 & 0.03 & $<0.1$ & out$^{\mathrm a}$ & -3.12 & .... & \\	
	215 & 1200 & 0.09 & 3100 (3020) & 7.11 & 0.11 & 0.17 & 0.97 & -9.74 & 1.06 & \\	

	\noalign{\vspace{1.0mm}}
	\hline

	\end{tabular}

	\begin{list}{}{}
		\item[$^{\mathrm{a}}$] Outside model grid (mass lower than 0.10\,M$_{\sun}$)
		\item[$^{\mathrm{b}}$] P Cygni profile
		\item[$^{\mathrm{c}}$] Double source not resolved in $J$ or $K_S$
		\item[$^{\mathrm{d}}$] Close to MS
	\end{list}

\end{table*}


\begin{figure*}
	\centering
	\includegraphics[width=15cm]{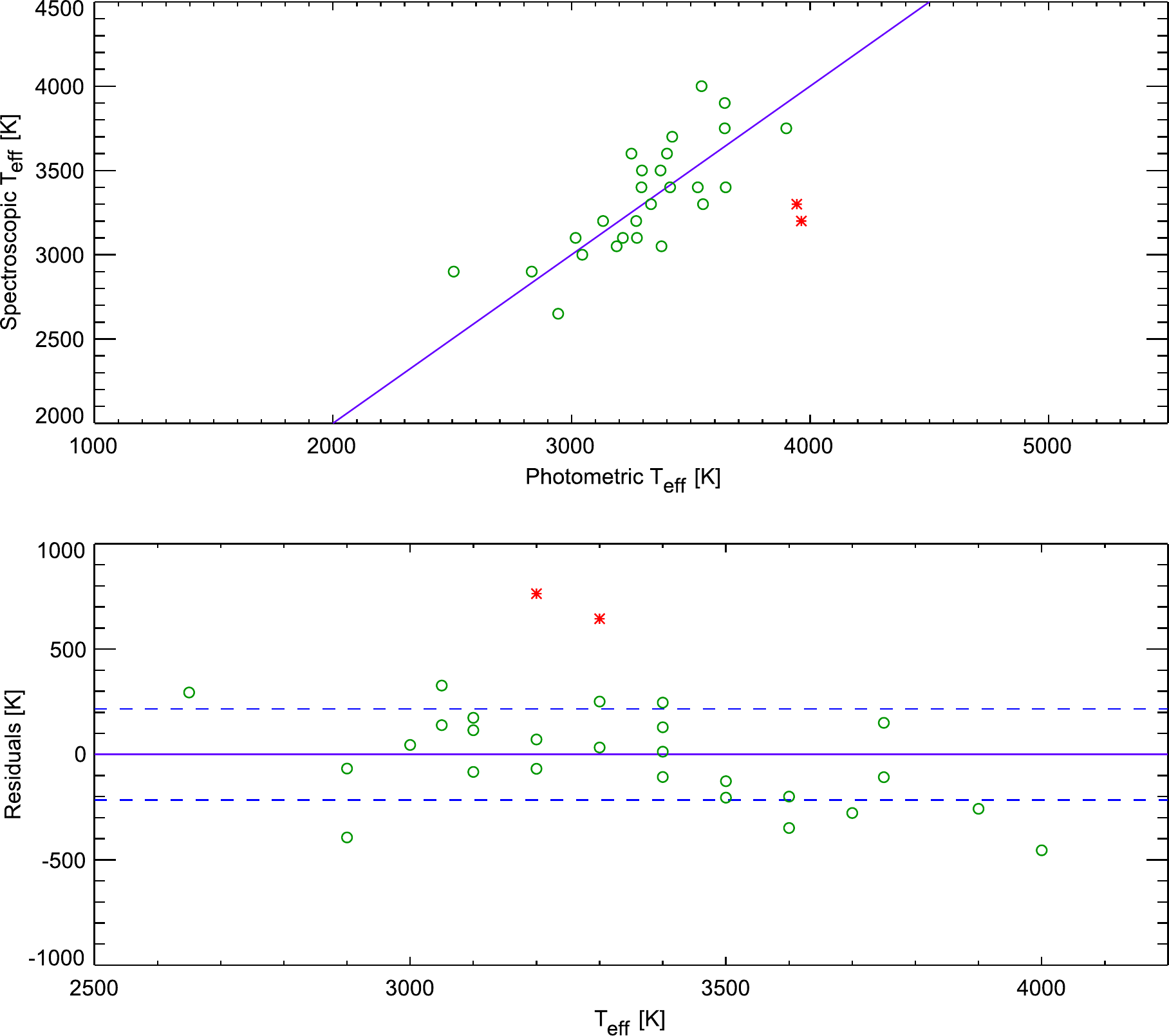}
	\caption{Comparison of spectroscopic and photometrically derived effective temperatures. The solid line
	indicates where T$_{\textnormal{eff}}$(spec) = T$_{\textnormal{eff}}$(phot). As can be seen in the lower panel, the residuals are
	fairly well centred on the solid line with a standard deviation of a few hundred Kelvins. The red asterisks represent two
	sources, L1641N-120 and 121, having larger residuals than the other stars. This is probably caused by a small amount of intrinsic excess
	emission in the K$_S$ band (enough to keep them within the reddening band in Fig.\,\ref{colourdiag_1} but yielding higher photometric temperatures).
	}
	\label{Tcompare}
\end{figure*}


In order to compare our observations to evolutionary models, we need the effective temperature, T$_{\textnormal{eff}}$. For this determination we use
synthetic spectra provided by P.~Hauschildt (personal communication) and in Table~\ref{tab_spectra} we give the results, where we also include the
photometrically determined temperatures. The agreement is relatively good (see Fig.\ref{Tcompare}) and the scatter is about 200\,K (1\,$\sigma$).
These GAIA-V2.0 LTE model spectra cover $T_{\textnormal{eff}} = 2000-4000\,K$ in steps of 100\,K (late K and M spectral types).
Smoothing was used to adapt the grid of high-resolution model spectra (0.1\,\AA~resolution) to the resolution of the observations.
We assumed a surface gravity of $log(g) = 4.5$ for all stars and used the extinctions, A$_J$, obtained from photometry to deredden the spectra.

For the temperature fits, we used TiO (Valenti et al. \cite{valenti}) absorption band heads (spectral types earlier than M6), VO absorption
features (later than M6) and the general shape of the spectra. If a photometrically obtained extinction differed markedly from that observed in
a spectrum, we re-fitted A$_J$ spectroscopically using the same extinction law as in the photometry.
In the visual comparison, we normalized the flux between the models and the spectra using a plateau around 7500\,\AA.
The fit in temperature (and thus the spectral type) has roughly the same accuracy as the model steps themselves (100\,K).
Our effective temperatures agree fairly well with those also found from $K$ band spectra in Hodapp \& Deane (\cite{hodapp93}), especially for early spectral types. A few of the late-type stars in our sample (L1641N-73, 145 and 192) have much lower temperatures (about 1000\,K) than those found from the $K$ band spectra.

The luminosity of the spectroscopic sample is calculated in the same way as for the photometric sample, by de-reddening the stars in the
$I-J$ vs $J-K_S$ diagram (yielding the extinction A$_J$ and intrinsic colour $(I-J)_0$). However, if the extinction found photometrically does not
fit the spectrum, A$_J$ is instead estimated from a spectroscopic fit. Then follows calculations of the bolometric $J$ band correction, BC($J$), and
finally the luminosity is calculated (assuming a distance of $\sim$450\,pc).

\begin{figure*}
	\centering
	\includegraphics[width=17cm]{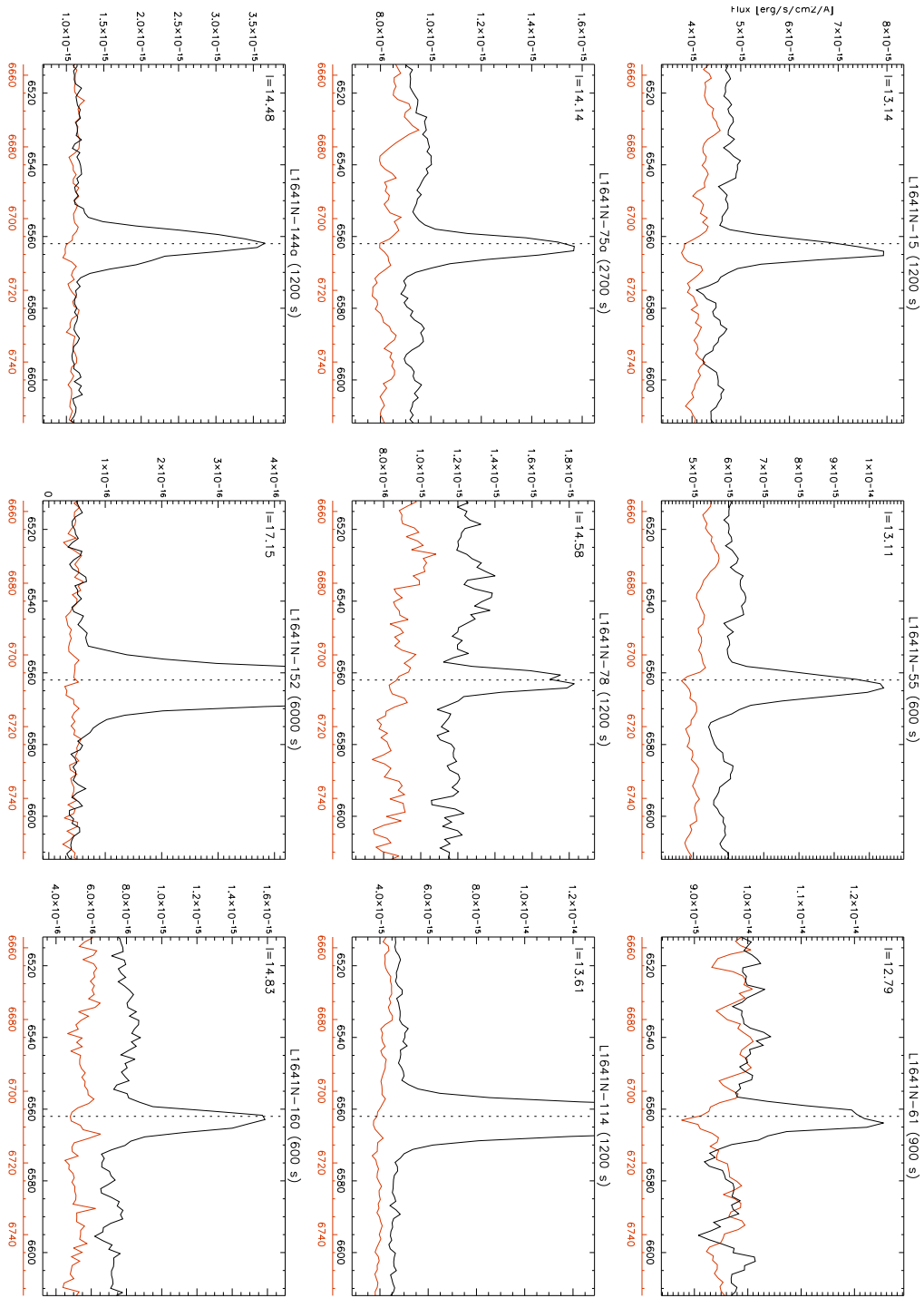}
	\caption{Evidence of youth - H$\alpha$ in emission and Li\,I\,$\lambda6707$ in absorption. The shifts in wavelength for each source is an
	artefact caused by each star being differently positioned within the wide slit (1\farcs2).
		}
	\label{Spectra_LiHa}
\end{figure*}

Most of the stars clearly have late type (mostly M) spectra typical of young dwarfs.
One way to confirm their youth, and thus that they belong to the cluster, is the presence of H$\alpha$ in
emission (tracing accretion) and Li\,I\,$\lambda6707$ in absorption (lithium has been destroyed in more evolved pre-main sequence stars).
In Fig.~\ref{Spectra_LiHa} we show some cases that have both, and in Table~\ref{tab_spectra} we give the equivalent widths
of these lines where applicable. 
Even though the spectral resolution of our spectra (R=830) is marginal for detecting the Li\,6707 line, in particular as the S/N is in general poor
due to the faintness of the stars, it was found present in many stars.

With the exception of the four brightest stars, almost all of the spectra show H$\alpha$ in emission and are of M-type.
Since also main sequence M dwarfs frequently show H$\alpha$ in emission, this by itself is not enough to classify them as YSOs.
On the other hand, the combination of intrinsic IR excess and H$\alpha$ emission is a reliable indication of YSO status. In addition, if the star
exhibits interstellar reddening, it must be at least at the distance of the cloud, and we can judge from its luminosity if it is a PMS star.


\subsection{Contamination and multiplicity in the YSO sample}


\subsubsection{Background sources}

An optical (or $I$ band) image of L1641N shows that the extinction is high across the whole cluster.
Our $I$ band image shows a total of 87 sources down to $I$~$\sim$22.5\,mag while the $K_S$ images show most of the Spitzer sources and
many additional sources (limiting magnitude $K_S$~$\sim$19.5\,mag). It is very likely that a number of $K_S$ sources in fact are
extragalactic objects. However, as can be seen in the comment column in Table~\ref{L1641N_ground} none of the $I$ band sources appear to be
extragalactic. As we need $I$ band photometry (or spectra at similar wavelengths) to calculate the luminosities and effective temperatures used
in the H-R diagram, most likely we have no extragalactic objects in the sample we used to calculate the MF.

The situation is similar for the $[4.5]$ and $[8.0]$ observations, which are the main filters we use to find YSO candidates through
intrinsic IR excess. The extinction is much less at these wavelengths. Even so, given the very high background (and high contrast) in these filters
due to PAH emission (Fig.~\ref{Spitzer124}), the sensitivity limit is severely affected. It is therefore unlikely that any extragalactic sources
are included as YSO candidates. This is supported by the $[8.0]$ vs $[4.5]-[8.0]$ diagram (Fig.~\ref{Colcol1}) in which just a few sources are
seen below the 'extragalactic contamination' line. In addition, all these sources have a point-like appearance in the $K_S$ band.

Background giants is another possible source of contamination. However, for all YSO candidates with intrinsic IR
excess (Table~\ref{YSO_candidates_IR}) and for most of the non-excess YSO candidates (Table~\ref{YSO_candidates_noIR}) we have other evidence
for youth and cluster membership. Even though we cannot rule out the possibility of background giants among the non-excess YSO candidates, for
which we lack spectra and for which the extinction A$_J$ is high, these should be very few.


\subsubsection{Early type and foreground stars}

\label{foreground_sec}


\begin{table*}
	\caption{The six earliest stars in our spectroscopic sample.}
	\label{foreground_stars}
	\begin{tabular}{ccccccccccl}
	  \hline
        \noalign{\vspace{0.5mm}}
        No. & T$_{\textnormal{eff}}$ & MS Type & M(V)$^{\mathrm a}$  & V-J$^{\mathrm b}$ & M(J)  & m(J)  & A$_J$ & Distance modulus & MS distance$^{\mathrm c}$ & Comment\\
            &                        &         & (mag) & (mag) & (mag) & (mag) & (mag) & $\mu$            & (pc)     & \\
 	\noalign{\vspace{0.5mm}}
        \hline

	\noalign{\vspace{1.0mm}}


	5  &	 6500 & F5    & 3.68 & 0.80 & 2.88 & 9.78 & 0.00 & 6.90 & 240 & Foreground star \\
	22 &	 5800 & G4    & 4.70 & 1.15 & 3.55 & 8.89 & 0.05 & 5.29 & 114 & Foreground star \\
	32 &	 6200 & F8    & 4.00 & 1.00 & 3.00 & 9.97 & 0.00 & 6.97 & 248 & Foreground star \\
	\bf{60} & 4600 & K4   & 6.92 & 1.95 & 4.97 & 14.30 & 1.23 & 8.10 & 417 & Belongs to the cluster \\
	98 &	 5700 & G6    & 4.90 & 1.18 & 3.72 & 9.76 & 0.00 & 6.04 & 161 & Foreground star \\
	\bf{210 a} & 5000 & K1 & 6.13 & 1.57 & 4.56 & 11.06 & 0.24 & 6.26 & 179$^{\mathrm d}$ & Belongs to the cluster \\

	\noalign{\vspace{1.0mm}}
	\hline

	\end{tabular}

	 \begin{list}{}{}
		\item[$^{\mathrm{a}}$]From Table 15.7 in Allen's astrophysical quantities (\cite{allens}).
		\item[$^{\mathrm{b}}$]From Kenyon \& Hartmann (\cite{kenyon}).
		\item[$^{\mathrm{c}}$]Luminosity distance; assuming that all the stars have reached the MS.
		\item[$^{\mathrm{d}}$]This is a young pre-MS star and thus much brighter than assumed for the MS luminosity distance.
	\end{list}
\
\end{table*}


Given the distance to L1641N (d$\sim$450\,pc), and the limiting magnitudes of our I, J and K$_S$ observations, we expect that all MS foreground stars
are seen in the observations (see Fig.~\ref{IJ_completeness}) . Cluster membership status for the late type stars in our sample is found through
a combination of intrinsic IR excess (from dust), spectral lines (H$\alpha$ in emission, Li\,6707 in absorption) and extinction. This method is less
applicable for earlier, more massive young stars (which have a much quicker star formation process).

The earliest stars in our sample (F and G stars) have virtually no extinction and are much brighter than the rest of the stars
(predominantly M-type stars). Since there are no Hipparcos data for these sources, we could not make distance estimates from
parallaxes.

The six earliest (and therefore among the most luminous) stars in our spectroscopic sample are listed in Table~\ref{foreground_stars}.
Their high apparent brightness could be caused by a combination of nearby distance, low extinction and high intrinsic luminosity.
Using spectra to calculate their extinctions and distances, cluster membership status can be investigated.
We assume in the calculations that these stars have all reached the MS (because of their relatively high temperatures and thus higher masses), however,
the coolest of these are more likely to yield closer MS luminosity distances than their actual distances since
they are probably still pre-MS stars and thus brighter than assumed.

Both L1641N-60 and 210A are YSO candidates, with measureable J band extinction, and thus belong to the cluster even though the MS
luminosity distance of 210A is only 179\,pc (a bright PMS star). We will return to this later, as it is supported by their positions
in the H-R diagram.

From the Wainscoat et al. (\cite{wainscoat}) model of the galaxy, we should expect about four foreground stars for spectral types F0--K5. This agrees
with the four F-G stars we see in the sample. However, we should then also expect roughly nine M-type foreground stars, whereas we see virtually
no late-type foreground stars. These would show up as M stars with almost no extinction, but for most of these stars we instead have evidence for
them being YSOs. For the four early type stars, we find no signature of youth. From this result, combined with the lack of extinction, we must
assume that these F and G stars are foreground stars.


\subsubsection{Double sources}

\label{binary_sec}


\begin{table*}
	\caption{All known double$^{\mathrm a}$ sources in L1641N.}
	\label{binaries}
	\begin{tabular}{rcclllcl}
	  \hline
        \noalign{\vspace{0.5mm}}
        No. & RA & Dec & \multicolumn{1}{c}{I}& \multicolumn{1}{c}{J} & \multicolumn{1}{c}{K$_S$} & Separation & Comment \\
 	\noalign{\vspace{0.5mm}}
            & (2000)    & (2000)     & \multicolumn{1}{c}{(mag)}   & \multicolumn{1}{c}{(mag)}   & \multicolumn{1}{c}{(mag)}   & (arcsec) & \\
 	\noalign{\vspace{0.5mm}}
        \hline

	\noalign{\vspace{1.0mm}}

	30 a & 05:36:06.943 & -06:18:53.23 & 13.383 $\pm$ 0.009 & 11.500 $\pm$ 0.023$^{\mathrm b}$ & 10.496 $\pm$ 0.002 & 6.19 & High contrast pair \\
	   b & 05:36:06.598 & -06:18:49.76 & 21.694 $\pm$ 0.242 & 				   & 16.637 $\pm$ 0.170 &      & \\

	\noalign{\vspace{1.0mm}}

	32 a & 05:36:06.870 & -06:23:34.94 & 11.010 $\pm$ 0.110$^{\mathrm b}$ & 09.965 $\pm$ 0.021$^{\mathrm b}$ & 09.787 $\pm$ 0.001 & 1.93 	& \\
	   b & 05:36:06.991 & -06:23:35.63 &                                  &                                  & 13.161 $\pm$ 0.030 & & \\

	\noalign{\vspace{1.0mm}}

	{\bf 66 a} & 05:36:11.474 & -06:22:21.82 & 16.078 $\pm$ 0.010 & 12.931 $\pm$ 0.004 & 10.761 $\pm$ 0.010 & 3.31 & \\
	{\bf    b} & 05:36:11.446 & -06:22:25.10 & 17.386 $\pm$ 0.011 & 14.121 $\pm$ 0.020 & 12.123 $\pm$ 0.040 & & \\  

	\noalign{\vspace{1.0mm}}

	{\bf 75 a} & 05:36:12.969 & -06:23:32.13 & 14.139 $\pm$ 0.025 & 11.762 $\pm$ 0.002 & 10.181 $\pm$ 0.010 & 2.78 & \\
	{\bf    b} & 05:36:12.984 & -06:23:29.35 & 15.980 $\pm$ 0.060 & 13.001 $\pm$ 0.040 & 11.162 $\pm$ 0.010  & & \\  

	\noalign{\vspace{1.0mm}}

	{\bf 110 a} & 05:36:17.859 & -06:22:28.55 & & & 15.650 $\pm$ 0.070 & 0.65 & \\
	{\bf     b} & 05:36:17.906 & -06:22:28.07 & & & 16.399 $\pm$ 0.100 &      & \\  

	\noalign{\vspace{1.0mm}}

	{\bf 138 a} & 05:36:21.044 & -06:21:53.88 & 19.004 $\pm$ 0.050 & 13.749 $\pm$ 0.010 & 10.315 $\pm$ 0.010 & 1.61 & observed in L' band\\
	{\bf     b} & 05:36:20.961 & -06:21:52.84 & 19.934 $\pm$ 0.100 & 14.221 $\pm$ 0.010 & 11.443 $\pm$ 0.010 & & \\  

	\noalign{\vspace{1.0mm}}

	{\bf 144 a} & 05:36:21.880 & -06:26:01.71 & 14.430 $\pm$ 0.030 & 12.267 $\pm$ 0.026$^{\mathrm b}$ & 10.262 $\pm$ 0.028$^{\mathrm b}$ & 1.03 & \\
	{\bf     b} & 05:36:21.814 & -06:26:02.08 & 15.107 $\pm$ 0.150 &                                  &                                  &      & \\  

	\noalign{\vspace{1.0mm}}

	{\bf 210 a} & 05:36:32.384 & -06:19:19.85 & 12.440 $\pm$ 0.100 & 11.059 $\pm$ 0.024 & 09.860 $\pm$ 0.027 & 5.27 & High contrast pair \\
	{\bf     b} & 05:36:32.687 & -06:19:22.54 & 17.170 $\pm$ 0.150 & 14.973 $\pm$ 0.117 & 14.140 $\pm$ 0.030 & & \\  
	
	\noalign{\vspace{1.0mm}}
	\hline

	\end{tabular}

	 \begin{list}{}{}
		\item[$^{\mathrm{a}}$]Unresolved with IRAC.
		\item[$^{\mathrm{b}}$]Total flux of both components.
	\end{list}
\end{table*}



\begin{figure*}
	\centering
	\includegraphics[width=15cm]{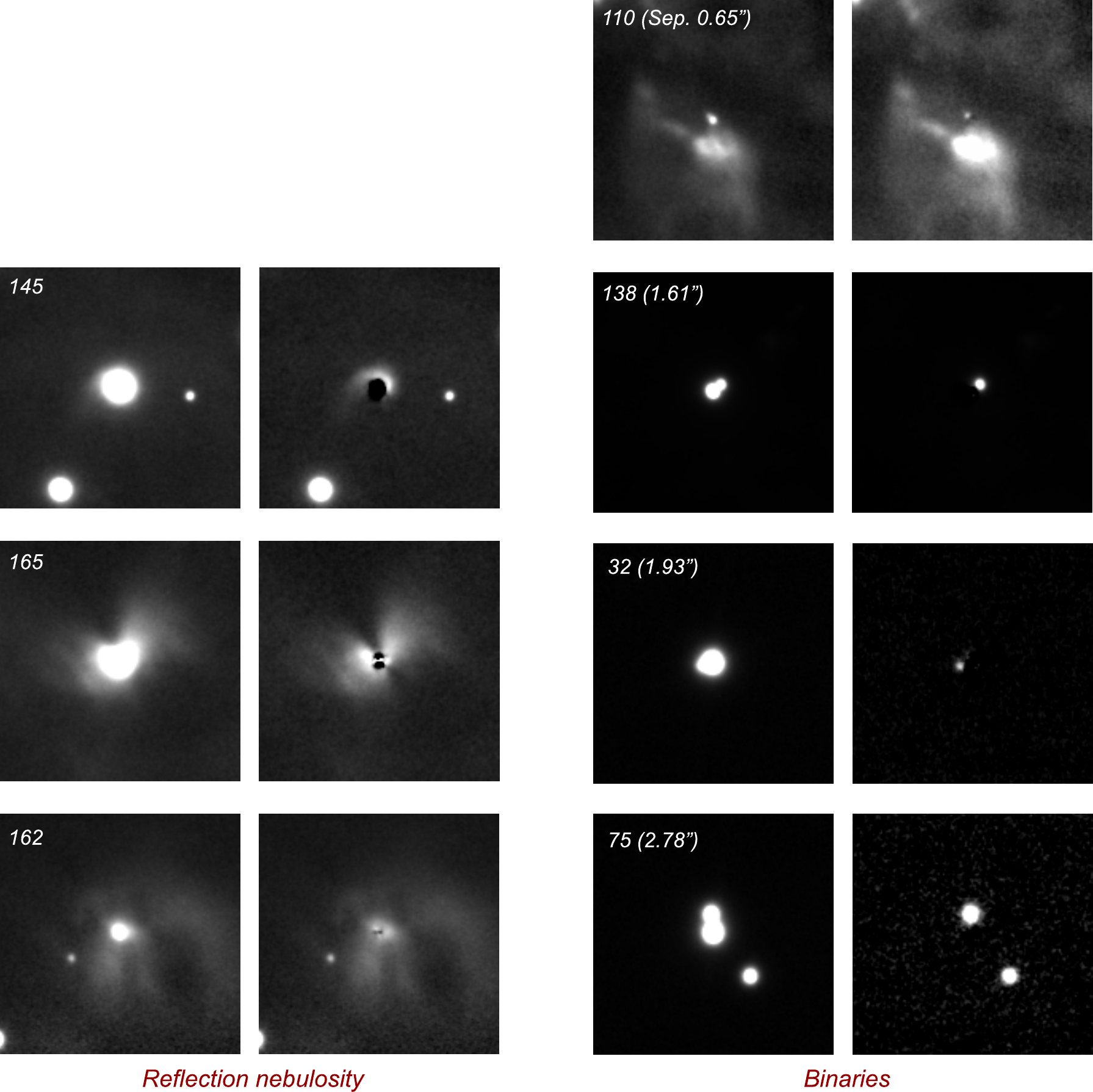}
	\caption{$K_S$~band images of reflection nebulosity (left panel) and double sources (right panel) with and without the main
	source (illuminating source and primary, respectively) removed using a normalized PSF from an isolated star.
	Each image has a field size of 34\farcs7 $\times$ 34\farcs7.
	}
	\label{PSFremoved}
\end{figure*}


Using our $I$, $J$, $K_S$ and $L'$ images we have visually detected eight double sources in the sample. These are listed
in Table~\ref{binaries} including coordinates, separations and component photometry.
We do not claim that these are physically bound binaries, although the ones with small separations most probably are. The criterion we use is
that they are not resolved in the Spitzer observations.

Most of the doubles with fairly small separations were found by subtracting a fitted PSF from
all sources in the sample, sometimes revealing a second component. In Figure~\ref{PSFremoved} some of these subtractions
are shown for both double sources (primary subtracted) and reflection nebulae (illuminating source removed to decrease
contrast).

In all Tables we have used the suffix {\it a} (primary) and {\it b} (secondary) when we denote the components of a double source,
and no suffix when a double is treated as a single source. We find a double frequency of $\sim$9\,\% using all our YSO candidates, however,
since we can only detect fairly bright and wide doubles (separations $\ga 0\farcs6$), this means that a lot of the multiple YSO systems are
most likely not resolved. Also, the very young, deeply embedded (Class 0) sources are not detected in the $I$, $J$ or $K_S$ bands, which we use
for our (wide) double source statistics.

There is clear evidence in literature that the binary frequency decreases with stellar mass. Duquennoy et al. (\cite{duquennoy}) found a binary frequency of $\sim$57\,\% for G-dwarf primaries and Marchal et al. (\cite{marchal}) a binary frequency of $\sim$27\,\% for M0--M6 dwarfs. 
Ahmic et al. (\cite{ahmic}) report a binary frequency of $\sim$11\,\% for young very low mass objects ($\sim$0.03--0.15\,M$_{\sun}$) in the Chamaeleon I star formation region using adaptive optics on the VLT, confirming the trend for a lower binary frequency with decreasing mass. This agrees well with the study by Bouy et al. (\cite{bouy}) that uses HST to look for BD binaries in the Pleiades ($\sim$13.3\,\%).

We thus expect a binary frequency in our YSO sample of 20--30\,\%, suggesting that there are unresolved binaries in our sample.
We do not detect any visually unresolved doubles in our spectroscopy, all stars can be fit with a single
model - this does however not mean that these are all single stars. If the effective temperatures of both stars are roughly similar or the
luminosity contrast is high, a double source could easily go unnoticed.


\subsection{The resulting YSO sample and H-R diagram}

The full list of YSO candidates is divided into two parts, YSOs with intrinsic IR excess (Table~\ref{YSO_candidates_IR}) and
YSOs that lack IR excess (Table~\ref{YSO_candidates_noIR}). In total, we find 89 YSO candidates. 

In addition, we have divided the sources into five categories, depending on the main method for classifying them as YSO candidates. These are
given in the Tables for each source. The first three categories are used for sources where we have spectra.
Category I is for YSO candidates where H$\alpha$ is seen in emission and Li\,I\,$\lambda6707$ in absorption.
Category II for H$\alpha$ in emission together with IR excess.
Category III are YSOs with non-zero extinction, implying that they are located inside the cloud. In addition, they are located
on a PMS track in the H-R diagram when this extinction has been corrected for.

For stars that are too faint for optical spectroscopy, we use category IV for sources with non-zero extinction, a point-like appearance in the $I$, $J$
and $K_S$ images and a H-R position on a PMS track. And finally, category V are YSO candidates with intrinsic IR excess as the only evidence of youth.

There are two additional YSOs in the list, L1641N-124 and 172. Both have intrinsic IR excess but are also the suggested sources for the two giant
H$_2$ flows in L1641N. There are probably many more YSOs in L1641N than in our YSO lists, since we have to detect each candidate in either
the $[5.8]$ and $[8.0]$ filters (for evidence of IR excess) or the $I$ and $J$ band filters (for extinction estimates). These Spitzer filters have
a high contrast background from PAH emission bands that effectively masks faint point sources at these wavelengths, and the $I$ and $J$ bands are
limited by interstellar extinction even at fairly low extinctions (see Fig.~\ref{IJ_completeness}).


\begin{figure*}
	\centering
	\includegraphics[width=18cm]{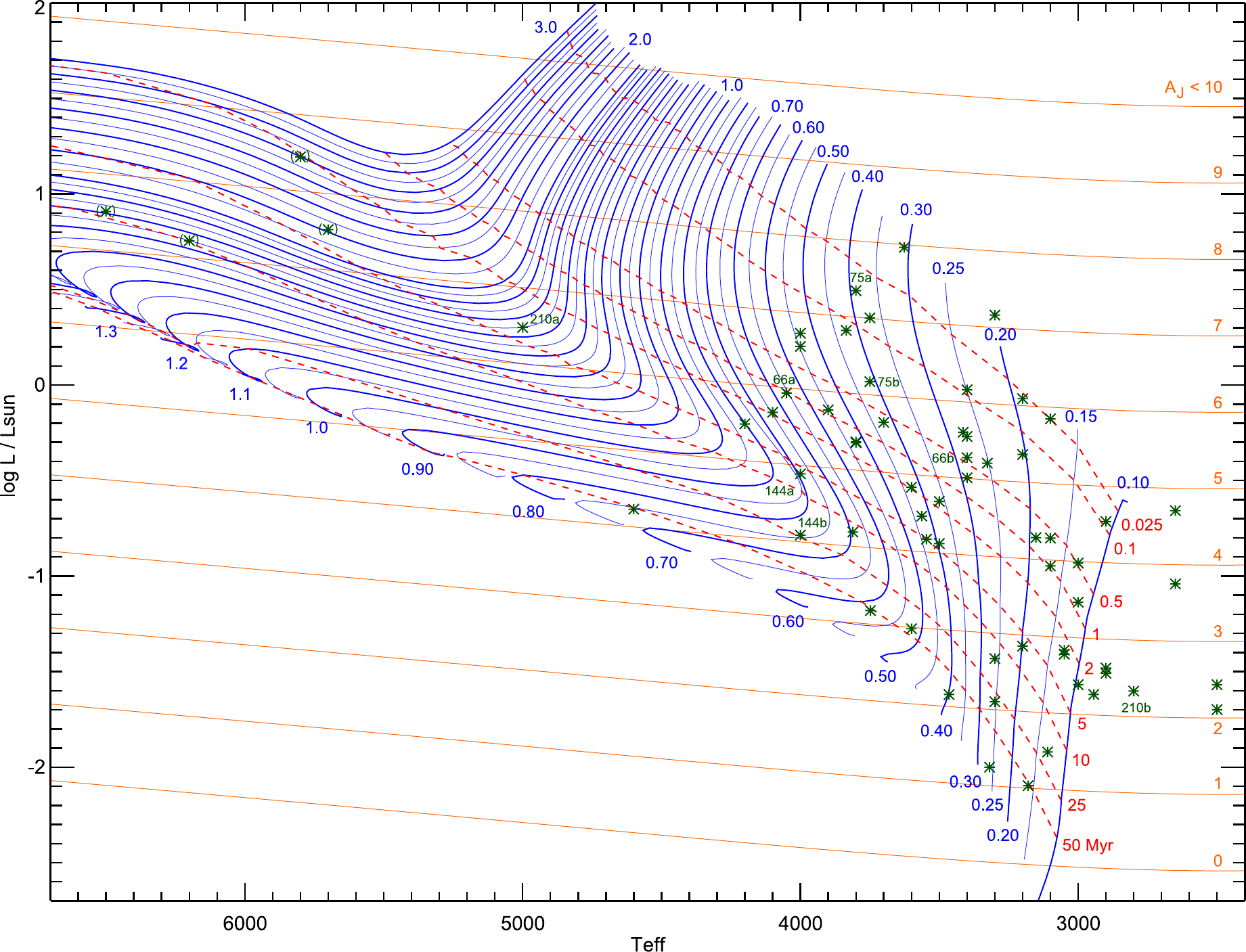}
	\caption{H-R diagram of the full (spectroscopic and photometric) sample for which we could obtain effective temperatures and extinctions.
	The evolution track models (blue curves with masses indicated) were obtained from the Dartmouth Stellar Evolution
	Database (Dotter et al. \cite{dotter}). Isochrones are drawn with red dashed lines and their ages shown on the right side of the grid.
	Observations are marked using green asterisks, with parentheses for the suggested four early foreground stars and source numbers given
	for double sources. The orange curves and numbers illustrate the completeness of the survey across the grid in terms of the A$_J$ extinction.
		}
	\label{HR_diag}
\end{figure*}

The resulting H-R diagram for all YSO candidates, for which extinctions could be calculated is presented in Figure~\ref{HR_diag}.
Evolutionary tracks (blue curves) are shown for 0.10--1.80\,M$_{\sun}$ in steps of 0.05 and for 1.80--3.00\,M$_{\sun}$ in
steps of 0.10 (Dotter et al. \cite{dotter}).
Isochrones are marked by red dashed curves and the extinction (A$_J$) to which the survey is complete is illustrated by orange curves.
All YSO candidates are plotted with green asterisks, with the addition of parentheses for the four suggested (early type)
foregrund stars and source numbers for double sources.

Using this H-R diagram we have determined the mass and age of each YSO candidate (Tables~\ref{tab_phot_mass_age} and \ref{tab_spectra}).
The luminosity function, mass function, effective temperature and age distributions for this sample are presented
in Section~\ref{IMF_sec}.



\subsection{Uncertainties in the H-R diagram}

There are several complicated sources of uncertainty in the age and mass determinations using a H-R diagram.
The two most important observational challenges are variability and multiplicity.

As mentioned previously, young stars are often variable
because of variable accretion onto the star, large star spots and variability in the extinction towards the star - colours could
therefore be affected by this and observations should ideally be made in all filters within a very short time frame.
For fairly bright sources we have used 2MASS photometry for $J$--$K_S$ colours since these are simultaneous observations. Our $I$ band
observations, however, were made at a different epoch, adding an uncertainty to the $I$--$J$ colours. Also, our $J$ and $K_S$ band observations were
made at two separate epochs.

It is generally believed that many young stars are binary systems (e.g. 79-86\,\% in sources driving giant HH flows, see
Reipurth\,\cite{reipurth00}), and by using all our observations at different
wavelengths we have detected 9\,\% of the sources in our YSO sample to be double sources. If one could detect and include all true binaries in the
H-R diagram this would of course increase the number of available sources, but also increase the stellar age estimates due to the components being fainter than the total flux from a binary. Observationally, this
would require very high resolution imaging using adaptive optics or interferometry, since spectroscopy probably would fail because
of spectral variability in these young sources.

Since the components of a binary star are expected to have formed at roughly the same time (except for binaries formed in gravitational
encounters between two pre-MS stars) we expect that the ages are similar for both components. As can be seen in the H-R diagram, although some of the
double sources (especially L1641N-66) have components on isochrones of similar age, there are double sources with components that do not appear coeval.
The fact that L1641N-210b is located at a lower mass than what is covered by the evolution tracks and the decreased accuracy for older stars (such as L1641N-144) makes a comparison difficult.

Another uncertainty in the H-R diagram is the distance module. The distance to the Orion GMC, and therefore Orion A which L1641N
belong to, is poorly determined, in part due to its extension. E.g. Brown et al.\,(\cite{brown}) finds a distance of 320\,pc to the near edge
of the Orion A~\&~B clouds and 500\,pc to the far edge. We have used the canonical distance of L1641N, which is 450\,pc.
A closer distance would mean that all sources are moved further down in the diagram. This would
not affect the mass determinations very much since the evolution tracks are almost vertical in the H-R diagram, except for older or more
massive stars.

Regarding the models, neither increased luminosity due to accretion nor excess emission from possible disks are included. Reflected
light in the atmospheres of the stars could also play a role.
The inclination of an YSO with a disk could be another source of uncertainty in the luminosity estimate, since the extinction correction assumes
that the disk and interstellar extinction laws are the same.

Most of the uncertainties mentioned in this section affects the luminosity determination and therefore mainly the age estimate of a star, while
the mass estimate is much more accurate (especially for low mass stars) because of the general direction of the evolution tracks. The uncertainty
in mass is particularly small for the spectroscopic sample where the effective temperatures have been determined to within 100\,K.


\subsection{The Mass Function}

\label{IMF_sec}

We have used both our spectroscopic and photometric sample to calculate the cluster mass function in L1641N. Stars for which we have spectra have more
accurate estimates of stellar mass, given that T$_{\textnormal{eff}}$ is fit directly from the spectra.


\begin{figure*}
	\centering
	\includegraphics[width=18cm]{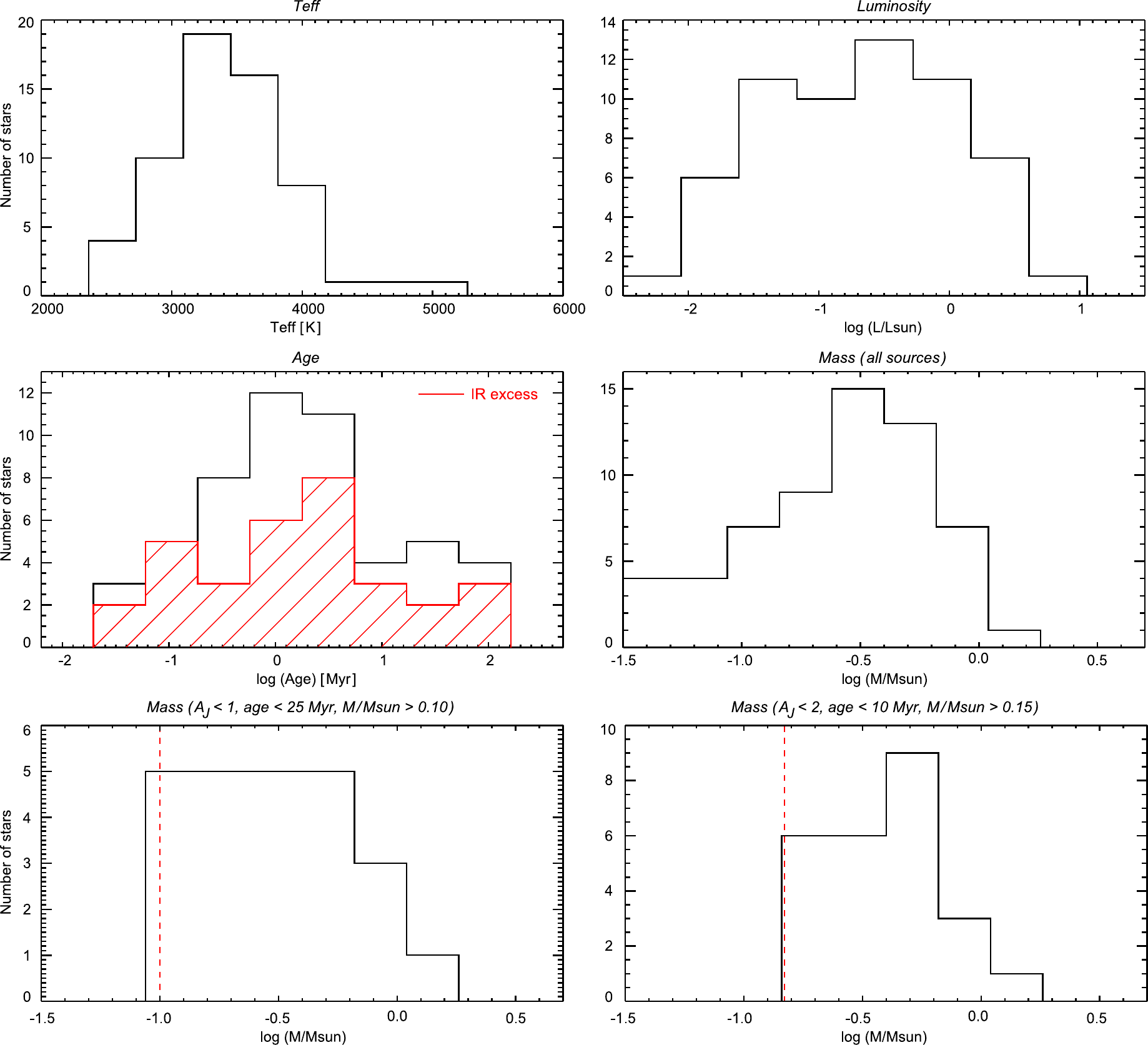}
	\caption{Resulting temperature, luminosity, age and mass functions for the YSO sample shown in Fig.~\ref{HR_diag}. The first four
	panels include all sources in the H-R diagram while the last two panels show the MF for the completeness limits given above the respective
	panels (the limiting mass is also illustrated by the red dotted lines). The red, hashed region in the age plot illustrates stars with
	intrinsic IR excess (disk stars). It is interesting to note that even though the excess YSOs are present at all ages, seven out of the eight
	youngest YSOs in the sample have IR excess.
		}
	\label{IMF}
\end{figure*}


In Figure~\ref{IMF} we have plotted the resulting Luminosity Function (LF), effective temperature distribution, age distribution and
Mass Function (MF) for the full sample, and two MFs with different completeness in A$_J$, minimum mass and maximum age (given in the Figure).
From the T$_{\textnormal{eff}}$ distribution it is clear that most stars in the sample are M-type stars.

Looking at the age distribution, it is clear that intrinsic IR excess (disk) stars are present at all ages. We note that seven out of
the eight youngest YSOs in the sample have IR excess. The median age of all YSOs is found to be $\sim$\,1\,Myr.

The MF that includes all sources in the sample is biased by bright (young and/or hot) sources being seen at higher extinctions than
faint sources.
The two additional MFs are un-biased within the completeness limits given in the last two panels in Fig.~\ref{IMF}, and illustrated by the
red dashed lines, marking  the minimum mass for which the MFs are complete.
The mass distribution peaks at about 0.3--0.4\,M$_{\sun}$ for all three MFs, and is fairly flat for masses lower than the peak until the
mass completeness limits are reached in both un-biased MFs.
The mass function is thus broadly consistent with IMFs found elsewhere in literature, showing a turnover or plateau at a few times 0.1\,M$_{\sun}$
(Elmegreen \cite{elmegreen1} ; Elmegreen et al. \cite{elmegreen2}) - keeping in mind the limitations set by small number statistics, limited mass range, unresolved binaries, more low- than high-mass members leaving
the cluster due to close encounters but also a possible counter-acting selection effect of low-mass YSOs being easier to find than higher mass YSOs due to varying disk lifetimes.

The methods we use to find our YSO candidates are differently sensitive to extinction (e.g. mid-IR excess detects YSOs at much higher extinctions than optical spectroscopy) and the processes involved (e.g. YSOs that have dispersed their disks are harder to find due to no IR excess). But, we need $I$, $J$ and $K_S$ photometry for all sources that we include in the MF, since this is used to calculate the absolute extinction A$_J$ and in turn the stellar luminosity (also the effective temperature for sources where we lack spectra). Therefore, both un-biased MFs are complete to the limits in A$_J$, age and mass given in Fig.~\ref{IMF}. These limits were found using the (orange) completeness curves in Fig.~\ref{HR_diag} which in turn where calculated using a grid of stellar temperatures and luminosities as input parameters (combined with the filter transmission curves and limiting magnitudes). The K$_S$ observations are deeper and the least sensitive to extinction of the three filters, therefore the $I$ and $J$ photometry sets the completeness limits (as illustrated in Fig.~\ref{IJ_completeness}).


\begin{figure*}
	\centering
	\includegraphics[width=15cm]{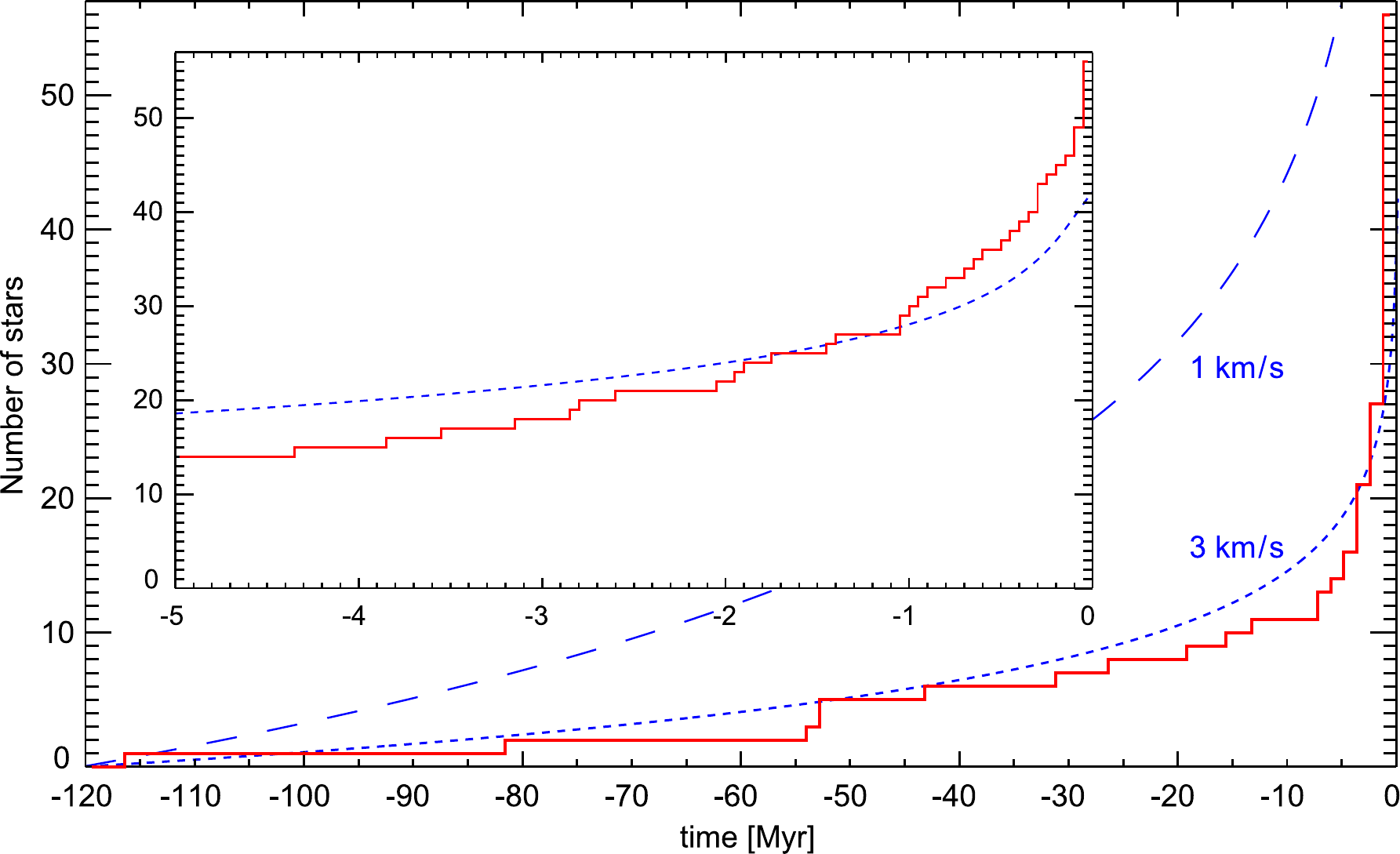}
	\caption{Cumulative star formation history plot for the sample where ages could be determined.
	The blue (dashed) curves are simple theoretical models assuming a constant star formation rate, an unbound cluster and velocity dispersions
	of 1 and 3\,km/s, respectively.
		}
	\label{SFhistory}
\end{figure*}


Given the large amount of sources confirmed as YSOs from our spectra, there is probably a large number of deeply embedded YSOs in the full L1641N
source catalogue that we miss in our MF (since we need $I$ and $J$ photometry to include them). In other words, we cannot go much deeper than a few magnitudes in A$_J$ if we want to include sufficiently low mass YSOs in the statistics. We also note that the number of sources in each mass bin is fairly small and that there may be a problem in general with cluster IMFs if the velocity dispersion of stars in a cluster (as will be discussed in the following section) depends on stellar mass.



\begin{figure*}
	\centering
	\includegraphics[width=15cm]{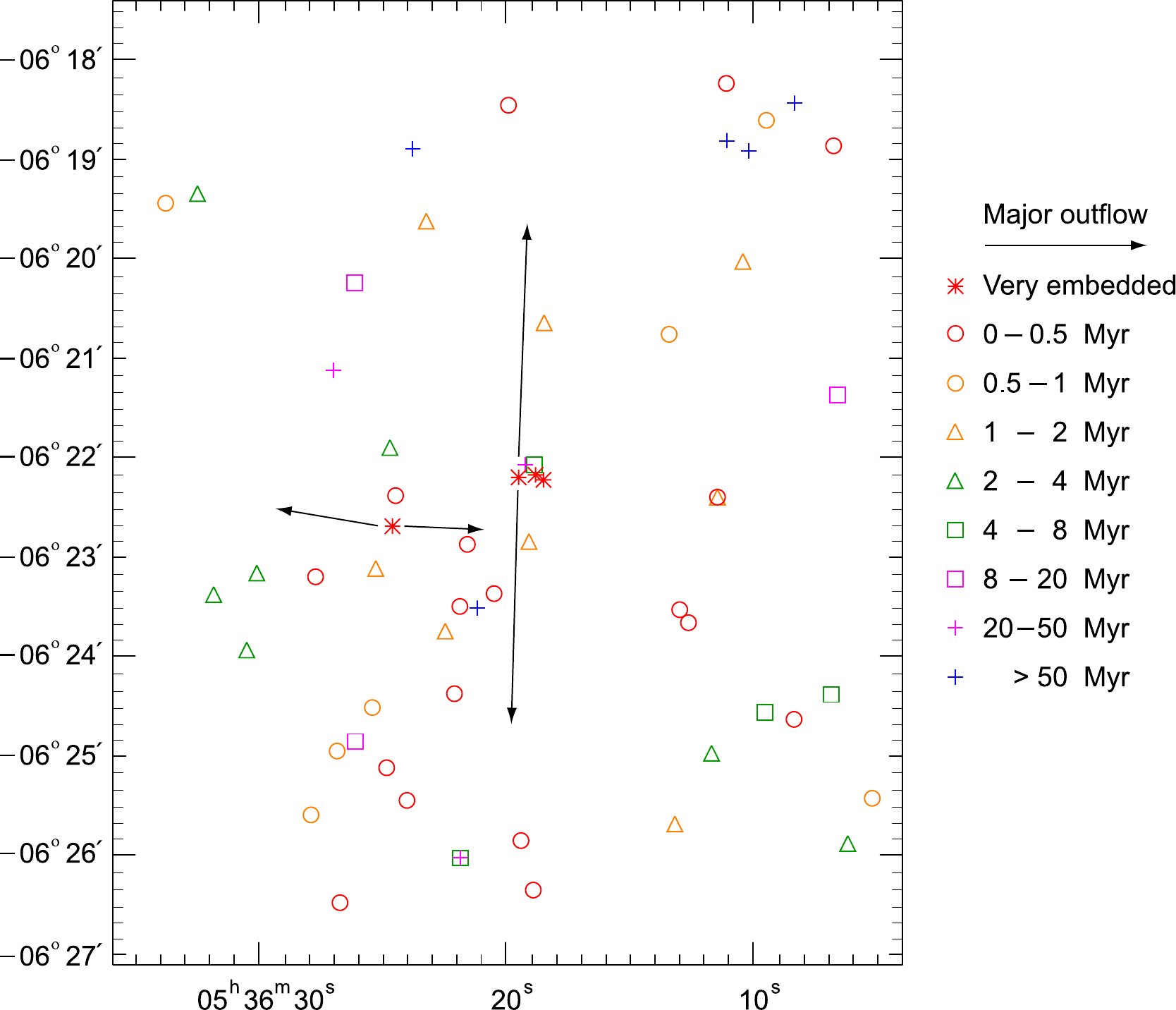}
	\caption{Spatial distribution of stellar ages. The two major bi-polar outflows are illustrated by arrows, with the corresponding outflow sources (and two additional deeply embedded
	sources, probably very young) marked by red asterisks. 
		}
	\label{SFspatial}
\end{figure*}



\begin{figure*}
	\centering
	\includegraphics[width=15cm]{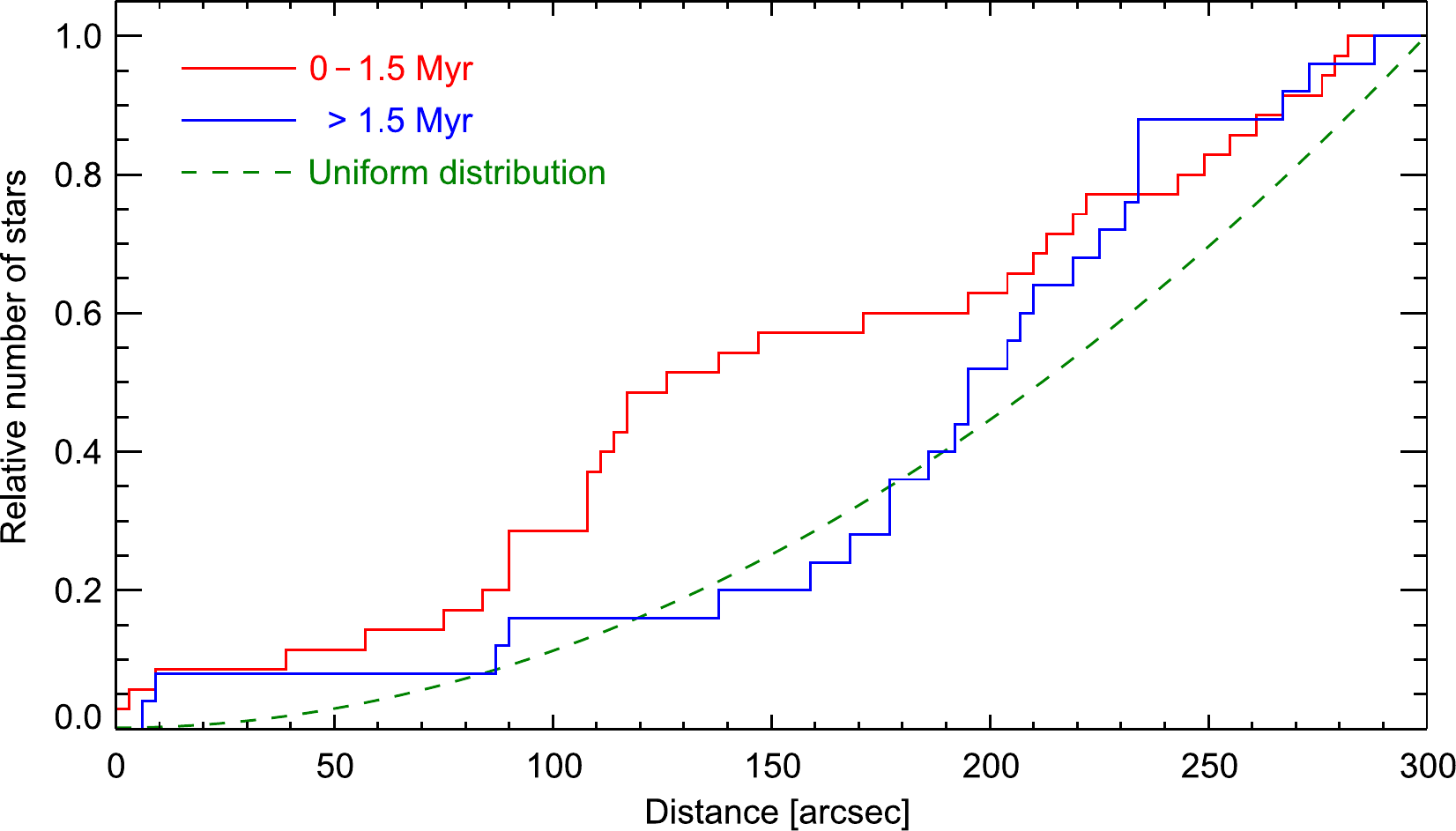}
	\caption{Cumulative radial distribution of YSOs. The relative number of sources versus distance from the
	2\,mm continuum dust peak (Chen et al. \cite{chen95}) at the centre of L1641N. The younger group (0--1.5\,Myr) is clearly more
	concentrated towards the centre while the older group ($>1.5$\,Myr) is more spread out (similar to a uniform distribution, dashed curve).
	This supports the age determinations and
	shows that the velocity dispersion of the cluster members is large enough that the older YSOs have had time to move away significantly from
	their birth sites.
		}
	\label{SFspatial2}
\end{figure*}


\subsection{Star formation history}

From the H-R diagram (Fig.~\ref{HR_diag}) we can obtain the stellar ages and thus the star formation (SF) history of most stars in the H-R sample.
The SF history is shown at two scales in Figure~\ref{SFhistory} in a cumulative manner, the number of stars formed in the sample versus time. The time spans are 120 and 5\,Myr with 100 bins, meaning bin sizes of 1.2 and 0.05\,Myr, respectively. 
The blue dashed curves represent simple theoretical models that assume a gravitationally unbound cluster with constant star formation rate
(one star in 3.7\,$\times$\,10$^4$\,yr, based on the last 1\,Myr of our sample) and stellar velocity dispersions of 1 and 3\,km/s, respectively.
For simplicity, all stars are assumed to form at the centre of the cluster in this model.

Given that the uncertainty in the H-R age determination increases drastically with age, due to the slow luminosity change with time close to the main
sequence, we cannot say with confidence that the first few stars in the plot really are older than $\sim$50\,Myr.
However, below this age the estimates are more certain and we see an accelerating increase in the number of stars with time.
It is interesting to note the large number of old stars (for Orion), we detect 11 stars older than 10\,Myr. There are at least three possible reasons for this; the star formation process
could be very slow to start, some older stars could have drifted into L1641N or star formation has actively been going on for a very long time
but the velocity dispersion has moved most of the older stars out of the cluster while the younger stars ($<$1\,Myr) are still close to their birth sites.
There is however an alternative, observational explanation for these old stars. Some of the YSOs with disks might be seen almost edge-on, so that mostly reflected light is detected, making them look fainter and thus older (as hinted by the fact that some of the oldest YSOs show IR excess).
Evidence for old members in the Orion cluster, using both H-R diagrams and models of Li burning can be found in e.g. Palla et al. (\cite{palla}) where several stars appear to have ages of 15--30\,Myr.

Figure~\ref{SFspatial} shows the spatial age distribution. Four deeply embedded (and presumably very young) sources, two of these being the outflow
sources for the major outflows, are marked by asterisks. Different symbols and colours (red being the youngest and blue the oldest) are used to illustrate
eight different age groups. The arrows show the direction and outflow sources of the two major bi-polar outflows in the region (G\aa lfalk \& Olofsson \cite{galfalk07}).
There is no very clear relation between age and spatial distribution seen in this plot, but it is evident that the most embedded (and presumably youngest) sources, including  the two outflow sources, are located close to the centre of the cluster.

In Figure~\ref{SFspatial2} we have divided the YSOs into two groups (younger and older than 1.5\,Myr, respectively) and plotted their relative cumulative radial distribution versus distance from the centre of L1641N (using the 2\,mm continuum dust peak of Chen et al. (\cite{chen95}) as the centre point). The sources in the younger group are clearly found closer to central L1641N than the older group. A spatial separation is also found
if e.g. 0.5 or 1.0\,Myr is used to divide the two groups. The exact centre of the cluster is difficult to pinpoint without a wide-field study (Fig.~\ref{SFspatial} suggests that the centre might be somewhat to the
southeast of the IRAS source), thus the best we can do is to assume that the dust peak (very close to the IRAS source) is located at the exact centre.

Madsen et al. (\cite{madsen}) have used Hipparcos data to calculate the velocity dispersion of stars in young clusters and associations. They find typical velocity dispersions of roughly 1\,km/s. For L1641N, a 50\,Myr star with such a velocity would (at most) have moved about 40 times the field of view
in Fig.~\ref{SFspatial} and a 1\,Myr star at most about one field size (if the velocity is perpendicular to the line of sight).

As can be seen in Fig.~\ref{SFhistory} the star formation rate is roughly constant during the last 1\,Myr (the youngest age bin includes the four very embedded sources marked by asterisks in Fig.~\ref{SFspatial}).
We estimate a star formation rate of one star in 3.7\,$\times$\,10$^4$\,yr.
The high current star formation rate agrees very well with the presence of two major outflows (the outflow phase has a duration of
$\sim$10$^{5}$\,yr ; Reipurth et al.~\cite{reipurth_HHreview}) and the very high concentration of HH objects in L1641N. These results agree with
the scenario suggested by Hodapp et al.~(\cite{hodapp93}) of ongoing successive formation of individual stars (although over a larger time span than the
suggested 1\,Myr) with no clear sign of a star burst event.

Our results support the possibility that star formation in L1641N could have been going on at a fairly constant rate for a very long time, but
only appears to have an increased rate with time and a current maximum because of velocity dispersion, making more and more YSOs leave the cluster as they age. The measured curve in Fig.\ref{SFhistory} has a shape similar to a simple model curve that assumes a constant star formation rate over a very long time with constant velocity dispersion (3 km/s) and an unbound cluster. If there is a relation between stellar mass and the probability of a star to leak out of a cluster with time, e.g. low-mass stars being more easily thrown out due to close encounters (see e.g. Delgado et al.~\cite{delgado}, Sterzik and Durisen~\cite{sterzik}), this could affect cluster IMFs in general (unless the surveyed region is very large) in addition to the star formation history. 


\section{Summary and conclusions}

In this detailed study of the L1641N cluster, a nearby star formation region in Orion, we have used
space based (Spitzer and ISO 3.6--14.3\,$\mu$m) and ground based ($I$, $J$, $K_S$ and $L'$) imaging, as well as
optical spectroscopy (5780--8340 \AA) of a large number of sources in an effort to find YSO candidates. YSO candidate status is found through a combination of mid-IR excess (presence of a disk), H$\alpha$ emission (accretion), Li non-depletion (sign of youth) and extinction (cluster membership).

We have calculated colours of normal stars for the IRAC, $I$, $J$ and $K_S$ filters and used a $I-J$ vs $J-K_S$ diagram
(and spectroscopy) to calculate extinctions, luminosities and effective temperatures.
We investigate statistical properties of this YSO sample through double sources, extinction, effective temperatures, age distribution, spatial
distribution, star formation history and the Mass Function (MF) for different completeness limits. The following results were obtained:

\vspace{3mm}
1. We detect a total of 216 Spitzer and $I$ band sources within the region of our ISOCAM survey.

2. Using the observations and criteria mentioned above we have found 89 YSO candidates in the sample.

3. Most of the spectra show M-type stars, with H$\alpha$ strongly in emission, that belong to the cluster.
Although the spectral resolution is somewhat low (R=830) we also detect Li\,I\,$\lambda6707$ in absorption in many of the sources.

4. The four brightest stars in the $I$ band observations, which are also shown to have the earliest spectral
types (F and G stars), are very likely foreground stars to L1641N.

5. We find that the interstellar extinction is well fit by a power law (index = 1.58) in the optical and near-IR, although in
the mid-IR (Spitzer) the observed extinctions are higher than expected from theory.

6. Extended PAH emission is seen in high contrast all over L1641N, we suggest that the illuminating source
is $\iota$~Orionis (an O9~III star 30$\arcmin$ to the north, distance $\sim$400\,pc).

7. Using Spitzer, $K_S$ and deep 2.12\,$\mu$m H$_2$ observations we suggest that IRAS 05338-0624 (the only IRAS source in L1641N) is in fact the
combined flux of at least four sources (and possibly some reflection nebulosity).

8. We find a double source frequency of $\sim$9\,\% using all our YSO candidates, however,
we can only detect fairly bright and wide doubles (separations $\ga 0\farcs6$), meaning that a large number of double/multiple YSO systems are
probably not resolved.

9. The median age of our YSO sample is $\sim$1\,Myr.

10. The unbiased MFs peak at about 0.3--0.4 M$_{\sun}$ with an essentially flat distribution for lower masses.

11. We find a star formation history with an accelerating increase in the number of stars with time, but note that this could
be caused by migration of stars with time from the cluster and that a velocity dispersion of a few km/s (which is typical in star formation
regions) is enough to account for this effect.
We detect some very old stars (for Orion) of up to $\sim$50\,Myr (11 of the stars are older than 10\,Myr).

12. A more or less constant star formation rate is found during the last $\sim$1\,Myr and estimated to be one star in 3.7\,$\times$\,10$^4$\,yr.
This agrees with the presence of two major bi-polar outflows (the outflow phase has an expected duration of $\sim$10$^{5}$\,yr ; Reipurth et al.~\cite{reipurth_HHreview}).

13. We find a spatial separation between older and younger YSOs (e.g. using a dividing age of 0.5, 1.0 or 1.5\,Myr) from the distribution of the
two groups. This shows that velocity dispersion is an important factor to consider when investigating the star formation history and possibly also the IMF (hence we calculate the MF). In fact, a star with the typical velocity of $\sim$1\,km/s could leave L1641N in a few Myr.

14. These results agree with a star formation process that has been slow to start. Another possibility is a roughly constant star formation rate over a very long time scale, where the star formation rate only appears to increase with time and have a current maximum because of the velocity dispersion, making more and more YSOs leave the cluster as they age. If there is a relation between stellar mass and velocity disperson, cluster IMFs in general could be affected unless the surveyed region is very large.

\begin{acknowledgements}
	The Swedish participation in this research is funded by the Swedish National Space Board.
	This publication made use of the NASA/IPAC Infrared Science Archive, which is operated by the Jet Propulsion
	Laboratory, California Institute of Technology, under contract with the National Aeronautics and Space
	Administration, and data products from the Two Micron All Sky Survey, which is a joint project of the University
	of Massachusetts and the Infrared Processing and Analysis Center/California Institute of Technology, funded
	by the National Aeronautics and Space Administration and the National Science Foundation.
	The authors would like to thank Peter Hauschildt for providing us with the dwarf model spectra we used in our
	classifications.
\end{acknowledgements}


\onecolumn \onecolumn			
\begin{scriptsize}
\begin{longtable}[l]{cccrrrrrr}
\multicolumn{9}{l} {\small{\bfseries \tablename\ \thetable{}.}		
Photometry of all sources - space based} \\
\label{L1641N_space} \\

\hline
\noalign{\vspace{0.5mm}}
\multicolumn{1}{c}{No.} & \multicolumn{1}{c}{RA} & \multicolumn{1}{c}{Dec} & \multicolumn{1}{c}{$F_{3.6}$} & \multicolumn{1}{c}{$F_{4.5}$} & \multicolumn{1}{c}{$F_{5.8}$} & \multicolumn{1}{c}{$F_{6.7}$} & \multicolumn{1}{c}{$F_{8.0}$} & \multicolumn{1}{c}{$F_{14.3}$} \\
\noalign{\vspace{0.5mm}}
\multicolumn{1}{c}{} &\multicolumn{1}{c}{(2000)} &\multicolumn{1}{c}{(2000)} &\multicolumn{1}{c}{mJy} &\multicolumn{1}{c}{mJy} &\multicolumn{1}{c}{mJy} &\multicolumn{1}{c}{mJy} &\multicolumn{1}{c}{mJy} &\multicolumn{1}{c}{mJy} \\
\noalign{\vspace{0.5mm}}
\hline
\noalign{\vspace{0.5mm}}
\endfirsthead

\multicolumn{9}{l} {\small{\bfseries \tablename\ \thetable{}.} -- continued from previous page} \\
\noalign{\vspace{2.5mm}}
\hline
\noalign{\vspace{0.5mm}}
\multicolumn{1}{c}{No.} & \multicolumn{1}{c}{RA} & \multicolumn{1}{c}{Dec} & \multicolumn{1}{c}{$F_{3.6}$} & \multicolumn{1}{c}{$F_{4.5}$} & \multicolumn{1}{c}{$F_{5.8}$} & \multicolumn{1}{c}{$F_{6.7}$} & \multicolumn{1}{c}{$F_{8.0}$} & \multicolumn{1}{c}{$F_{14.3}$} \\
\noalign{\vspace{0.5mm}}
\multicolumn{1}{c}{} &\multicolumn{1}{c}{(2000)} &\multicolumn{1}{c}{(2000)} &\multicolumn{1}{c}{mJy} &\multicolumn{1}{c}{mJy} &\multicolumn{1}{c}{mJy} &\multicolumn{1}{c}{mJy} &\multicolumn{1}{c}{mJy} &\multicolumn{1}{c}{mJy} \\
\noalign{\vspace{0.5mm}}
\hline
\noalign{\vspace{0.5mm}}
\endhead

\hline \noalign{\vspace{1.5mm}} \multicolumn{9}{l}{{\small Continued on next page...}} \\
\endfoot

\hline
\endlastfoot

1 & 05:36:02.90 & -06:26:10.7 & 0.068 $\pm$ 0.023 & 0.040 $\pm$ 0.015 & \multicolumn{1}{c}{...} & \multicolumn{1}{c}{...} & \multicolumn{1}{c}{...} & \multicolumn{1}{c}{...} \\
2 & 05:36:02.90 & -06:25:55.7 & 0.110 $\pm$ 0.012 & 0.086 $\pm$ 0.010 & \multicolumn{1}{c}{...} & \multicolumn{1}{c}{...} & \multicolumn{1}{c}{...} & \multicolumn{1}{c}{...} \\
3 & 05:36:03.61 & -06:25:28.4 & 0.073 $\pm$ 0.016 & 0.056 $\pm$ 0.008 & \multicolumn{1}{c}{...} & \multicolumn{1}{c}{...} & \multicolumn{1}{c}{...} & \multicolumn{1}{c}{...} \\
4 & 05:36:04.19 & -06:24:55.1 & 0.023 $\pm$ 0.004 & 0.034 $\pm$ 0.003 & \multicolumn{1}{c}{...} & \multicolumn{1}{c}{...} & \multicolumn{1}{c}{...} & \multicolumn{1}{c}{...} \\
5 & 05:36:04.22 & -06:20:44.0 & 36.840 $\pm$ 0.076 & 25.158 $\pm$ 0.065 & 16.222 $\pm$ 0.278 & \multicolumn{1}{c}{...} & 8.036 $\pm$ 0.350 & \multicolumn{1}{c}{...} \\
6 & 05:36:04.41 & -06:22:24.9 & 0.062 $\pm$ 0.004 & 0.061 $\pm$ 0.004 & \multicolumn{1}{c}{...} & \multicolumn{1}{c}{...} & \multicolumn{1}{c}{...} & \multicolumn{1}{c}{...} \\
7 & 05:36:04.56 & -06:23:59.4 & 0.070 $\pm$ 0.010 & 0.053 $\pm$ 0.010 & \multicolumn{1}{c}{...} & \multicolumn{1}{c}{...} & \multicolumn{1}{c}{...} & \multicolumn{1}{c}{...} \\
8 & 05:36:04.60 & -06:25:01.7 & 0.031 $\pm$ 0.005 & 0.018 $\pm$ 0.005 & \multicolumn{1}{c}{...} & \multicolumn{1}{c}{...} & \multicolumn{1}{c}{...} & \multicolumn{1}{c}{...} \\
\end{longtable}
\end{scriptsize}


\onecolumn \onecolumn			
\begin{scriptsize}
\begin{longtable}[l]{cccrrrl}
\multicolumn{7}{l} {\small{\bfseries \tablename\ \thetable{}.}		
Photometry of all sources - ground based} \\
\label{L1641N_ground} \\

\hline
\noalign{\vspace{0.5mm}}
\multicolumn{1}{c}{No.} & \multicolumn{1}{c}{RA} & \multicolumn{1}{c}{Dec} & \multicolumn{1}{c}{$I$} & \multicolumn{1}{c}{$J$} & \multicolumn{1}{c}{$K_S$} & \multicolumn{1}{c}{} \\
\noalign{\vspace{0.5mm}}
\multicolumn{1}{c}{} &\multicolumn{1}{c}{(2000)} &\multicolumn{1}{c}{(2000)} &\multicolumn{1}{c}{mag} &\multicolumn{1}{c}{mag} &\multicolumn{1}{c}{mag} &\multicolumn{1}{c}{Comment} \\
\noalign{\vspace{0.5mm}}
\hline
\noalign{\vspace{0.5mm}}
\endfirsthead

\multicolumn{7}{l} {\small{\bfseries \tablename\ \thetable{}.} -- continued from previous page} \\
\noalign{\vspace{2.5mm}}
\hline
\noalign{\vspace{0.5mm}}
\multicolumn{1}{c}{No.} & \multicolumn{1}{c}{RA} & \multicolumn{1}{c}{Dec} & \multicolumn{1}{c}{$I$} & \multicolumn{1}{c}{$J$} & \multicolumn{1}{c}{$K_S$} & \multicolumn{1}{c}{} \\
\noalign{\vspace{0.5mm}}
\multicolumn{1}{c}{} &\multicolumn{1}{c}{(2000)} &\multicolumn{1}{c}{(2000)} &\multicolumn{1}{c}{mag} &\multicolumn{1}{c}{mag} &\multicolumn{1}{c}{mag} &\multicolumn{1}{c}{Comment} \\
\noalign{\vspace{0.5mm}}
\hline
\noalign{\vspace{0.5mm}}
\endhead

\hline \noalign{\vspace{1.5mm}} \multicolumn{7}{l}{{\small Continued on next page...}} \\
\endfoot

\hline
\endlastfoot

1 & 05:36:02.90 & -06:26:10.7 & 21.08 $\pm$ 0.18 & \multicolumn{1}{c}{...} & \multicolumn{1}{c}{...} &  \\
5 & 05:36:04.22 & -06:20:44.0 & \multicolumn{1}{c}{Sat.} & 9.78 $\pm$ 0.03 & 9.60 $\pm$ 0.02 &  \\
7 & 05:36:04.56 & -06:23:59.4 & \multicolumn{1}{c}{...} & \multicolumn{1}{c}{...} & 17.07 $\pm$ 0.12 &  \\
8 & 05:36:04.60 & -06:25:01.7 & \multicolumn{1}{c}{...} & \multicolumn{1}{c}{...} & 17.55 $\pm$ 0.20 &  \\
12 & 05:36:04.93 & -06:25:54.0 & 22.64 $\pm$ 0.70 & \multicolumn{1}{c}{...} & \multicolumn{1}{c}{...} & Artefact? \\
13 & 05:36:05.01 & -06:24:01.7 & 20.89 $\pm$ 0.10 & \multicolumn{1}{c}{...} & 14.24 $\pm$ 0.02 &  \\
14 & 05:36:05.09 & -06:23:01.8 & \multicolumn{1}{c}{...} & \multicolumn{1}{c}{...} & 17.11 $\pm$ 0.27 &  \\
15 & 05:36:05.16 & -06:25:25.4 & 13.14 $\pm$ 0.01 & 11.82 $\pm$ 0.03 & 10.90 $\pm$ 0.01 &  \\
\end{longtable}
\end{scriptsize}



\begin{table*}
\begin{scriptsize}
	\caption{List of all YSO candidates with intrinsic IR excess. (65 sources)}
	\label{YSO_candidates_IR}
	\begin{tabular}{lcccccccccll}
	  \hline
        \noalign{\vspace{0.5mm}}
	No.$^{\mathrm a}$ & \multicolumn{6}{c}{Intrinsic IR excess} & \multicolumn{2}{c}{Spectral features} & Extinction / HR-pos & YSO method$^{\mathrm c}$ & Remarks \\
	    & 3.6 & 4.5 & 5.8 & 8.0 & 14.3 & $K_S$    & H$\alpha$ & Li                        & A$_J$ (mag)         &            &         \\
	\noalign{\vspace{0.5mm}}
        \hline

	\noalign{\vspace{1.0mm}}

   {\bf 23}   & .... & x    & x    & x    &      & .... &      &      & 1.69 / PMS & IV  & \\
   {\bf 29}  &      & x    &      &      &      &      &      &      &            &  V &  \\
   {\bf 33}  &      & x    &      &      &      &      &      &      &            &  V &  \\
   {\bf 46}   & .... & x    & x    & x    & x    & .... & em   & .... & 0.00 / PMS & II  & \\
   {\bf 56}   & .... & x    & x    & x    &      & .... & em   & .... & 0.24 / PMS & II  & \\
   {\bf 60}   & .... & .... & .... & x    &      & .... & .... & .... & 1.23 / MS  & III & ZAMS ? \\
\hline

	\end{tabular}

	\begin{list}{}{}
		\item[$^{\mathrm{a}}$] Bold source numbers are used to indicate intrinsic IR excess.
		\item[$^{\mathrm{b}}$] See G\aa lfalk \& Olofsson (\cite{galfalk07}).
		\item[$^{\mathrm{c}}$] Reliability class of YSO candidate: \\
		I:~~~H$\alpha$ in emission and Li\,I\,$\lambda6707$ in absorption. \\
		II:~~H$\alpha$ in emission and intrinsic IR excess. \\
		III:~Non-zero extinction, stellar spectra and located on a PMS track. \\
		IV:~Non-zero extinction, point-like and located on a PMS track. \\
		V:~~~Intrinsic IR excess only. \\
	\end{list}

\end{scriptsize}
\end{table*}



\begin{table*}
\begin{scriptsize}
	\caption{List of all additional YSO candidates (no intrinsic IR excess). (24 sources)}
	\label{YSO_candidates_noIR}
	\begin{tabular}{lcccccccccll}
	  \hline
        \noalign{\vspace{0.5mm}}
	No. & \multicolumn{6}{c}{Intrinsic IR excess} & \multicolumn{2}{c}{Spectral features} & Extinction / HR-pos & YSO method$^{\mathrm a}$ & Remarks \\
	    & 3.6 & 4.5 & 5.8 & 8.0 & 14.3 & $K_S$    & H$\alpha$ & Li                        & A$_J$ (mag)         &            &         \\
	\noalign{\vspace{0.5mm}}
        \hline

	\noalign{\vspace{1.0mm}}

	15   & .... & .... & .... & .... &      & .... & em   & abs  & 0.20 / PMS & I    & \\
 	28   & .... & .... &      &      &      & .... &      &      & 1.37 / PMS & IV  & \\
	30 a  & .... & .... & .... & .... &      & .... & em   & .... & 0.33 / PMS & III & \\
	31   & .... & .... & .... & .... &      & .... &      &      & 2.79 / PMS & IV  & \\
	47   & .... & .... & .... &      &      & .... &      &      & 1.46 / MS  & IV  & ZAMS? \\
	55   & .... & .... & .... & .... &      & .... & em   & abs  & 0.03 / PMS & I   & \\
	61   & .... & .... & .... & .... &      & .... & em   & abs  & 0.01 / PMS & I   & \\
	65   & .... & .... & .... & .... &      & .... & em   & .... & 1.38 / PMS & III & \\
	78   & .... & .... & .... & .... &      & .... & em   & abs  & 0.07 / PMS & I  & \\
	80   & .... & .... & .... & .... &      & .... &      &      & 3.41 / PMS & IV & \\
	92   & .... & .... & .... &      &      & .... &      &      & 2.11 / PMS & IV & \\
	118  & .... & .... & .... & .... &      & .... &      &      & 2.28 / PMS & IV & \\
	120  &      &      &      &      &      & .... & em   & .... & 0.46 / PMS & III & Close to MS \\
	128  & .... & .... & .... & .... &      & .... &      &      & 3.45 / PMS & IV & \\
 	148  & .... & .... & .... & .... &      & .... &      &      & 5.05 / PMS & IV & \\
 	160  & .... & .... & .... & .... &      & .... & em   & abs  & 0.68 / PMS & I & \\
 	168  & .... & .... & .... &      & .... & .... &      &      & 4.37 / PMS & IV & \\
	170  & .... & .... & .... &      &      & .... &      &      & 1.87 / PMS & IV & \\
	177  & .... & .... &      &      &      & .... & em   & abs  & 0.06 / PMS & I & \\
	184  & .... & .... & .... & .... &      & .... & em   & .... & 0.73 / PMS & III & \\
	186  & .... & .... & .... &      &      & .... &      &      & 4.12 / PMS & IV & \\
	187  & .... & .... & .... & .... &      & .... & em   & abs  & 0.39 / PMS & I & \\
	190  & .... & .... & .... &      &      & .... & .... & .... & 1.16 / PMS & III & Close to MS \\
 	215  & .... & .... & .... & .... &      & .... & em   & abs  & 0.09 / PMS & I & \\	

	\noalign{\vspace{1.0mm}}
	\hline

	\end{tabular}

	\begin{list}{}{}
		\item[$^{\mathrm{a}}$] Reliability class of YSO candidate: \\
		I:~~~H$\alpha$ in emission and Li\,I\,$\lambda6707$ in absorption. \\
		III:~Non-zero extinction, stellar spectra and located on a PMS track. \\
		IV:~Non-zero extinction, point-like and located on a PMS track. \\
	\end{list}

\end{scriptsize}
\end{table*}


\begin{figure*}
	\centering
	\includegraphics[width=17cm]{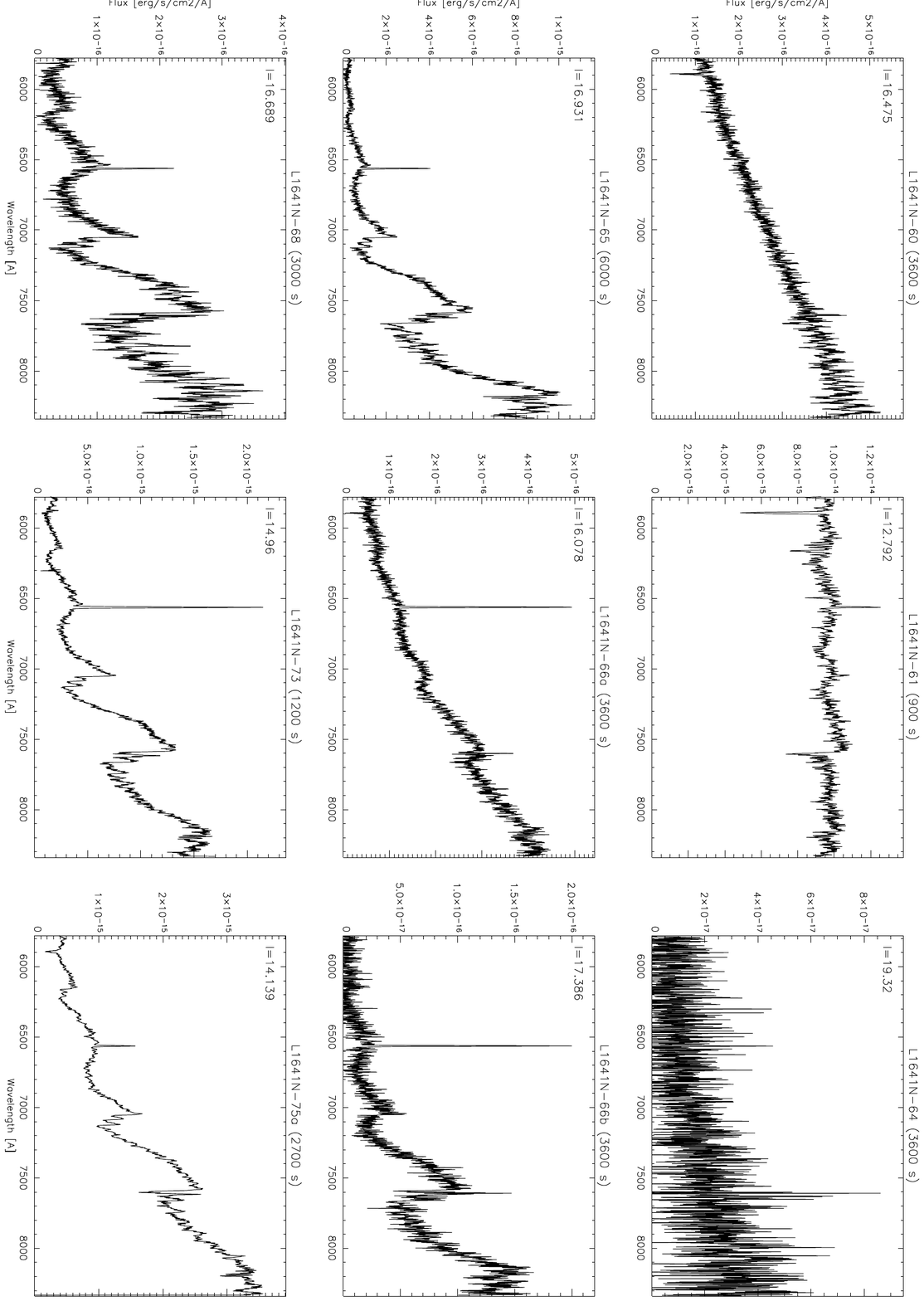}
\end{figure*}
\begin{figure*}
	\centering
	\includegraphics[width=17cm]{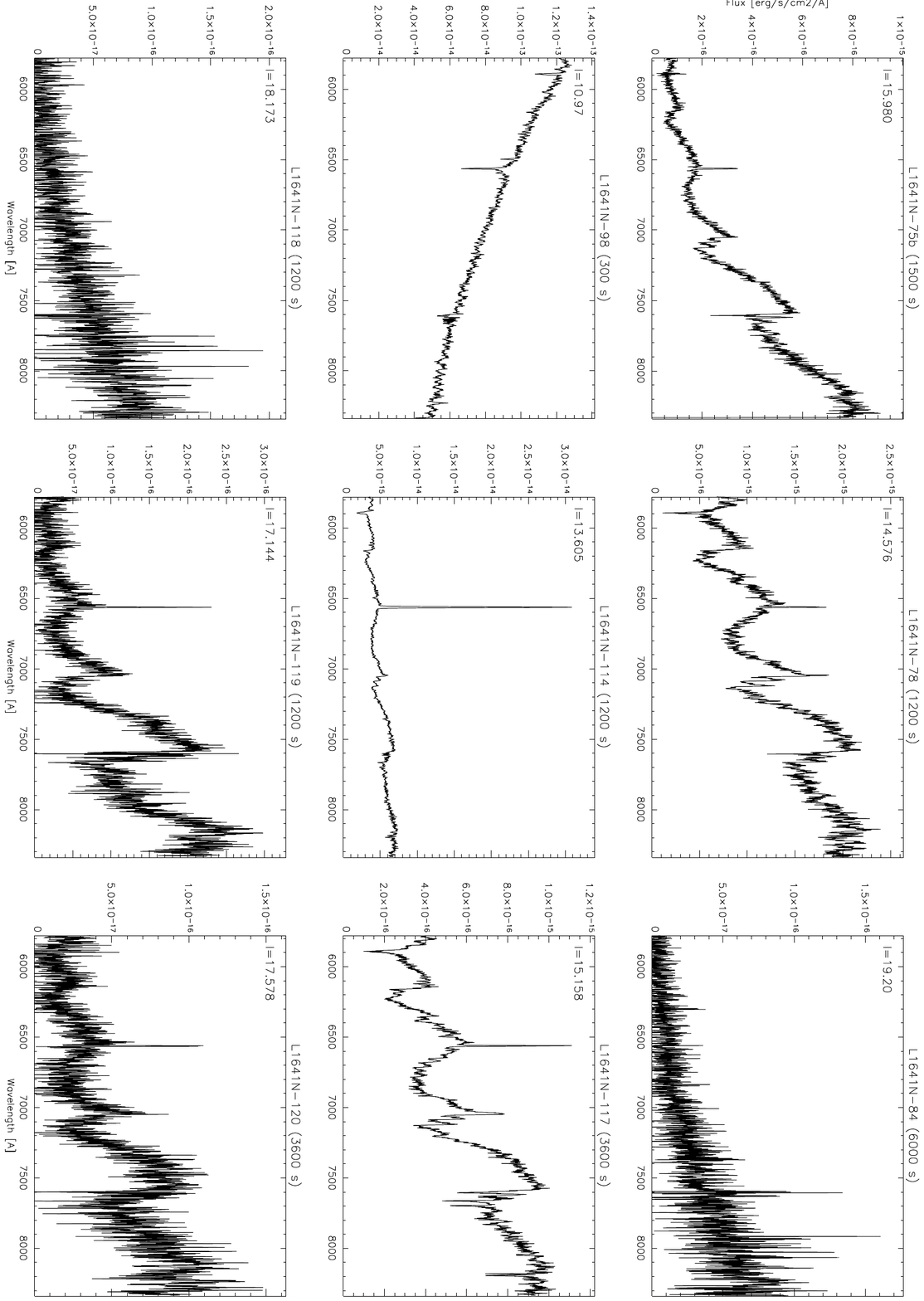}
\end{figure*}
\begin{figure*}
	\centering
	\includegraphics[width=17cm]{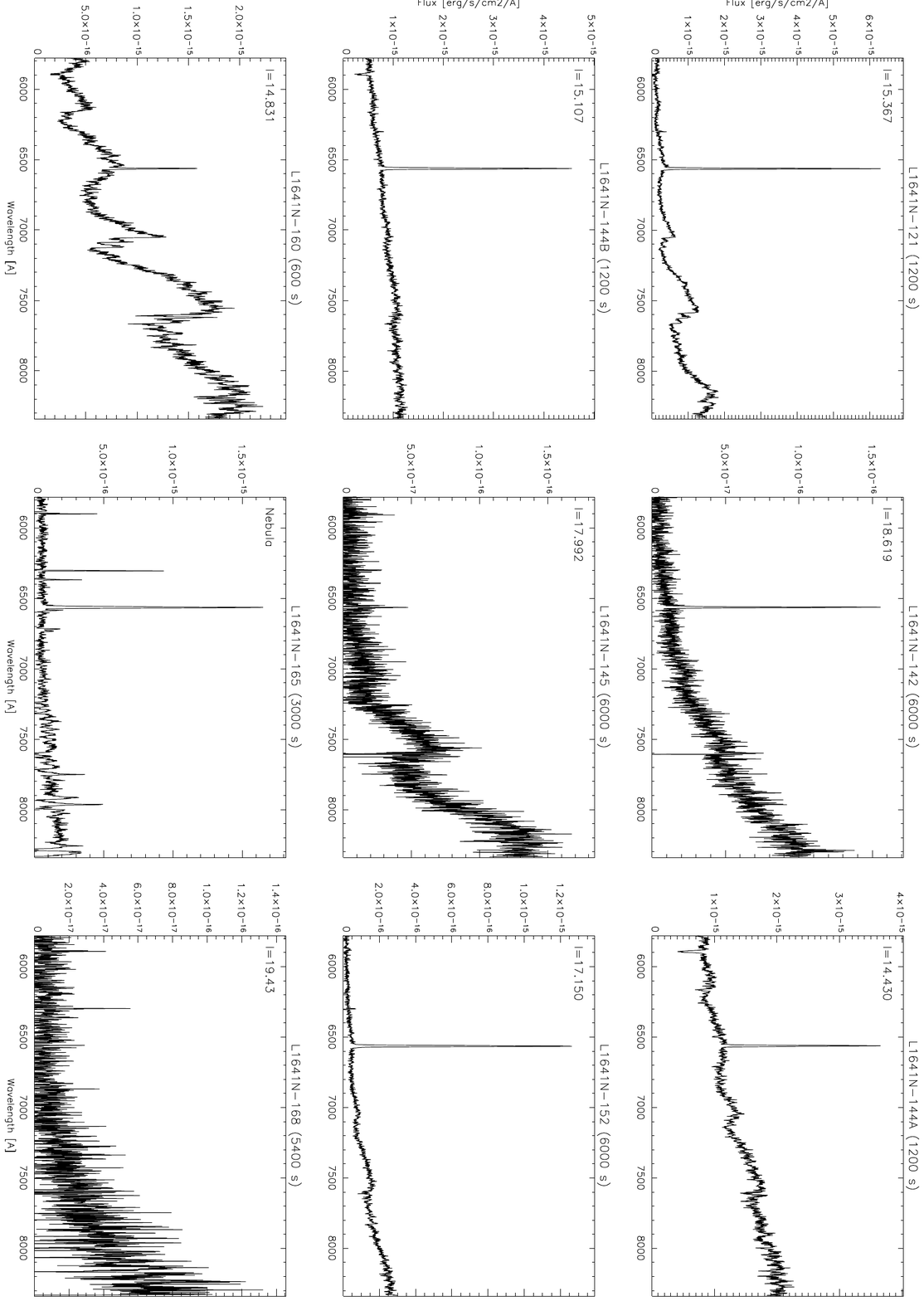}
\end{figure*}
\begin{figure*}
	\centering
	\includegraphics[width=17cm]{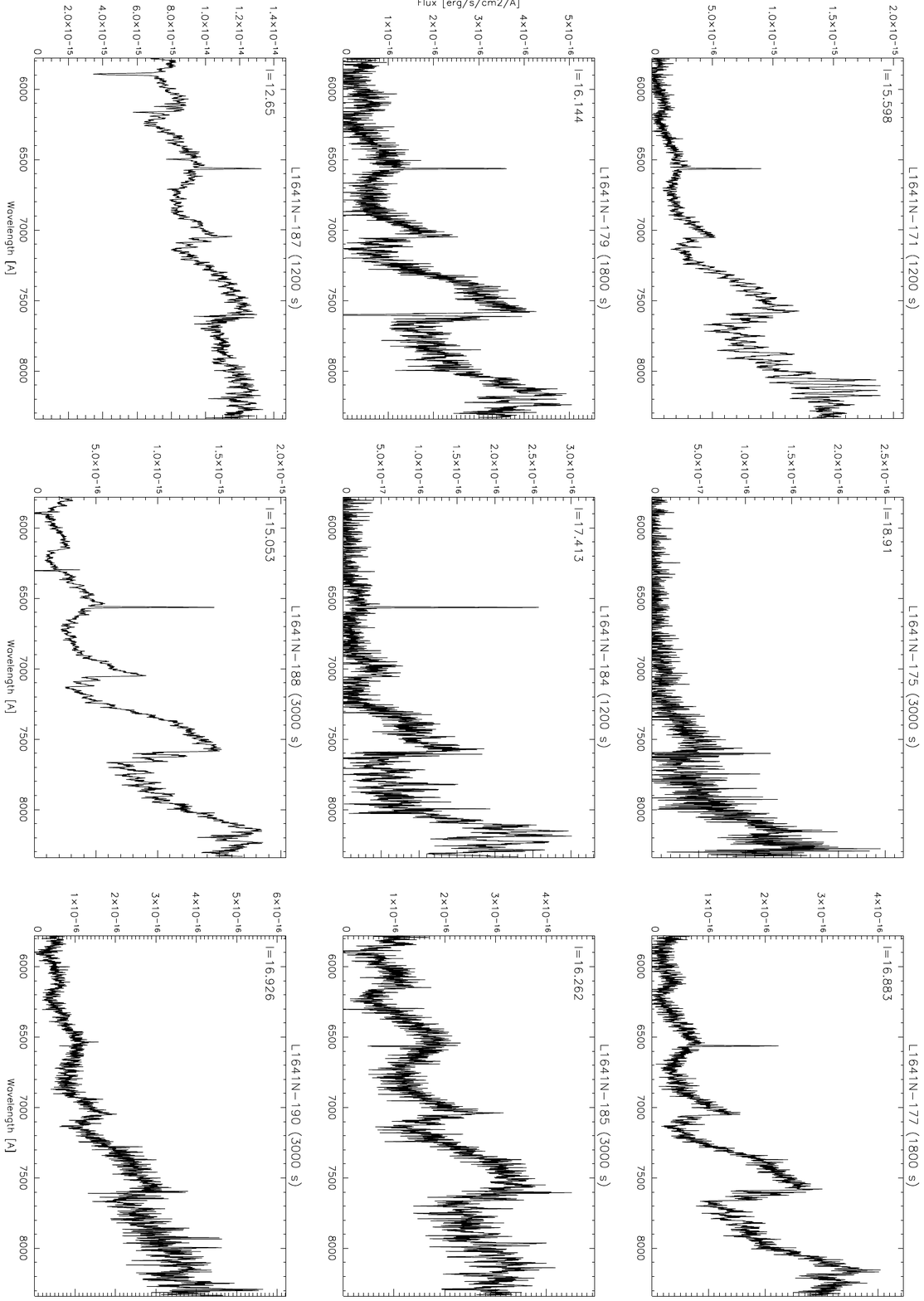}
\end{figure*}
\begin{figure*}
	\centering
	\includegraphics[width=17cm]{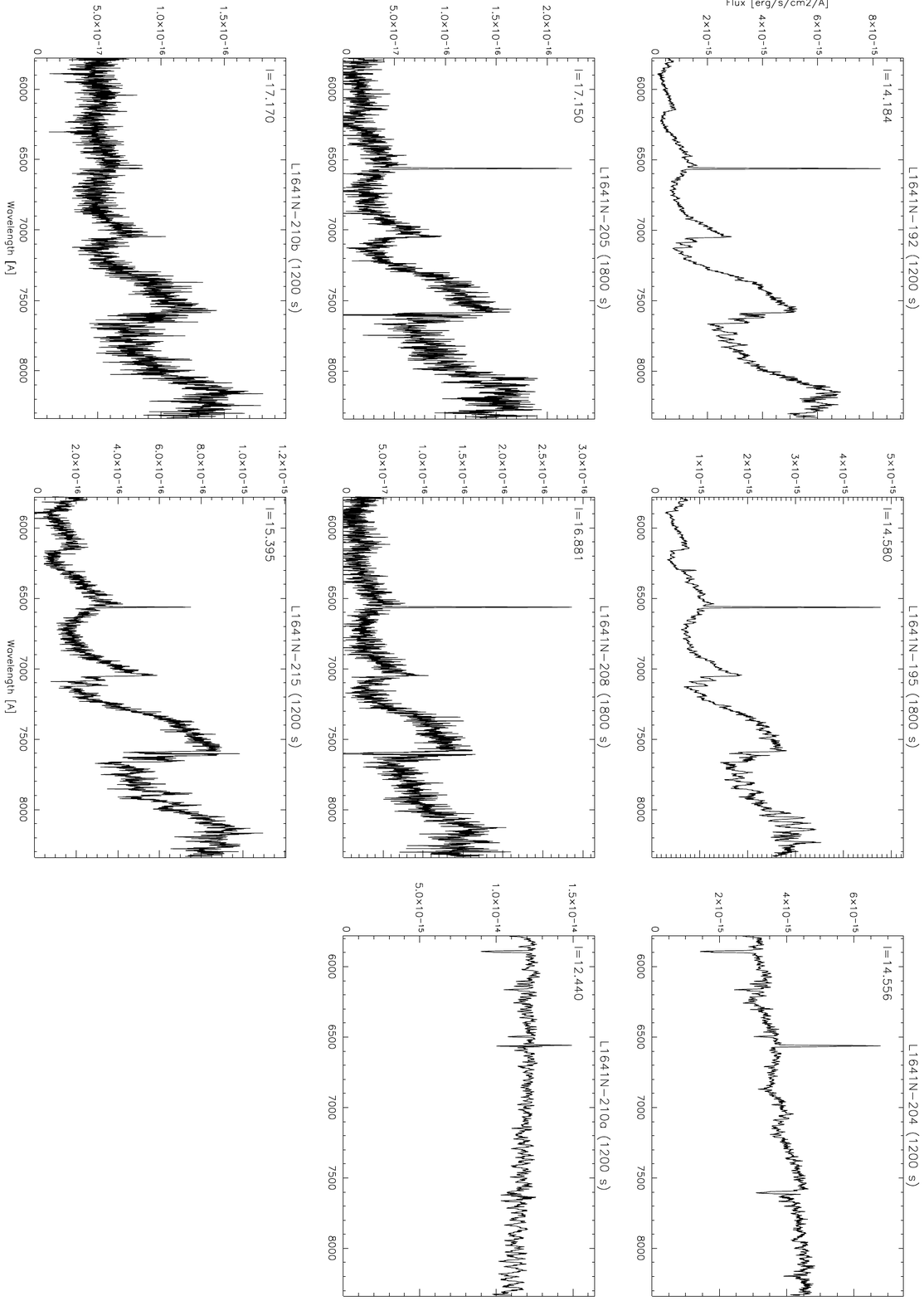}
\end{figure*}

\begin{figure*}
	\centering
	\includegraphics[width=17cm]{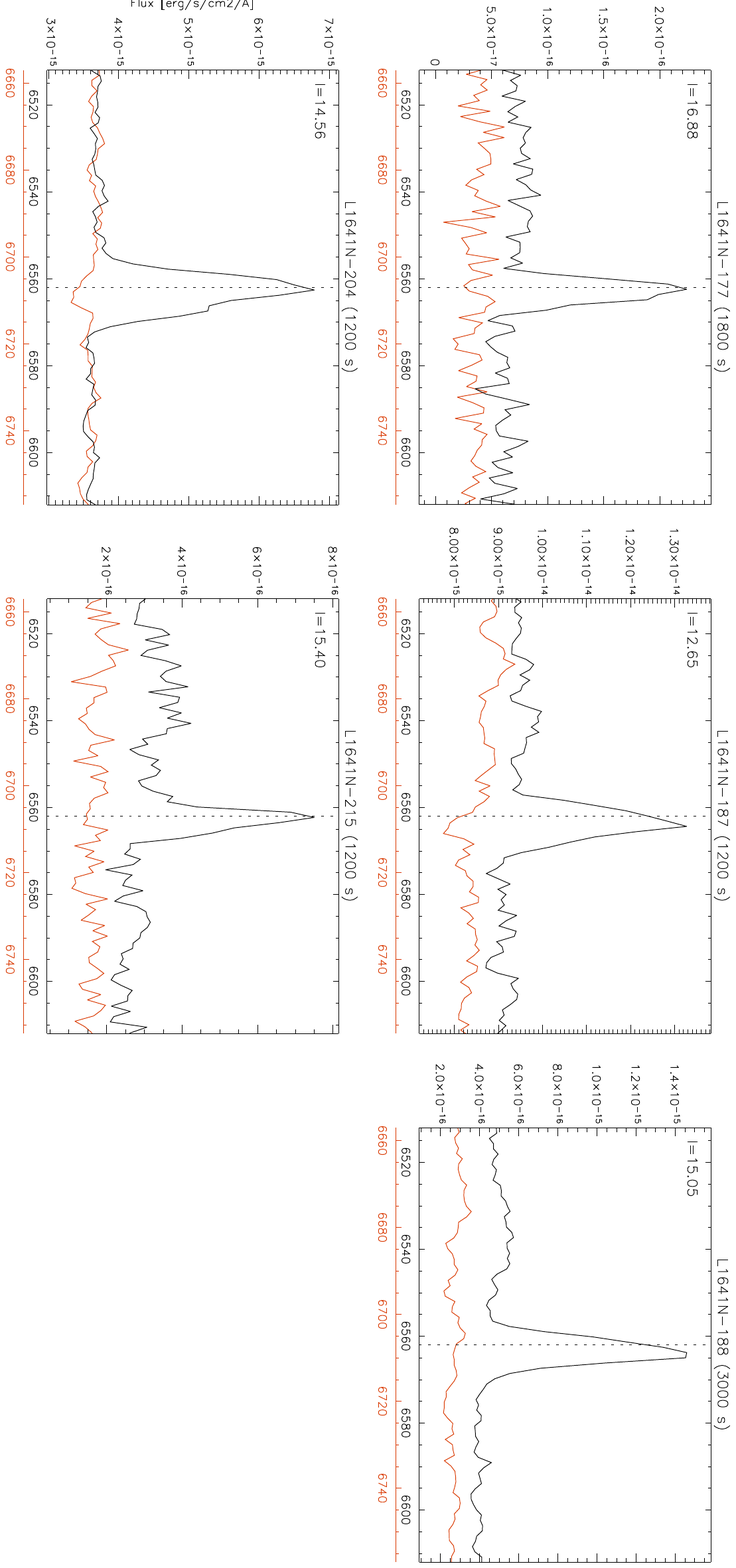}
\end{figure*}


\begin{thebibliography}{}

\bibitem[2007]{ahmic} Ahmic M., Jayawardhana R., Brandeker A., et al., 2007, ApJ 671, 2074

\bibitem[2004]{ali} Ali B., Noriega-Crespo A., 2004, ApJ 613, 374

\bibitem[2000]{allens} Allen's astrophysical quantities, 2000, 4th edition. Editor Cox A.N.


\bibitem[2000]{biviano} Biviano A., Sauvage M., Gallais P., et al., 2000, Experimental
			Astronomy, v. 10, Issue 2/3, p. 255-277

\bibitem[2003]{blommaert} Blommaert J., et al., 2003, The ISO Handbook, Vol. II: CAM—-The ISO Camera, ver. 2.0 (ESA SP-1262; Noordwijk: ESA)

\bibitem[2001]{bontemps} Bontemps S., Andr\'{e} P., Kaas A.A., et al., 2001, A\&A 372, 173

\bibitem[1994]{brown} Brown A.G.A., de Geus E.J., de Zeeuw P.T., 1994, A\&A 289, 101

\bibitem[2006]{bouy} Bouy H., Moraux E., Bouvier J., et al., 2006, ApJ 637, 1056

\bibitem[1993]{chen93} Chen H., Tokunaga T., Strom K.M., Hodapp K.W., 1993, ApJ 407, 639
\bibitem[1995]{chen95} Chen H., Zhao J.H., Ohashi N., 1995, ApJ 450, L71
\bibitem[1996]{chen96} Chen H., Ohashi N., Umemoto T., 1996, AJ 112, 717

\bibitem[2000]{coulais} Coulais A., Abergel A., 2000, A\&AS 141, 533

\bibitem[2003-2006]{cutri} Cutri R.M., Skrutskie M.F., Van Dyk, S., et al, 2003-2006, Explanatory Supplement to the 2MASS All Sky Data Release, www.ipac.caltech.edu/2mass/releases/allsky/doc/explsup.html


\bibitem[2000]{delaney} Delaney M.(ed.), 2000, ISOCAM Interactive Analysis User's Manual, Version 
			4.0, ESA Document, Reference Number SAI/96-5226/Dc

\bibitem[2003]{delgado} Delgado-Donate E.J., Clarke C.J., Bate M.R., 2003, MNRAS 342, 926

\bibitem[2003]{draine} Draine B.T., 2003, ARA\&A 41, 241.

\bibitem[2007]{dotter} Dotter A., Chaboyer B., Jevremovi\'{c} D., et al., 2007, AJ 134, 376

\bibitem[1991]{duquennoy} Duquennoy A., Mayor M., 1991, A\&A 248, 485

\bibitem[2008]{elmegreen1} Elmegreen B.G., 2008, arXiv:0803.3154
\bibitem[2008]{elmegreen2} Elmegreen B.G., Klessen R.S., Wilson C.D., 2008, arXiv:0803.4411

\bibitem[1986]{fukui86} Fukui Y., Sugitani K., Takaba H., et al., 1986, ApJ 311, L85
\bibitem[1988]{fukui88} Fukui Y., Takaba H., Iwata T., et al. 1988, ApJ 325, L13

\bibitem[2005]{notcamdist} G\aa lfalk M., 2005, NOT Annual report 2004, p18-19
\bibitem[2004]{galfalk} G\aa lfalk M., Olofsson G., Kaas A.A., et al., 2004, A\&A 420, 945
\bibitem[2007]{galfalk07} G\aa lfalk M., Olofsson G., 2007, A\&A 466, 579

\bibitem[2006]{harvey} Harvey P.M., Chapman N., Lai S.-P., et al., 2006, ApJ 644, 307

\bibitem[1993]{hodapp93} Hodapp K.W., Deane J., 1993, ApJS 88, 119

\bibitem[2006]{jorgensen} J\o rgensen J.K, Harvey P.M., Evans N.J.II, et al., 2006, ApJ 645, 1246

\bibitem[1999]{kaas99} Kaas A.A., Olofsson G., Bontemps S., et al., 1999, in: The Universe
		     as Seen by ISO. Eds. P.Cox \& M.F.Kessler. ESA-SP 427

\bibitem[2004]{kaas04} Kaas A.A., Olofsson G., Bontemps S., et al., 2004, A\&A 421, 623

\bibitem[1995]{kenyon} Kenyon S.J., Hartmann L., 1995, ApJS 101, 117

\bibitem[1998]{lejeune98} Lejeune T., Cuisinier F., Buser R., 1998, A\&AS 130, 65
\bibitem[2002]{lejeune02} Lejeune T., 2002, ASP Conference Series, Vol. 274, p159-165

\bibitem[1996]{lutz} Lutz D., Feuchtgruber H., Genzel R., et al., 1996, A\&A 315, L269

\bibitem[1999]{mader} Mader S.L., Zealey W.J., Parker Q.A et al., 1999, MNRAS 310, 331

\bibitem[2002]{madsen} Madsen S., Dravins D., Lindegren L., 2002, A\&A 381, 446

\bibitem[2003]{marchal} Marchal L., Delfosse X., Forveille T., et al., 2003, in SF2A-2003: Semaine de l'Astrophysique Francaise

\bibitem[1999]{olofsson} Olofsson G., Huldtgren M., Kaas A.A., et al., 1999, A\&A 350, 883

\bibitem[1997]{ott} Ott S., Abergel A., Altieri B., et al., 1997, In: Hunt G., Payne H.E. (eds.) 
		    Astronomical Data Analysis Software and Systems. ASP Conf. Ser. 125, 34

\bibitem[2007]{palla} Palla F., Randich S., Pavlenko Ya.V., et al., ApJ 659, L41

\bibitem[2000]{persi} Persi P., Marenzi A.R., Olofsson G., et al., 2000, A\&A 357, 219

\bibitem[2005]{reach} Reach W.T., Megeath T., Martin C., et al., 2005, PASP 117, 978

\bibitem[1998]{reipurth} Reipurth B., Devine D., Bally J., 1998, AJ 116, 1396
\bibitem[2000]{reipurth00} Reipurth B., 2000, AJ, 120, 3177
\bibitem[2001]{reipurth_HHreview} Reipurth B., Bally J., 2001, Annu.Rev. A\&A 39, 403

\bibitem[1997]{sakamoto} Sakamoto S., Hasegawa T., Hayashi M., et al, 1997, ApJ 481, 302

\bibitem[1998]{stanke} Stanke T., McCaughrean M.J., Zinnecker H., 1998, A\&A 332, 307 
\bibitem[2000]{stanke2000} Stanke T., McCaughrean M.J., Zinnecker H., 2000, A\&A 355, 639
\bibitem[2007]{stanke07} Stanke T., Williams J.P., 2007, AJ 133, 1307

\bibitem[1996]{starck} Starck J.L., Murtagh F., Pirenne B., Albrecht M., 1996, PASP 108, 446

\bibitem[2003]{sterzik} Sterzik M.F. \& Durisen R.H., 2003, A\&A 400, 1031

\bibitem[1989]{strom89} Strom K., Margulis M., Strom S., 1989, ApJ 346, L33

\bibitem[1998]{valenti} Valenti J.A., Piskunov N., Johns-Krull C.M., 1998, ApJ 498, 851

\bibitem[1992]{wainscoat} Wainscoat R., Cohen M., Volk K., 1992, ApJS 83, 111

\end{thebibliography}
\end{document}